\documentclass{IEEEtran}
\usepackage[T1]{fontenc}
\usepackage[latin9]{inputenc}
\usepackage{color}
\usepackage{amsmath}
\usepackage{amsthm}
\usepackage{amssymb}
\usepackage{graphicx}
\usepackage[unicode=true,
 bookmarks=true,bookmarksnumbered=true,bookmarksopen=true,bookmarksopenlevel=3,
 breaklinks=false,pdfborder={0 0 0},pdfborderstyle={},backref=false,colorlinks=true]
 {hyperref}
\hypersetup{pdftitle={Your Title},
 pdfauthor={Your Name},
 pdfborderstyle={},pdfborderstyle={},pdfpagelayout=OneColumn,pdfnewwindow=true,pdfstartview=XYZ,plainpages=false,linkcolor=blue,urlcolor=blue,citecolor=red,anchorcolor=blue,linkcolor=blue,urlcolor=blue,citecolor=red,anchorcolor=blue}

\makeatletter
\theoremstyle{plain}
\newtheorem{thm}{\protect\theoremname}
\theoremstyle{definition}
\newtheorem{defn}[]{\protect\definitionname}
\theoremstyle{plain}
\newtheorem{lem}[]{\protect\lemmaname}
\theoremstyle{remark}
\newtheorem{rem}[]{\protect\remarkname}
\theoremstyle{plain}
\newtheorem{cor}[]{\protect\corollaryname}

\usepackage{bm}

\usepackage{subfigure}
\usepackage{epstopdf}
\usepackage{float}
\usepackage{subfloat}
\usepackage{array}
\usepackage{tabularx}
\usepackage{multirow}
\usepackage{tikz}
\usepackage{algorithmic}

\allowdisplaybreaks

\tikzstyle{arw}=[->,>=latex]
\tikzstyle{node}=[draw,rectangle,rounded corners, minimum width=1cm,minimum height =.75 cm]

\usepackage{color}

\usepackage{adjustbox}

\usepackage{algorithmic}

\providecommand{\corollaryname}{Corollary}
\providecommand{\definitionname}{Definition}
\providecommand{\lemmaname}{Lemma}

\providecommand{\theoremname}{Theorem}

\providecommand{\corollaryname}{Corollary}
\providecommand{\definitionname}{Definition}
\providecommand{\lemmaname}{Lemma}
\providecommand{\remarkname}{Remark}
\providecommand{\theoremname}{Theorem}

\makeatother

\providecommand{\corollaryname}{Corollary}
\providecommand{\definitionname}{Definition}
\providecommand{\lemmaname}{Lemma}
\providecommand{\remarkname}{Remark}
\providecommand{\theoremname}{Theorem}

\begin{document}

\title{Distortion Bounds for Source Broadcast Problems}

\author{Lei Yu, Houqiang Li, \IEEEmembership{Senior Member,~IEEE} and Weiping
Li, \IEEEmembership{Fellow,~IEEE} \thanks{Manuscript received May 22, 2016; revised March 17, 2018; accepted
June 5, 2018. This work was supported in part by NSFC under Grant
61390514, in part by the 973 Program under Grant 2015CB351803, and
in part by NSFC under Grant 61631017. This paper was presented in
part at IEEE ISIT 2016 \cite{Yu2016}.} \thanks{L. Yu is with the Department of Electrical and Computer Engineering,
National University of Singapore, 117583 Singapore (e-mail: leiyu@nus.edu.sg).
H. Li and W. Li are with the Department of Electronic Engineering
and Information Science, University of Science and Technology of China,
Hefei 230027, China (e-mail: \{lihq,wpli\}@ustc.edu.cn).} \thanks{Communicated by S. S. Pradhan, Associate Editor for Shannon Theory.}
\thanks{Copyright (c) 2017 IEEE. Personal use of this material is permitted.
However, permission to use this material for any other purposes must
be obtained from the IEEE by sending a request to pubs-permissions@ieee.org. }}
\maketitle
\begin{abstract}
This paper investigates the joint source-channel coding problem of
sending a memoryless source over a memoryless broadcast channel. An
inner bound and several outer bounds on the admissible distortion
region are derived, which respectively generalize and unify several
existing bounds. As a consequence, we also obtain an inner bound and
an outer bound for the degraded broadcast channel case. When specialized
to the Gaussian or binary source broadcast, the inner bound and outer
bound not only recover the best known inner bound and outer bound
in the literature, but also generate some new results. Besides, we
also extend the inner bound and outer bounds to the Wyner-Ziv source
broadcast problem, i.e., source broadcast with side information available
at decoders. Some new bounds are obtained when specialized to the
Wyner-Ziv Gaussian and Wyner-Ziv binary cases.
\end{abstract}


\section{Introduction}

As stated in Shannon's source-channel separation theorem \cite{Shannon48},
cascading source coding and channel coding does not lose the optimality
for point-to-point communication systems. This separation theorem
does not only suggest a simple system architecture in which source
coding and channel coding are separated by a universal digital interface,
but also guarantees that such architecture does not incur any asymptotic
performance loss. Consequently, it forms the basis of the architecture
of today's communication systems. However, for many multi-user communication
systems, the optimality of such a separation does not hold any more
\cite{Goblick65,Gastpar03}. Therefore, an increasing amount of literature
focus on joint source-channel coding (JSCC) in multi-user setting.

One of the most classical problems in this area is JSCC of transmitting
a Gaussian source over a $K$-user Gaussian broadcast channel with
average transmitting power constrained. Goblick \cite{Goblick65}
observed that when the source bandwidth and the channel bandwidth
are matched (i.e., one channel use per source sample) linear uncoded
transmission (symbol-by-symbol mapping) is optimal. However, the optimality
of such a simple linear scheme cannot be extended to the bandwidth
mismatch case. One way to approximately characterize the admissible
distortion region is finding its inner bound and outer bound. For
inner bound, analog coding or hybrid coding has been studied in a
vast body of literature \cite{Gastpar03,Shannon49,Shamai98,Prabhakaran11}.
For 2-user Gaussian broadcast communication, Prabhakaran \emph{et
al.} \cite{Prabhakaran11} gave the best known inner bound, which
is achieved by a hybrid digital-analog (HDA) scheme. On the other
hand, Reznic \emph{et al.} \cite{Reznic06} derived a non-trivial
outer bound (tighter than the single-user bound) for 2-user Gaussian
broadcast problem with bandwidth expansion (i.e., more than one channel
uses per source sample) by introducing an auxiliary random variable
(or a remote source). Tian \emph{et al.} \cite{Tian11} extended this
outer bound to the $K$-user case by introducing more than one auxiliary
random variables. Similar to the results of Reznic \emph{et al.},
the outer bound given by Tian \emph{et al.} is also nontrivial only
for the bandwidth expansion case \cite{Yu2015comments}. Beyond broadcast
communication, Minero \emph{et al.} \cite{Minero} considered sending
memoryless correlated sources over a memoryless multi-access channel,
and derived an inner bound using a unified framework of hybrid coding.
Lee \emph{et al.} \cite{Lee} derived a unified achievability result
for memoryless network communication.

Besides, in \cite{Shamai98,Nayak,Gao} the Wyner-Ziv source communication
problem was investigated, in which side information correlated with
the source is available at decoder(s). Shamai \emph{et al.} \cite{Shamai98}
studied the problem of sending a Wyner-Ziv source over a \emph{point-to-point}
channel, and proved that for such communication system, the separate
coding (which cascades Wyner-Ziv coding with channel coding) does
not incur any loss of optimality. Nayak \emph{et al.} \cite{Nayak}
and Gao \emph{et al.} \cite{Gao} investigated the Wyner-Ziv source
\emph{broadcast} problem, and obtained the single-user outer bound
by simply applying the cut-set bound (the minimum distortion achieved
in point-to-point setting) for each receiver.

In this paper, we consider JSCC of transmitting a memoryless source
over a $K$-user memoryless broadcast channel, and derive an inner
bound and two outer bounds on the admissible distortion region. The
inner bound is derived by using a unified framework of hybrid coding
inspired by \cite{Minero}, and the outer bounds are derived by introducing
remote sources  at the sender side or introducing remote channels
at receiver sides. The proof method of introducing remote sources
at the sender side can be found in \cite{Reznic06} and \cite{Tian11},
and hence it is not new. However, the introducing remote channels
method is different from existing methods. The existing method of
introducing auxiliary random variables at receiver sides, named the
genie-aided method, can be found in \cite[Section 6.4.3]{Tian11,Khezeli14,El Gamal}.
Both the genie-aided method and the introducing remote channels method
convert the original communication system into a new one. However,
the genie-aided method constructs a  ``stronger'' network such
that the original network is a degraded version of it, while the introducing
remote channels method  constructs a  degraded (``weaker'') version
of the original network. Hence our introducing remote channels method
is different from the existing genie-aided method. Furthermore, our distortion
bounds are generalizations and unifications of several existing bounds
in the literature. Besides, as a consequence, we also obtain an inner
bound and an outer bound for the degraded broadcast channel case.
When specialized to the Gaussian source broadcast and binary source
broadcast, our inner bound could recover the best known performance
achieved by hybrid coding, and our outer bound could recover the best
known outer bounds given by Tian \emph{et al.} \cite{Tian11} and
Khezeli \emph{et al. }\cite{Khezeli14}. Moreover, for these cases,
our bounds can be also used to generate some new results. Besides,
we also extend the inner bound and outer bounds to the Wyner-Ziv source
broadcast problem, i.e., source broadcast with side information at
decoders. When specialized to the Wyner-Ziv Gaussian and binary cases,
our bounds reduce to some new bounds.

The rest of this paper is organized as follows. Section II summarizes
basic notations, definitions and preliminaries, and formulates the
problem. Section III gives the main results for the source broadcast
problem, including the discrete, discrete and degraded, binary, and
Gaussian cases. Section IV extends the results to the Wyner-Ziv source
broadcast problem. Finally, Section V gives the concluding remarks.

\section{Problem Formulation and Preliminaries}

\subsection{Notation}

Throughout this paper, we follow the notation in \cite{El Gamal}.
For example, for a discrete random variable $X\sim p_{X}$ on alphabet
$\mathcal{X}$ and $\epsilon\in\left(0,1\right)$, the set of $\epsilon$-typical
$n$-sequences $x^{n}$ (or the typical set in short) is defined as
\begin{align*}
 & \mathcal{T}_{\epsilon}^{\left(n\right)}\left(X\right)\\
 & =\left\{ x^{n}:\left|\frac{|\{i:x_{i}=x\}|}{n}-p_{X}(x)\right|\leq\epsilon p_{X}(x),\forall x\in\mathcal{X}\right\} .
\end{align*}
When it is clear from the context, we will use $\mathcal{T}_{\epsilon}^{\left(n\right)}$
to denote $\mathcal{T}_{\epsilon}^{\left(n\right)}\left(X\right)$.

In addition, we use $X_{\mathcal{A}}$ to denote the vector $(X_{j}:j\in\mathcal{A})$,
use $[i:j]$ to denote the set $\left\{ \left\lfloor i\right\rfloor ,\left\lfloor i\right\rfloor +1,\cdots,\left\lfloor j\right\rfloor \right\} $,
and use $\mathbf{1}$ to denote an all-one vector (similarly, use
$\mathbf{2}$ to denote an all-2 vector). We say vector $m_{[1:N]}$
is smaller than vector $m'_{[1:N]}$ if $m_{j}=m'_{j}$ for $k<j\leq K$
and $m_{k}<m'_{k}$ for some $k$. For two vectors $m_{\mathcal{I}}$
and $m_{\mathcal{I}}^{\prime}$, we say $m_{\mathcal{I}}$ is component-wise
unequal to $m_{\mathcal{I}}^{\prime}$, if $m_{i}\neq m_{i}^{\prime}$
for all $i\in\mathcal{I}$, and denote it as $m_{\mathcal{I}}\nLeftrightarrow m_{\mathcal{I}}^{\prime}$.
Besides, we use $1\left\{ \mathcal{A}\right\} $ to denote the indicator
function of an event $\mathcal{A}$, i.e.,
\[
1\left\{ \mathcal{A}\right\} =\begin{cases}
1, & \textrm{if }\mathcal{A}\textrm{ is true};\\
0, & \textrm{if }\mathcal{A}\textrm{ is false}.
\end{cases}
\]

We use $\exp$ and $\log$ to respectively denote the exponential
and logarithmic functions with the base $2$.

\subsection{Problem Formulation}

\begin{figure}[t]
\centering\includegraphics[width=0.95\columnwidth]{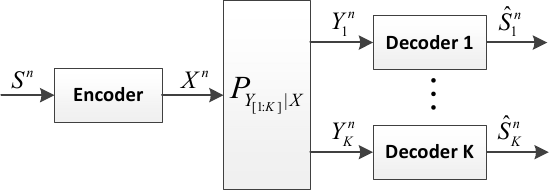} \caption{\label{fig:broadcast communication system }Source broadcast system. }
\end{figure}

Consider the source broadcast system shown in Fig. \ref{fig:broadcast communication system }.
A discrete memoryless source (DMS) $S^{n}$ is first coded into $X^{n}$
using a source-channel code, then transmitted to $K$ receivers through
a discrete memoryless broadcast channel (DM-BC) $p_{Y_{[1:K]}|X}$,
and finally, the receiver $k$ produces a source reconstruction $\hat{S}_{k}^{n}$
from the received signal $Y_{k}^{n}$.

\begin{defn}[Source]
\label{def:source} A discrete memoryless source (DMS) is specified
by a probability mass function (pmf) $p_{S}$ on a finite alphabet
${\mathcal{S}}$. The DMS $p_{S}$ generates an i.i.d. random process
$\left\{ S_{i}\right\} $ with $S_{i}\sim p_{S}$.
\end{defn}
\begin{defn}[Broadcast Channel]
\label{def:DBC}A $K$-user discrete memoryless broadcast channel
(DM-BC) is specified by a collection of conditional pmfs $p_{Y_{[1:K]}|X}$
on a finite output alphabet ${\mathcal{Y}_{1}\times\cdots\times\mathcal{Y}_{K}}$
for each $x$ in a finite input alphabet $\mathcal{X}$.
\end{defn}
\begin{defn}[Degraded Broadcast Channel]
A DM-BC $p_{Y_{[1:K]}|X}$ is stochastically degraded (or simply
degraded) if there exists a random vector $\tilde{Y}_{[1:K]}$ such
that $p_{\tilde{Y}_{k}|X}=p_{Y_{k}|X},1\leq k\leq K$ (i.e., $\tilde{Y}_{[1:K]}$
has the same conditional marginal pmfs as $Y_{[1:K]}$ given $X$),
and $X\rightarrow\tilde{Y}_{K}\rightarrow\tilde{Y}_{K-1}\rightarrow\cdots\rightarrow\tilde{Y}_{1}$\footnote{For brevity, the Markov chain is assumed in this direction.This differs
from that in the conference version \cite{Yu2016}.} form a Markov chain. In addition, as a special case, if $X\rightarrow Y_{K}\rightarrow Y_{K-1}\rightarrow\cdots\rightarrow Y_{1}$,
i.e., $\tilde{Y}_{k}=Y_{k},1\le k\le K$, then $p_{Y_{[1:K]}|X}$
is physically degraded.
\end{defn}
\begin{defn}
An $n$-length source-channel code is defined by a encoding function
$x^{n}:{\mathcal{S}}^{n}\mapsto{\mathcal{X}}^{n}$ and $K$ decoding
functions $\hat{s}_{k}:\mathcal{Y}_{k}^{n}\mapsto\hat{\mathcal{S}_{k}^{n}},1\leq k\leq K$,
where $\hat{{\mathcal{S}_{k}}}$ is the alphabet of source reconstruction
at the receiver $k$.
\end{defn}
For any $n$-length source-channel code, the induced distortion is
defined as
\begin{equation}
\mathbb{E}d_{k}\left(S^{n},\hat{S}_{k}^{n}\right)=\frac{1}{n}\sum_{t=1}^{n}\mathbb{E}d_{k}\left(S_{t},\hat{S}_{k,t}\right),
\end{equation}
for $1\le k\le K$, where $d_{k}\left(s,\hat{s}_{k}\right):{\mathcal{S}}\times\hat{{\mathcal{S}_{k}}}\mapsto\left[0,+\infty\right]$
is a distortion measure function for the receiver $k$.
\begin{defn}
For transmitting a source $S$ over a channel $p_{Y_{[1:K]}|X}$,
if there exists a sequence of source-channel codes such that
\begin{equation}
\mathop{\limsup}\limits _{n\to\infty}\mathbb{E}d_{k}\left(S^{n},\hat{S}_{k}^{n}\right)\le D_{k},
\end{equation}
then we say that the distortion tuple $D_{[1:K]}$ is achievable.
\end{defn}
\begin{defn}
For transmitting a source $S$ over a channel $p_{Y_{[1:K]}|X}$,
the admissible distortion region is defined as
\begin{align}
\mathcal{D}\triangleq & \left\{ D_{[1:K]}:D_{[1:K]}\textrm{ is achievable}\right\} .
\end{align}
\end{defn}
The admissible distortion region $\mathcal{R}$ only depends on the
marginal distributions of $p_{Y_{[1:K]}|X}$, hence for the stochastically
degraded channel case, it suffices to only consider the corresponding
physically degraded channel case.

In addition, Shannon's source-channel separation theorem shows that
the minimum distortion for transmitting a source over a point-to-point
channel satisfies
\begin{equation}
R_{k}\left(D_{k}\right)=C_{k},
\end{equation}
where $R_{k}\left(\cdotp\right)$ is the rate-distortion function
of the source with the distortion measure $d_{k}$, and $C_{k}$ is
the capacity for the receiver $k$. Therefore, the optimal distortion
(Shannon limit) is
\begin{equation}
D_{k}^{*}=R_{k}^{-1}\left(C_{k}\right).\label{eq:-28-1}
\end{equation}
Obviously,
\begin{equation}
\mathcal{D}\subseteq\mathcal{D}^{*}\triangleq\left\{ D_{[1:K]}:D_{k}\geq D_{k}^{*},1\le k\le K\right\} ,\label{eq:trivialouterbound}
\end{equation}
where $\mathcal{D}^{*}$ is named \emph{single-user outer bound}.

In the system above, the source bandwidth and channel bandwidth are
matched. In this paper, we also consider the communication system
with bandwidth mismatch, whereby $m$ samples of a DMS are transmitted
through $n$ uses of a DM-BC. For this case, the bandwidth mismatch
factor is defined as $b=\frac{n}{m}$.

\subsection{Multivariate Covering/Packing Lemma}

We first introduce the following multivariate covering and packing
lemmas which are important in the achievability part in this work.

Let $(U,V_{[0:k]})\sim p_{U,V_{[0:k]}}$. Let $\left(U^{n},V_{0}^{n}\right)\sim p_{U^{n},V_{0}^{n}}$
be a random vector sequence. For each $j\in[1:k]$, let $\mathcal{A}_{j}\subseteq[1:j-1]$.
Assume $\mathcal{A}_{j}$ satisfies if $i\in\mathcal{A}_{j}$, then
$\mathcal{A}_{i}\subseteq\mathcal{A}_{j}$. For each $j\in[1:k]$
and each $m_{\mathcal{A}_{j}}\in\prod_{i\in\mathcal{A}_{j}}[1:2^{nr_{i}}]$,
let $V_{j}^{n}(m_{\mathcal{A}_{j}},m_{j}),m_{j}\in[1:2^{nr_{j}}],$
be pairwise conditionally independent random sequences, each distributed
according to $\prod_{i=1}^{n}p_{V_{j}|V_{\mathcal{A}_{j}},V_{0}}(v_{j,i}|v_{\mathcal{A}_{j},i}(m_{\mathcal{A}_{j}}),v_{0,i})$.
Hence for each $j\in[1:k]$, $\mathcal{A}_{j}\cup\left\{ 0\right\} $
denotes the index set of the random variables on which the codeword
$V_{j}^{n}$ is superposed. Based on the notations above, we have
the following multivariate covering and packing lemmas.
\begin{lem}[Multivariate Covering Lemma]
\label{lem:Covering} Let $\epsilon'<\epsilon$. If $\lim_{n\rightarrow\text{\ensuremath{\infty}}}\mathbb{P}\left(\left(U^{n},V_{0}^{n}\right)\in\mathcal{T}_{\epsilon'}^{\left(n\right)}\right)=1$,
then there exists $\delta(\epsilon)$ that tends to zero as $\epsilon\rightarrow0$
such that
\begin{align}
 & \lim_{n\rightarrow\text{\ensuremath{\infty}}}\mathbb{P}\left((U^{n},V_{0}^{n},V_{[1:k]}^{n}(m_{[1:k]}))\in\mathcal{T}_{\epsilon}^{\left(n\right)}\textrm{ for some }m_{[1:k]}\right)\nonumber \\
 & =1,
\end{align}
if $\sum_{j\in\mathcal{J}}r_{j}>\sum_{j\in\mathcal{J}}H\left(V_{j}|V_{\mathcal{A}_{j}}V_{0}\right)-H\left(V_{\mathcal{J}}|V_{0}U\right)+\delta(\epsilon)$
for all $\mathcal{J}\subseteq[1:k]$ such that $\mathcal{J}\neq\emptyset$
and if $j\in\mathcal{J}$, then $\mathcal{A}_{j}\subseteq\mathcal{J}$.
\end{lem}
\begin{lem}[Multivariate Packing Lemma]
\label{lem:Packing} There exists $\delta(\epsilon)$ that tends
to zero as $\epsilon\rightarrow0$ such that
\begin{align}
 & \lim_{n\rightarrow\text{\ensuremath{\infty}}}\mathbb{P}\left((U^{n},V_{0}^{n},V_{[1:k]}^{n}(m_{[1:k]}))\in\mathcal{T}_{\epsilon}^{\left(n\right)}\textrm{ for some }m_{[1:k]}\right)\nonumber \\
 & =0,
\end{align}
if $\sum_{j\in\mathcal{J}}r_{j}<\sum_{j\in\mathcal{J}}H\left(V_{j}|V_{\mathcal{A}_{j}}V_{0}\right)-H\left(V_{\mathcal{J}}|V_{0}U\right)-\delta(\epsilon)$
for some $\mathcal{J}\subseteq[1:k]$ such that $\mathcal{J}\neq\emptyset$
and if $j\in\mathcal{J}$, then $\mathcal{A}_{j}\subseteq\mathcal{J}$.
\end{lem}
Note that all the existing covering and packing lemmas such as \cite[Lem. 8.2]{El Gamal}
and \cite[Lem. 4]{Shayevitz}, involve only single-layer codebook.
Our multivariate covering and packing lemmas generalize them to the
case of multilayer codebooks.

\section{Source Broadcast }

\subsection{Discrete Memoryless Broadcast }

\begin{figure}[t]
\centering \subfigure[]{\includegraphics[width=0.35\columnwidth]{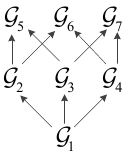}
}\centering \qquad{}\subfigure[]{\includegraphics[width=0.35\columnwidth]{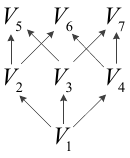}}

\caption{\label{fig:codebook}(a) The inclusion relation of the sets $\mathcal{G}_{j},1\leq j\leq N$
with $K=3$; (b) The structure of the codebook for the unified hybrid
coding with $K=3$. }
\end{figure}

Now, we bound the distortion region for the source broadcast communication.
To write the inner bound, for $1\leq j\leq N\triangleq2^{K}-1$, we
first introduce an auxiliary random variable $V_{j}$ for each of
the $2^{K}-1$ nonempty subsets $\mathcal{G}_{j}\subseteq[1:K]$,
and let $V_{j}$ denote a common message transmitted from the sender
to all the receivers in $\mathcal{G}_{j}$. The $V_{j}$ corresponds
to a subset $\mathcal{G}_{j}$ by the following one-to-one mapping.
Sort all the nonempty subsets $\mathcal{G}_{j}\subseteq[1:K]$ in
the decreasing order\footnote{We say a set $\mathcal{G}$ is larger than another $\mathcal{H}$
if $|\mathcal{G}|>|\mathcal{H}|$, or $|\mathcal{G}|=|\mathcal{H}|$
and there exists some $1\leq i\leq|\mathcal{G}|$ such that $\mathcal{G}\left[i\right]>\mathcal{H}\left[i\right]$
and $\mathcal{G}\left[l\right]=\mathcal{H}\left[l\right]$ for all
$1\leq l\leq i-1$, where $\mathcal{G}\left[i\right]$ (or $\mathcal{H}\left[i\right]$)
denotes the $i$th largest element in $\mathcal{G}$ (or $\mathcal{H}$). }. Map the $j$th subset in the resulting sequence to $j$. Obviously
this mapping is one-to-one corresponding. For example, if $K=3$,
then $\mathcal{G}_{1}=\left\{ 1,2,3\right\} ,\mathcal{G}_{2}=\left\{ 2,3\right\} ,\mathcal{G}_{3}=\left\{ 1,3\right\} ,\mathcal{G}_{4}=\left\{ 1,2\right\} ,\mathcal{G}_{5}=\left\{ 3\right\} ,\mathcal{G}_{6}=\left\{ 2\right\} ,\mathcal{G}_{7}=\left\{ 1\right\} $;
see Fig. \ref{fig:codebook}.

Besides, let
\begin{align}
\mathcal{A}_{j} & \triangleq\left\{ i\in[1:N]:\mathcal{G}_{j}\subsetneqq\mathcal{G}_{i}\right\} ,1\leq j\leq N,\\
\mathcal{B}_{k} & \triangleq\left\{ i\in[1:N]:k\in\mathcal{G}_{i}\right\} ,1\leq k\leq K.
\end{align}
If $K=3$, then $\mathcal{A}_{1}=\emptyset,\mathcal{A}_{2}=\left\{ 1\right\} ,\mathcal{A}_{3}=\left\{ 1\right\} ,\mathcal{A}_{4}=\left\{ 1\right\} ,\mathcal{A}_{5}=\left\{ 1,2,3\right\} ,\mathcal{A}_{6}=\left\{ 1,2,4\right\} ,\mathcal{A}_{7}=\left\{ 1,3,4\right\} $
and $\mathcal{B}_{1}=\left\{ 1,3,4,7\right\} ,\mathcal{B}_{2}=\left\{ 1,2,4,6\right\} ,\mathcal{B}_{3}=\left\{ 1,2,3,5\right\} $.
Later we will show that $\mathcal{A}_{j}$ and $\mathcal{B}_{k}$
respectively correspond to the index set of the random variables on
which the codeword $V_{j}^{n}$ is superposed, and the index set of
decodable codewords $V_{j}^{n}$'s for the receiver $k$ in the proposed
hybrid coding scheme; see Appendix \ref{sub:Inner-Bound}. The decoder
$k$ is able to recover correctly the $V_{\mathcal{B}_{k}}^{n}$ with
probability approaching 1 as $n\rightarrow\infty$. In addition, it
is easy to verify that if $j\in\mathcal{B}_{k},\textrm{ then }\mathcal{A}_{j}\subseteq\mathcal{B}_{k}$.
It means that the proposed codebook satisfies that if the information
$V_{j}^{n}$ can be recovered correctly by the receiver $k$ (i.e.,
$j\in\mathcal{B}_{k}$), then $V_{\mathcal{A}_{j}}^{n}$ can also
be recovered correctly by it.

Based on the notations above, we define a distortion region (inner
bound)
\begin{align}
\mathcal{D}^{(i)}= & \bigl\{ D_{[1:K]}:\exists p_{V_{[1:N]}|S},r_{[1:N]},x\left(v_{[1:N]},s\right),\hat{s}_{k}\left(v_{\mathcal{B}_{k}},y_{k}\right),\nonumber \\
 & k\in[1:K]\text{ s.t. }\mathbb{E}d_{k}(S,\hat{S}_{k})\le D_{k},k\in[1:K],\nonumber \\
 & \sum_{j\in\mathcal{J}}r_{j}>\sum_{j\in\mathcal{J}}H\left(V_{j}|V_{\mathcal{A}_{j}}\right)-H\left(V_{\mathcal{J}}|S\right)\nonumber \\
 & \textrm{for all }\mathcal{J}\subseteq[1:N]\text{ s.t.}\mathcal{J}\neq\emptyset\textrm{ and }\mathcal{A}_{j}\subseteq\mathcal{J},\forall j\in\mathcal{J},\nonumber \\
 & \sum_{j\in\mathcal{J}^{c}}r_{j}<\sum_{j\in\mathcal{J}^{c}}H\left(V_{j}|V_{\mathcal{A}_{j}}\right)-H\left(V_{\mathcal{J}^{c}}|Y_{k}V_{\mathcal{J}}\right),\nonumber \\
 & k\in[1:K]\textrm{ for all }\mathcal{J}\subseteq\mathcal{B}_{k}\text{ s.t. }\nonumber \\
 & \mathcal{J}^{c}\triangleq\mathcal{B}_{k}\backslash\mathcal{J}\neq\emptyset\textrm{ and }\mathcal{A}_{j}\subseteq\mathcal{J},\forall j\in\mathcal{J}\bigr\}.\label{eq:-28}
\end{align}

Besides, for any positive integer $L$, define a distortion region
(outer bound achieved by introducing remote sources at the sender)
\begin{align}
\mathcal{D}_{1}^{(o)}= & \bigl\{ D_{[1:K]}:\exists p_{\hat{S}_{[1:K]}|S}\text{ s.t.}\nonumber \\
 & \mathbb{E}d_{k}\left(S,\hat{S}_{k}\right)\le D_{k},k\in[1:K],\nonumber \\
 & \text{and for any }p_{U_{[1:L]}|S},\text{ one can find }p_{\tilde{Y}_{[1:K]}\tilde{U}_{[1:L]}X}\text{ s.t.}\nonumber \\
 & I\left(\hat{S}_{\mathcal{A}};U_{\mathcal{B}}|U_{\mathcal{C}}\right)\leq I\left(Y_{\mathcal{A}};\tilde{U}_{\mathcal{B}}|\tilde{U}_{\mathcal{C}}\tilde{Y}_{\mathcal{A}}\right)\nonumber \\
 & \textrm{for any }\mathcal{A}\subseteq\left[1:K\right],\mathcal{B},\mathcal{C}\subseteq\left[1:L\right]\bigr\},
\end{align}
and another distortion region (outer bound achieved by introducing
remote channels at receivers)
\begin{align}
\mathcal{D}_{2}^{(o)}= & \bigl\{ D_{[1:K]}:\exists p_{X},\hat{s}_{k}\left(\tilde{y}_{k}\right),k\in[1:K]\text{ s.t.}\nonumber \\
 & \mathbb{E}d_{k}\left(S,\hat{S}_{k}\right)\le D_{k},k\in[1:K],\nonumber \\
 & \text{and for any }p_{U_{[1:L]}|Y_{[1:K]}},\nonumber \\
 & \text{one can find }p_{\tilde{Y}_{[1:K]}|S}p_{\tilde{U}_{[1:L]}|\tilde{Y}_{[1:K]}}\text{ s.t.}\nonumber \\
 & I\left(S;\tilde{Y}_{\mathcal{B}}\tilde{U}_{\mathcal{B}'}|\tilde{Y}_{\mathcal{C}}\tilde{U}_{\mathcal{C}'}\right)\leq I\left(X;Y_{\mathcal{B}}U_{\mathcal{B}'}|Y_{\mathcal{C}}U_{\mathcal{C}'}\right)\nonumber \\
 & \textrm{for any }\mathcal{B},\mathcal{C}\subseteq\left[1:K\right],\mathcal{B}',\mathcal{C}'\subseteq\left[1:L\right]\bigr\}.
\end{align}

Then we have the following theorem. The proof is given in Appendix
\ref{sec:broadcast}.
\begin{thm}
\label{thm:AdmissibleRegion-GBC} For transmitting a DMS $S$ over
a DM-BC $p_{Y_{[1:K]}|X}$,
\begin{equation}
\mathcal{D}^{(i)}\subseteq\mathcal{D}\subseteq\mathcal{D}_{1}^{(o)}\cap\mathcal{D}_{2}^{(o)}.
\end{equation}
\end{thm}
\begin{rem}
\label{rem:The-inner-bound}The inner bound of Theorem \ref{thm:AdmissibleRegion-GBC}
can be easily extended to Gaussian or any other well-behaved continuous-alphabet
source-channel pairs by the standard discretization method \cite[Thm. 3.3]{El Gamal}.
Moreover for these cases the outer bounds still hold. Theorem \ref{thm:AdmissibleRegion-GBC}
can be also extended to the bandwidth mismatch case, where $m$ samples
of a DMS are transmitted through $n$ uses of a DM-BC. This can be
accomplished by replacing the source and channel symbols in Theorem
\ref{thm:AdmissibleRegion-GBC} by supersymbols of lengths $m$ and
$n$, respectively. Besides, Theorem \ref{thm:AdmissibleRegion-GBC}
could be also extended to the problems of broadcasting correlated
sources (by modifying the distortion measures) and source broadcast
with channel input cost constraints (by introducing channel input
constraints).
\end{rem}
\begin{figure*}[t]
\centering\includegraphics[width=0.7\textwidth]{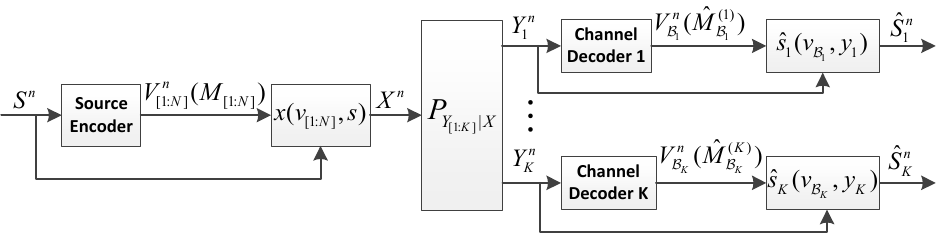}
\caption{\label{fig:hybridcoding}The unified hybrid coding used to prove the
inner bound in Theorem \ref{thm:AdmissibleRegion-GBC}. }
\end{figure*}

The inner bound $\mathcal{D}^{(i)}$ in Theorem \ref{thm:AdmissibleRegion-GBC}
is achieved by a unified hybrid coding scheme depicted in Fig. \ref{fig:hybridcoding}.
In this scheme, the codebook has a layered (or superposition) structure
(see Fig. \ref{fig:codebook}), and consists of randomly and independently
generated codewords $V_{[1:N]}^{n}(m_{[1:N]})$, $m_{[1:N]}\in\prod_{i=1}^{N}\left[1:2^{nr_{i}}\right]$,
where $r_{[1:N]}$ satisfies \eqref{eq:-28}. At the encoder side,
upon a source sequence $S^{n}$, the encoder produces digital messages
$M_{[1:N]}$ with $M_{i}$ meant for all the receivers $k$'s satisfying
$i\in\mathcal{B}_{k}$. Then, the codeword $V_{[1:N]}^{n}(M_{[1:N]})$
and the source sequence $S^{n}$ are used to generate a channel input
$X^{n}$ by a symbol-by-symbol mapping $x(v_{[1:N]},s)$. At the decoder
sides, upon the received signal $Y_{k}^{n}$, the decoder $k$ reconstructs
$M_{\mathcal{B}_{k}}$ (and also $V_{\mathcal{B}_{k}}^{n}(M_{\mathcal{B}_{k}})$)
losslessly, and then generates $\hat{S}_{k}^{n}$ by a symbol-by-symbol
mapping $\hat{s}_{k}(v_{\mathcal{B}_{k}},y_{k})$. Such a scheme could
achieve any $D_{[1:K]}$ in the inner bound $\mathcal{D}^{(i)}$.

To reveal essence of such hybrid coding, the digital transmission
part of this hybrid coding can be roughly understood as the cascade
of a $K$-user Gray-Wyner source-coding and a $K$-user Marton's broadcast
channel-coding, which share a common codebook. According to \cite[Thm. 13.3]{El Gamal},
the encoding operation of Gray-Wyner source-coding with rates $r_{[1:N]}$
is successful if $\sum_{j\in\mathcal{J}}r_{j}>\sum_{j\in\mathcal{J}}H\left(V_{j}|V_{\mathcal{A}_{j}}\right)-H\left(V_{\mathcal{J}}|S\right)\textrm{ for all }\mathcal{J}\subseteq[1:N]\textrm{ such that }\mathcal{J}\neq\emptyset\textrm{ and if }j\in\mathcal{J},\textrm{ then }\mathcal{A}_{j}\subseteq\mathcal{J}$,
and according to \cite[Thm 8.4]{El Gamal} the decoding operation
of Marton's broadcast channel-coding with rates $r_{[1:N]}$\footnote{Note that for Marton's broadcast channel-coding \cite[p. 212-213]{El Gamal},
$r_{[1:N]}$ here does not correspond to the regular broadcast-rates
$R_{[1:N]}$, but corresponds to the total rates $\widetilde{R}_{[1:N]}$
of each subcodebooks. In addition, here we only require the decoding
operation is successful, hence the condition that the chosen codewords
and the source sequence are jointly typical, which was required in
the encoding operation of Gray-Wyner source-coding, is not repeatedly
required here.} is successful if $\sum_{j\in\mathcal{J}^{c}}r_{j}<\sum_{j\in\mathcal{J}^{c}}H\left(V_{j}|V_{\mathcal{A}_{j}}\right)-H\left(V_{\mathcal{J}^{c}}|Y_{k}V_{\mathcal{J}}\right)\textrm{for all }1\le k\le K\textrm{ and for all }\mathcal{J}\subseteq\mathcal{B}_{k}\textrm{ such that }\mathcal{J}^{c}\neq\emptyset\textrm{ and if }j\in\mathcal{J},\textrm{ then }\mathcal{A}_{j}\subseteq\mathcal{J}$.
Since the proposed hybrid coding satisfies the two sufficient conditions
above, $V_{\mathcal{B}_{k}}^{n}(M_{\mathcal{B}_{k}})$ can be correctly
recovered by the receiver $k$. Note that such informal understanding
is inaccurate owing to the use of symbol-by-symbol mappings, but it
provides a rationale for our scheme. Besides, the design of such unified
hybrid coding is inspired by the hybrid coding scheme for sending
correlated sources over a multi-access channel in \cite{Minero}.

The outer bounds $\mathcal{D}_{1}^{(o)}$ and $\mathcal{D}_{2}^{(o)}$
in Theorem \ref{thm:AdmissibleRegion-GBC} are derived by introducing
auxiliary random variables $U_{[1:L]}^{n}$ at the sender side or
at  receiver sides. The proof method of introducing auxiliary random
variables (or remote sources) at sender side could be found in \cite{Tian11},
\cite[Thm. 2]{Khezeli14} and \cite[Lem. 1]{Khezeli}. In \cite{Tian11}
it was used to derive the outer bound for Gaussian source broadcast,
and in \cite[Thm. 2]{Khezeli14} and \cite[Lem. 1]{Khezeli} it was
used to derive the outer bounds for sending source over 2-user general
broadcast channel. This proof method generalizes the one used to derive
the single-user outer bound, but it does not always result in a strictly
tighter outer bound than the single-user one \cite{Yu2015comments}.
On the other hand, introducing remote channels method is different
from the existing genie-aided method \cite[Section 6.4.3]{Tian11,Khezeli14,El Gamal},
since the genie-aided method constructs a  ``stronger'' network
such that the original network is a degraded version of it, but the
introducing remote channels method  constructs a  degraded (``weaker'')
version of the original network. A deeper understanding of these
two proof methods has been given by Khezeli \emph{et al. }in \cite{Khezeli}.
$p_{\hat{S}{}_{[1:K]}|S}$ can be considered as a virtual broadcast
channel realized over physical broadcast channel $p_{Y_{[1:K]}|X}$,
and hence certain mutual informations (e.g., channel capacity region)
based on $p_{\hat{S}{}_{[1:K]}|S}$ are less than or equal to those
based on $p_{Y_{[1:K]}|X}$. This leads to the desired necessary conditions.
Besides, the necessary conditions can be also understood from the
perspective of virtual sources. $X$ and $Y_{[1:K]}$ respectively
can be considered as a virtual source and $K$ virtual reconstructions.
Then the physical source $S$ and the physical reconstructions $\hat{S}{}_{[1:K]}$
are correlated through the virtual source and virtual reconstructions.
Hence the physical source should be more ``tractable'' than the
virtual one, and certain mutual informations (e.g., source-coding
rate region) based on the physical source and reconstructions should
be less than or equal to those based on the virtual source and reconstructions.
The analysis above gives the reasons why $\mathcal{D}_{1}^{(o)}$
is expressed in form of comparison of the ``capacity regions'' of
virtual broadcast channel and physical broadcast channel, while $\mathcal{D}_{2}^{(o)}$
is expressed in form of comparison of the ``source-coding rate regions''
of virtual source and physical source.

For the 2-user broadcast case, the inner bound in Theorem \ref{thm:AdmissibleRegion-GBC}
reduces to
\begin{align}
\mathcal{D}^{(i)}= & \bigl\{\left(D_{1},D_{2}\right):\exists p_{V_{0},V_{1},V_{2}|S},x\left(v_{0},v_{1},v_{2},s\right),\nonumber \\
 & \hat{s}_{k}\left(v_{0},v_{k},y_{k}\right),k=1,2,\text{ s.t. }\mathbb{E}d_{k}\left(S,\hat{S}_{k}\right)\le D_{k},\nonumber \\
 & I(V_{0}V_{k};S)<I(V_{0}V_{k};Y_{k}),k=1,2,\nonumber \\
 & I(V_{0}V_{1}V_{2};S)+I(V_{1};V_{2}|V_{0})\nonumber \\
 & <\min\left\{ I(V_{0};Y_{1}),I(V_{0};Y_{2})\right\} \nonumber \\
 & \qquad+I(V_{1};Y_{1}|V_{0})+I(V_{2};Y_{2}|V_{0}),\nonumber \\
 & I(V_{0}V_{1};S)+I(V_{0}V_{2};S)+I(V_{1};V_{2}|V_{0}S)\nonumber \\
 & <I(V_{0}V_{1};Y_{1})+I(V_{0}V_{2};Y_{2})\bigr\}.
\end{align}
This inner bound was first given by Yassaee \emph{et. al} \cite{Yassaee}.
On the other hand, letting $K=2,L=1$, $\mathcal{D}_{1}^{(o)}$ and
$\mathcal{D}_{2}^{(o)}$ respectively reduce to
\begin{align*}
\mathcal{D}_{1}^{(o)}= & \bigl\{\left(D_{1},D_{2}\right):\exists p_{\hat{S}_{1}\hat{S}_{2}|S}\text{ s.t.}\\
 & \mathbb{E}d_{k}\left(S,\hat{S}_{k}\right)\le D_{k},k=1,2,\\
 & \text{and for any pmf }p_{U|S},\text{ one can find }p_{\tilde{U}X\tilde{Y}_{1}\tilde{Y}_{2}}\text{ s.t.}\\
 & I\left(\hat{S}_{1};U\right)\leq I\left(Y_{1};\tilde{U}|\tilde{Y}_{1}\right),\\
 & I\left(\hat{S}_{2};U\right)\leq I\left(Y_{2};\tilde{U}|\tilde{Y}_{2}\right),\\
 & I\left(\hat{S}_{1}\hat{S}_{2};U\right)\leq I\left(Y_{1}Y_{2};\tilde{U}|\tilde{Y}_{1}\tilde{Y}_{2}\right),\\
 & I\left(\hat{S}_{1};S|U\right)\leq I\left(Y_{1};X|\tilde{U}\tilde{Y}_{1}\right),\\
 & I\left(\hat{S}_{2};S|U\right)\leq I\left(Y_{2};X|\tilde{U}\tilde{Y}_{2}\right),\\
 & I\left(\hat{S}_{1}\hat{S}_{2};S|U\right)\leq I\left(Y_{1}Y_{2};X|\tilde{U}\tilde{Y}_{1}\tilde{Y}_{2}\right)\bigr\},
\end{align*}
or a simpler but possibly looser one,
\begin{align*}
\mathcal{D}_{1}^{(o)}= & \bigl\{\left(D_{1},D_{2}\right):\exists p_{\hat{S}_{1}\hat{S}_{2}|S}\text{ s.t.}\\
 & \mathbb{E}d_{k}\left(S,\hat{S}_{k}\right)\le D_{k},k=1,2,\\
 & \text{and for any pmf }p_{U|S},\text{ one can find }p_{X\tilde{Y}_{1}\tilde{Y}_{2}}\text{ s.t.}\\
 & I\left(\hat{S}_{1};U\right)\leq I\left(Y_{1};\tilde{Y}_{1}\right),\\
 & I\left(\hat{S}_{2};U\right)\leq I\left(Y_{2};\tilde{Y}_{2}\right),\\
 & I\left(\hat{S}_{1}\hat{S}_{2};U\right)\leq I\left(Y_{1}Y_{2};\tilde{Y}_{1}\tilde{Y}_{2}\right),\\
 & I\left(\hat{S}_{1};S|U\right)\leq I\left(Y_{1};X|\tilde{Y}_{1}\right),\\
 & I\left(\hat{S}_{2};S|U\right)\leq I\left(Y_{2};X|\tilde{Y}_{2}\right),\\
 & I\left(\hat{S}_{1}\hat{S}_{2};S|U\right)\leq I\left(Y_{1}Y_{2};X|\tilde{Y}_{1}\tilde{Y}_{2}\right)\bigr\},
\end{align*}
and
\begin{align*}
\mathcal{D}_{2}^{(o)}= & \bigl\{\left(D_{1},D_{2}\right):\exists p_{X},\hat{s}_{k}\left(\tilde{y}_{k}\right),k=1,2\text{ s.t.}\\
 & \mathbb{E}d_{k}\left(S,\hat{S}_{k}\right)\le D_{k},k=1,2,\\
 & \text{and for any pmf }p_{U|Y_{1}Y_{2}},\\
 & \text{one can find }p_{\tilde{Y}_{1}\tilde{Y}_{2}|S}p_{\tilde{U}|\tilde{Y}_{1}\tilde{Y}_{2}}\text{ s.t.}\\
 & I\left(S;\tilde{U}\right)\leq I\left(X;U\right),\\
 & I\left(S;\tilde{Y}_{1}|\tilde{U}\right)\leq I\left(X;Y_{1}|U\right),\\
 & I\left(S;\tilde{Y}_{2}|\tilde{U}\right)\leq I\left(X;Y_{2}|U\right),\\
 & I\left(S;\tilde{Y}_{1}|\tilde{Y}_{2}\tilde{U}\right)\leq I\left(X;Y_{1}|Y_{2}U\right),\\
 & I\left(S;\tilde{Y}_{2}|\tilde{Y}_{1}\tilde{U}\right)\leq I\left(X;Y_{2}|Y_{1}U\right),\\
 & I\left(S;\tilde{Y}_{1}\tilde{Y}_{2}|\tilde{U}\right)\leq I\left(X;Y_{1}Y_{2}|U\right)\bigr\}.
\end{align*}

Note that the outer bounds $\mathcal{D}_{1}^{(o)}$ and $\mathcal{D}_{2}^{(o)}$
are not trivial in general. The necessity of introducing nondegenerate
variable(s) for $\mathcal{D}_{1}^{(o)}$ can be concluded from some
special cases, e.g., source broadcast over a degraded channel, quadratic
Gaussian source broadcast, or Hamming binary source broadcast (see
the details in the subsequent three subsections). To show the necessity
of introducing a nondegenerate variable for $\mathcal{D}_{2}^{(o)}$,
we consider the first three inequalities on mutual information in
$\mathcal{D}_{2}^{(o)}$. Next we show that the necessary conditions
\begin{align}
 & I\left(S;\tilde{U}\right)\leq I\left(X;U\right),\label{eq:-29}\\
 & I\left(S;\tilde{Y}_{1}|\tilde{U}\right)\leq I\left(X;Y_{1}|U\right),\\
 & I\left(S;\tilde{Y}_{2}|\tilde{U}\right)\leq I\left(X;Y_{2}|U\right),\label{eq:-31}
\end{align}
with a nondegenerate $U$ results in a tighter bound than that with
a degenerate $U$ (for the latter case, the necessary conditions reduce
to the single-user bound).

Suppose the broadcast channel $P_{Y_{1}Y_{2}|X}$ satisfies $Y_{1}=(Y_{0},Y_{1}'),Y_{2}=(Y_{0},Y_{2}')$
for some $Y_{0},Y_{1}',Y_{2}'$. Consider a lossless transmission
case (Hamming distortion measure and $D_{1}=D_{2}=0$): $S=(S_{1},S_{2}),\hat{S}_{1}=S_{1},\hat{S}_{2}=S_{2}$,
and $H(S_{1})=C_{1}$ and $H(S_{2})=C_{2}$. Obviously, these conditions
do not violate the single-user outer bound. Now we show that for some
sources,these conditions violate the outer bound $\mathcal{D}_{2}^{(o)}$.
Set $U=Y_{0}$ in $\mathcal{D}_{2}^{(o)}$. Then it is easy to obtain
the following inequalities from \eqref{eq:-29}-\eqref{eq:-31}.
\begin{align}
 & I(S_{1}S_{2};V)\leq I(X;Y_{0}),\label{eq:-29-1}\\
 & H(S_{1}|V)\leq I(X;Y_{1}|Y_{0}),\\
 & H(S_{2}|V)\leq I(X;Y_{2}|Y_{0}),\label{eq:-31-1}
\end{align}
for some $p_{V|S_{1}S_{2}}$. Therefore, we further have
\begin{align}
H(S_{1}) & \leq H(S_{1})+I(S_{2};V|S_{1})\\
 & =I(S_{1}S_{2};V)+H(S_{1}|V)\\
 & \leq I(X;Y_{0})+I(X;Y_{1}|Y_{0})\\
 & =I(X;Y_{1})\leq C_{1}.
\end{align}
On the other hand, by the assumptions $H(S_{1})=C_{1}$ and $H(S_{2})=C_{2}$,
the equalities hold in all the inequalities above, which implies $I(X;Y_{1})=C_{1}$
(i.e., $P_{X}$ is a capacity-achieving distribution), $I(S_{1}S_{2};V)=I(X;Y_{0})$
, and $I(S_{2};V|S_{1})=0$, i.e., $S_{2}\rightarrow S_{1}\rightarrow V$.
Similarly, we have $S_{1}\rightarrow S_{2}\rightarrow V$. In addition,
the G\'acs-K{\"o}rner common information \cite{Gacs,Ahlswede} is defined
as
\begin{equation}
C_{\mathsf{GK}}(S_{1};S_{2})=\sup_{P_{V|S_{1}S_{2}}:S_{2}\rightarrow S_{1}\rightarrow V,S_{1}\rightarrow S_{2}\rightarrow V}I(S_{1}S_{2};V).
\end{equation}
Hence there exists $p_{V|S_{1}S_{2}}$ such that $S_{2}\rightarrow S_{1}\rightarrow V,S_{1}\rightarrow S_{2}\rightarrow V$
and $I(S_{1}S_{2};V)=I(X;Y_{0})$ , only if $C_{\mathsf{GK}}(S_{1};S_{2})\geq I(X;Y_{0})>0$
(suppose the channel $P_{Y_{0}|X}$ satisfies $I(X;Y_{0})>0$ for
the capacity-achieving distribution $P_{X}$). However the G\'acs-K{\"o}rner
common information does not always exist for all source pairs $(S_{1},S_{2})$,
e.g., $C_{\mathsf{GK}}(S_{1};S_{2})=0$ for a doubly symmetric binary
source. This implies the outer bound is tighter than the single-user
one, which in turn implies $\mathcal{D}_{2}^{(o)}$ are not trivial
in general.

\subsection{Discrete Memoryless Broadcast over Degraded Channel}

If the channel is degraded, define
\begin{align}
\mathcal{D}_{\mathsf{DBC}}^{(i)}= & \Bigl\{ D_{[1:K]}:\exists p_{V_{K}|S}p_{V_{K-1}|V_{K}}\cdots p_{V_{1}|V_{2}},x\left(v_{K},s\right),\nonumber \\
 & \hat{s}_{k}\left(v_{k},y_{k}\right),k\in[1:K]\text{ s.t.}\nonumber \\
 & \mathbb{E}d_{k}(S,\hat{S}_{k})\le D_{k},\nonumber \\
 & I\left(S;V_{k}\right)\leq\sum_{j=1}^{k}I\left(Y_{j};V_{j}|V_{j-1}\right),k\in[1:K],\nonumber \\
 & \text{ where }V_{0}\triangleq\emptyset\Bigr\},
\end{align}
and
\begin{align}
 & \mathcal{D}_{\mathsf{DBC}}^{(o)}=\nonumber \\
 & \Bigl\{ D_{[1:K]}:\exists p_{\hat{S}_{[1:K]}|S},p_{X}\text{ s.t. }\nonumber \\
 & \mathbb{E}d_{k}(S,\hat{S}_{k})\le D_{k},k\in[1:K],\nonumber \\
 & \left(I(\hat{S}_{[1:k]};U_{k}|U_{k-1}):k\in[1:K]\right)\in\mathcal{R}_{\mathsf{DBC}}(p_{X}p_{Y_{[1:K]}|X})\nonumber \\
 & \text{for any }p_{U_{K-1}|S}p_{U_{K-2}|U_{K-1}}\cdots p_{U_{1}|U_{2}},U_{0}\triangleq\emptyset,U_{K}\triangleq S,\nonumber \\
 & \textrm{and }\left(I(S;\hat{S}_{[1:k]}):k\in[1:K]\right)\in\mathcal{R}_{\mathsf{SRC}}(p_{X}p_{Y_{[1:K]}|X})\Bigr\},
\end{align}
where
\begin{align}
 & \mathcal{R}_{\mathsf{DBC}}\left(p_{X}p_{Y_{[1:K]}|X}\right)=\nonumber \\
 & \Bigl\{ R_{[1:K]}:R_{k}\geq0,\exists p_{V_{K-1}|X}p_{V_{K-2}|V_{K-1}}\cdots p_{V_{1}|V_{2}}\text{ s.t.}\nonumber \\
 & \sum_{j=1}^{k}R_{j}\leq\sum_{j=1}^{k}I\left(Y_{j};V_{j}|V_{j-1}\right),k\in[1:K],\nonumber \\
 & \text{ where }V_{0}\triangleq\emptyset,V_{K}\triangleq X\Bigr\}\label{eq:DBCcapacity}
\end{align}
denotes the capacity of the degraded broadcast channel $p_{Y_{[1:K]}|X}$
with the input $X$ following $p_{X}$, and
\begin{align}
 & \mathcal{R}_{\mathsf{SRC}}\left(p_{X}p_{Y_{[1:K]}|X}\right)=\nonumber \\
 & \Bigl\{ R_{[1:K]}:R_{k}\geq0,\sum_{j=1}^{k}R_{j}\geq I\left(X;Y_{[1:k]}\right),k\in[1:K]\Bigr\}\label{eq:DBCcapacity2}
\end{align}
denotes the successive refinement coding rate region of source $X$
with reconstructions $Y_{[1:K]}$ following $p_{Y_{[1:K]}|X}$.

Then as a consequence of Theorem \ref{thm:AdmissibleRegion-GBC},
the following theorem holds.
\begin{thm}
\label{thm:AdmissibleRegion-DBC} For transmitting a DMS $S$ over
a degraded DM-BC $p_{Y_{[1:K]}|X}$,
\begin{equation}
\mathcal{D}_{\mathsf{DBC}}^{(i)}\subseteq\mathcal{D}\subseteq\mathcal{D}_{\mathsf{DBC}}^{(o)}.
\end{equation}
\end{thm}
\begin{rem}
$\mathcal{D}_{\mathsf{DBC}}^{(o)}$ can be also expressed as
\begin{align}
\mathcal{D}_{\mathsf{DBC}}^{(o)}= & \Bigl\{ D_{[1:K]}:\exists p_{\hat{S}_{[1:K]}|S},p_{X}\text{ s.t.}\nonumber \\
 & \mathbb{E}d_{k}\left(S,\hat{S}_{k}\right)\le D_{k},k\in[1:K],\nonumber \\
 & \mathcal{R}_{\mathsf{DBC}}\left(p_{S}p_{\hat{S}_{[1:K]}^{\prime}|S}\right)\subseteq\mathcal{R}_{\mathsf{DBC}}\left(p_{X}p_{Y_{[1:K]}|X}\right),\nonumber \\
 & \mathcal{R}_{\mathsf{SRC}}\left(p_{S}p_{\hat{S}_{[1:K]}^{\prime}|S}\right)\supseteq\mathcal{R}_{\mathsf{SRC}}\left(p_{X}p_{Y_{[1:K]}|X}\right),\nonumber \\
 & \hat{S}_{k}^{\prime}\triangleq\hat{S}_{[1:k]},k\in[1:K]\Bigr\}.\label{eq:-1}
\end{align}
From the last constraint of $\mathcal{D}_{\mathsf{DBC}}^{(o)}$, one
can obtain an interesting conclusion: the single-user outer bound
$D_{[1:K]}^{*}$ can be achieved for source broadcast over a degraded
channel only if the source is successively refinable.
\end{rem}
\begin{rem}
If $p_{S}$ is Gaussian and all $d_{k}\left(s,\hat{s}_{k}\right),k\in[1:K]$
are the quadratic distortion function (i.e., $d_{k}\left(s,\hat{s}_{k}\right)=(s-\hat{s}_{k})^{2},k\in[1:K]$),
then the smallest capacity region (with input distribution restricted
to $p_{S}$) $\mathcal{R}_{\mathsf{DBC}}(p_{S}p_{\hat{S}_{[1:K]}^{\prime}|S})$
 over all $p_{\hat{S}_{[1:K]}|S}$ is obtained by setting $p_{\hat{S}_{[1:K]}|S}$
as the Gaussian broadcast channel such that for $k\in[1:K]$, $S=\hat{S}_{k}+E_{k}$
and $\hat{S}_{k}\sim\mathcal{N}(0,N_{S}-D_{k}),E_{k}\sim\mathcal{N}(0,D_{k})$
are independent. That is, setting $p_{\hat{S}_{[1:K]}|S}$ as the
Gaussian broadcast channel is ``optimal'' for this case. This point
is obtained by observing that: 1) the capacity region for this case
is
\begin{align}
 & \mathcal{C}=\bigcup_{+\infty=\tau_{0}\geq\tau_{1}\geq\cdots\geq\tau_{K}=0}\nonumber \\
 & \left\{ R_{[1:K]}:R_{k}\le\frac{1}{2}\log\frac{\left(D_{k}+\tau_{k-1}\right)\left(N_{S}+\tau_{k}\right)}{\left(D_{k}+\tau_{k}\right)\left(N_{S}+\tau_{k-1}\right)},k\in[1:K]\right\} ;
\end{align}
2) for any other channels $p_{\hat{S}_{[1:K]}|S}$ satisfying $\mathbb{E}d_{k}\left(S,\hat{S}_{k}\right)\le D_{k},k\in[1:K]$,
$\mathcal{C}$ is a subset (or an inner bound) of the capacity region
(with input distribution restricted to $p_{S}$) of $p_{\hat{S}_{[1:K]}|S}$
\cite{Tian11}. Therefore, Gaussian broadcast channels have the smallest
capacity regions given noise powers (here noise is not restricted
to be independent of the channel input).  This observation is consistent
with the point-to-point case \cite[Problem 10.8]{Cover91}.
\end{rem}
Note that the last constraint of $\mathcal{D}_{\mathsf{DBC}}^{(i)}$
can be understood as the intersection between the successive refinement
rate region of the source $S$ with reconstructions $V_{[1:K]}$ and
the capacity of the degraded broadcast channel $p_{Y_{[1:K]}|X}$
with the input $X$ and auxiliary random variables $V_{[1:K]}$, is
not empty. The second constraint of $\mathcal{D}_{\mathsf{DBC}}^{(o)}$
can be understood as the capacity of the virtual degraded broadcast
channel $p_{\hat{S}_{[1:K]}^{\prime}|S}$ with the input $S$ is included
in the capacity of the physical degraded broadcast channel $p_{Y_{[1:K]}|X}$
with the input $X$. Similarly, the last constraint of $\mathcal{D}_{\mathsf{DBC}}^{(o)}$
can be understood as the successive refinement rate region of the
physical source $S$ with reconstructions $\hat{S}_{[1:K]}^{\prime}$
includes the successive refinement rate region of the virtual source
$X$ with reconstructions $Y_{[1:K]}$.

\subsection{Hamming Binary Broadcast}

Consider sending a binary source $S\sim\textrm{Bern}\left(\frac{1}{2}\right)$
with the Hamming distortion measure $d_{k}(s,\hat{s})=d(s,\hat{s})\triangleq0,\textrm{ if }s=\hat{s};1,\textrm{ otherwise}$,
over a binary broadcast channel $Y_{k}=X\oplus W_{k},1\le k\le K$
with $W_{k}\sim\textrm{Bern}\left(p_{k}\right),\frac{1}{2}\geq p_{1}\geq p_{2}\geq\cdots\geq p_{K}\geq0$.
Assume the bandwidth mismatch factor is $b$.

We first consider the inner bound part. For bandwidth expansion ($b>1$)
case, as a special case of hybrid coding, systematic source-channel
coding (or Uncoded Systematic Coding) was first investigated in \cite{Shamai98}.
For any point-to-point lossless communication, such systematic coding
does not lose the optimality; however, for some lossy transmission
cases such as Hamming binary source communication, it is not optimal
any more \cite{Shamai98}. To retain the optimality, we can first
quantize the source $S$, and then transmit the quantized signal using
Uncoded Systematic Coding. The performance of such code can be obtained
directly from Theorem \ref{thm:AdmissibleRegion-DBC}.

\begin{figure}[t]
\centering\includegraphics[width=0.5\textwidth]{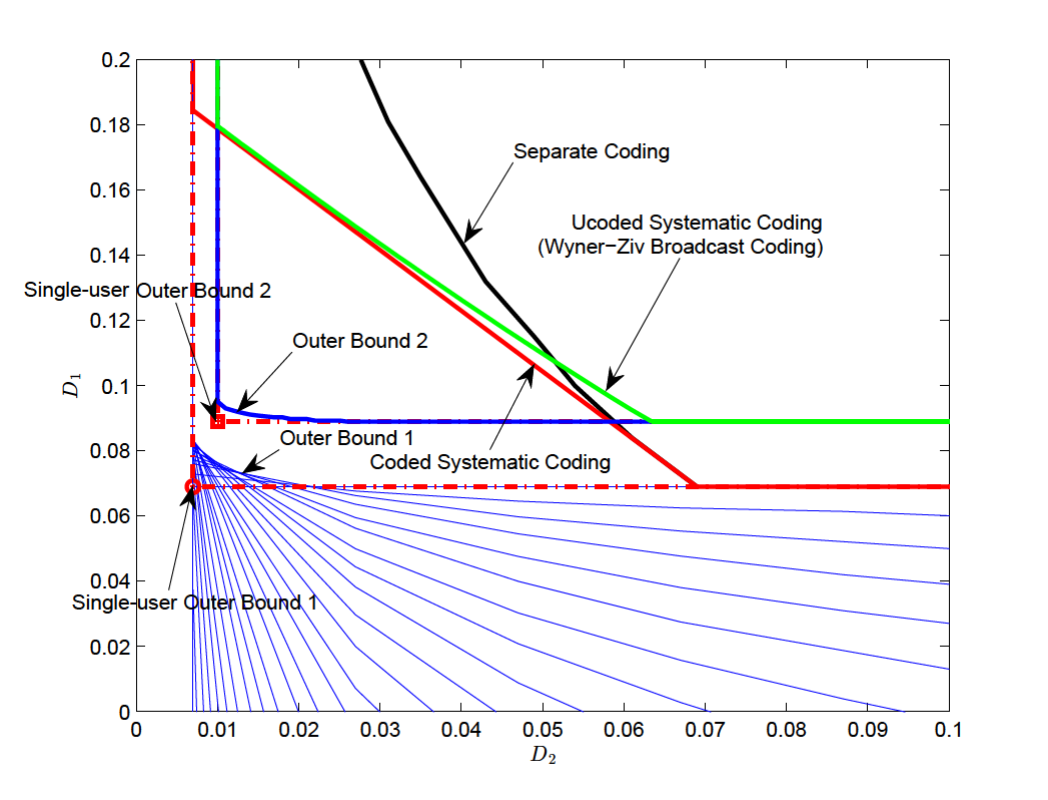} \protect\caption{\label{fig:HDAcoding-1}Distortion bounds for sending a binary source
over a binary broadcast channel with $b=2,p_{1}=0.18,p_{2}=0.12$.
Outer Bounds 1 and 2 respectively correspond to the outer bound of
Theorem \ref{thm:AdmissibleRegionBB} and the outer bound of Theorem
\ref{thm:AdmissibleRegionBBSI}. Separate Coding, Uncoded Systematic
Coding, and Coded Systematic Coding respectively correspond to the
separate scheme (combining successive-refinement code \cite[Example 13.3]{El Gamal}
with superposition code \cite[Example 5.3]{El Gamal}), the inner
bound in Corollary \ref{thm:bandwidthmismatch}, and the inner bound
in Corollary \ref{cor:Coded}. Single-user Outer Bounds 1 and 2 correspond
to the single-user outer bound \eqref{eq:trivialouterbound} and the
single-user Wyner-Ziv outer bound \eqref{eq:wzouterbound}, respectively.
Besides, Single-user Outer Bound 2, Outer Bound 2 and Uncoded Systematic
Coding can be considered as outer bounds and inner bound for the Wyner-Ziv
source broadcast problem with $b=1,\beta_{1}=p_{1},\beta_{2}=p_{2}$.}
\end{figure}

Specifically, let $U_{2}=S\oplus E_{2}$, $U_{1}=U_{2}\oplus E_{1}$
with $E_{2}\sim\textrm{Bern}(D_{2}),E_{1}\sim\textrm{Bern}(d_{1})$.
Let $V_{2}=\left(U_{2},X^{b-1}\right),V_{1}=\left(U_{1},X_{1}^{b-1}\right),X_{1}^{b-1}=X^{b-1}\oplus B^{b-1}$,
where $X_{1}^{b-1}$ and $X^{b-1}$ are independent of $U_{2}$ and
$U_{1}$, and $X^{b-1}$ and $B^{b-1}$ follow $b-1$ dimensional
$\textrm{Bern}(\frac{1}{2})$ and $\textrm{Bern}(\theta)$, respectively.
Let $x^{b}(v_{2},s)=\left(u_{2},x^{b-1}\right)$, $\hat{s}_{2}\left(v_{2},y_{2}^{b}\right)=u_{2}$,
and $\hat{s}_{1}\left(v_{1},y_{1}^{b}\right)=u_{1},\textrm{if }d_{1}<p_{1};y_{1},\textrm{otherwise}$,
where $y_{1}$ is the first letter of $y_{1}^{b}$. Substituting these
variables and functions into the inner bound $\mathcal{D}_{\mathsf{DBC}}^{(i)}$
in Theorem \ref{thm:AdmissibleRegion-DBC}, we get the following corollary.
\begin{cor}[Coded Systematic Coding]
\label{cor:Coded}For transmitting a binary source $S$ with the
Hamming distortion measure over a $2$-user binary broadcast channel
with the bandwidth mismatch factor $b$,
\begin{align}
 & \mathcal{D}\supseteq\mathcal{D}_{\mathsf{CSC}}^{(i)}\triangleq\mathrm{convexhull}\Bigl\{\left(D_{1},D_{2}\right):\nonumber \\
 & 0\leq\theta,d_{1}\leq\frac{1}{2},\nonumber \\
 & D_{1}\geq\min\left\{ d_{1}\star D_{2},p_{1}\star D_{2}\right\} ,\nonumber \\
 & r_{1}\triangleq1-H_{2}(d_{1}\star p_{1})+\left(b-1\right)[1-H_{2}(\theta\star p_{1})],\nonumber \\
 & r_{2}\triangleq H_{2}(d_{1}\star p_{2})-H_{2}(p_{2})+\left(b-1\right)[H_{2}(\theta\star p_{2})-H_{2}(p_{2})],\nonumber \\
 & 1-H_{2}(d_{1}\star D_{2})\leq r_{1},\nonumber \\
 & 1-H_{2}(D_{2})\leq r_{1}+r_{2}\Bigr\},
\end{align}
where $\star$ denotes an binary operation such that
\begin{equation}
x\star y=(1-x)y+x(1-y),\label{eq:star}
\end{equation}
and $H_{2}$ denotes the binary entropy function, i.e.,
\begin{equation}
H_{2}(p)=-p\log p-(1-p)\log(1-p).\label{eq:binaryentropy}
\end{equation}
\end{cor}
\begin{rem}
Coded Systematic Coding without timesharing does not always lead to
a convex distortion region, hence a timesharing mechanism is needed
to improve the performance. This is equivalent to adding a timesharing
variable $Q$ into $V_{2}$ and $V_{1}$, before substituting them
into the inner bound $\mathcal{D}_{\mathsf{DBC}}^{(i)}$. Besides,
note that unlike Uncoded Systematic Coding, the Coded Systematic Coding
could always achieve the optimal distortion for at least one of the
receivers. Moreover, unlike separate coding the Coded Systematic Coding
weakens the cliff effect, and results in a \emph{slope-cliff effect}.
\end{rem}
The outer bound of Theorem \ref{thm:AdmissibleRegion-DBC} reduces
to the following outer bound for the Hamming binary source broadcast
problem. This outer bound was first given in \cite[Eqn (41)]{Khezeli14}
for the 2-user case.
\begin{thm}
\label{thm:AdmissibleRegionBB}\cite[Eqn (41)]{Khezeli14} For transmitting
a binary source $S$ with the Hamming distortion measure over a $K$-user
binary broadcast channel with the bandwidth mismatch factor $b$,
\begin{align}
 & \mathcal{D}\subseteq\mathcal{D}_{\mathsf{DBC}}^{(o)}\triangleq\nonumber \\
 & \Bigl\{ D_{[1:K]}:\textrm{For any values }\frac{1}{2}=\tau_{0}\geq\tau_{1}\geq\cdots\geq\tau_{K}=0,\nonumber \\
 & \frac{1}{b}\left(H_{2}\left(\tau_{k-1}\star D_{k}\right)-H_{2}\left(\tau_{k}\star D_{k}\right):k\in[1:K]\right)\in\mathcal{R}_{\mathsf{BBC}}\Bigr\},
\end{align}
where $\mathcal{R}_{\mathsf{BBC}}$ denotes the capacity of the binary
broadcast channel given by
\begin{align}
 & \mathcal{R}_{\mathsf{BBC}}=\nonumber \\
 & \Bigl\{ R_{[1:K]}:\exists\textrm{ some values }\frac{1}{2}=\theta_{0}\geq\theta_{1}\geq\cdots\geq\theta_{K}=0\textrm{ \text{s.t. }}\nonumber \\
 & 0\leq R_{k}\leq H_{2}\left(\theta_{k-1}\star p_{k}\right)-H_{2}\left(\theta_{k}\star p_{k}\right),k\in[1:K]\Bigr\}.\label{eq:binarycapacity}
\end{align}
\end{thm}
The bounds in Corollary \ref{cor:Coded} and Theorem \ref{thm:AdmissibleRegionBB}
are illustrated in Fig. \ref{fig:HDAcoding-1}.

\subsection{Quadratic Gaussian Broadcast}

\begin{figure}[t]
\centering\includegraphics[width=0.5\textwidth]{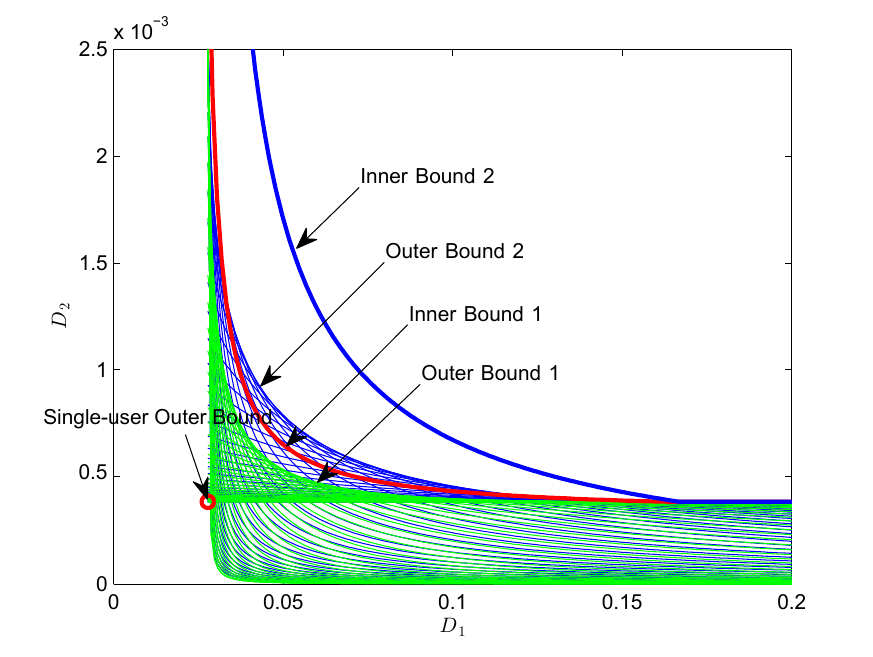} \protect\caption{\label{fig:HDAcoding-GG}Distortion bounds for sending a Gaussian
source over a Gaussian broadcast channel with $b=2,N_{S}=1,P=50,N_{1}=10,N_{2}=1$.
Outer Bounds 1 and 2 and Inner Bounds 1 and 2 respectively correspond
to the outer bound in \cite[Thm. 2]{Tian11}, the outer bound in Theorem
\ref{thm:AdmissibleRegionGGSI}, the inner bound in \cite[Thm. 5]{Prabhakaran11},
and the inner bound achieved by Wyner-Ziv separate coding (uncoded
systematic code) \cite[Lem. 3]{Nayak}. Single-user Outer Bound corresponds
to the single-user outer bound \eqref{eq:trivialouterbound}. Besides,
Outer Bound 2 and Inner Bound 2 can be considered as an outer bound
and an inner bound for the Wyner-Ziv source broadcast problem with
$b=1,\beta_{1}=\frac{N_{S}N_{1}}{P+N_{1}},\beta_{2}=\frac{N_{S}N_{2}}{P+N_{2}}$.
For this case, Single-user Outer Bound corresponds to the single-user
Wyner-Ziv outer bound \eqref{eq:wzouterbound}.}
\end{figure}

Consider sending a Gaussian source $S\sim\mathcal{N}\left(0,N_{S}\right)$
with the quadratic distortion measure $d_{k}(s,\hat{s})=d(s,\hat{s})\triangleq(s-\hat{s})^{2}$
over a power-constrained Gaussian broadcast channel $Y_{k}=X+W_{k},1\le k\le K$
with $\mathbb{E}\left(X^{2}\right)\leq P$ and$W_{k}\sim\mathcal{N}\left(0,N_{k}\right),N_{1}\geq N_{2}\geq\cdots\geq N_{K}$.
Assume the bandwidth mismatch factor is $b$. As stated in Remark
\ref{rem:The-inner-bound}, Theorem \ref{thm:AdmissibleRegion-GBC}
(or \ref{thm:AdmissibleRegion-DBC}) holds for the Gaussian case as
well. The inner bound $\mathcal{D}_{\mathsf{DBC}}^{(i)}$ in Theorem
\ref{thm:AdmissibleRegion-DBC} could recover the best known inner
bound \cite[Thm. 5]{Prabhakaran11} by setting the random variables
and symbol-by-symbol mappings to some suitable ones. On the other
hand, setting $U_{[1:K-1]}$ to be jointly Gaussian with $S$, the
outer bound $\mathcal{D}_{\mathsf{DBC}}^{(o)}$ in Theorem \ref{thm:AdmissibleRegion-DBC}
could recover the best known outer bound \cite[Thm. 2]{Tian11}. The
bounds in \cite[Thm. 5]{Prabhakaran11} and \cite[Thm. 2]{Tian11}
are illustrated in Fig. \ref{fig:HDAcoding-GG}.

\section{Wyner-Ziv Source Broadcast: Source Broadcast with Side Information}

\begin{figure}[t]
\centering\includegraphics[width=0.5\textwidth]{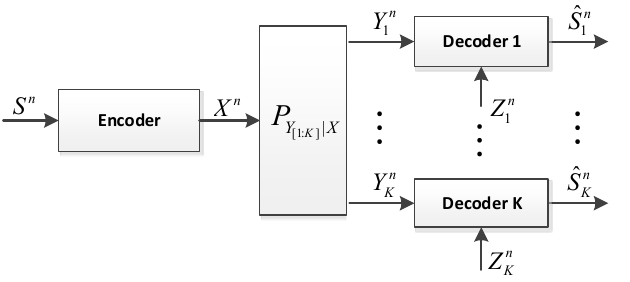}\caption{\label{fig:WZbroadcast communication system}Wyner-Ziv source broadcast
system: a broadcast communication system with side information at
decoders. }
\end{figure}

We now extend the source broadcast problem by allowing decoders to
access side information correlated with the source. As depicted in
Fig. \ref{fig:WZbroadcast communication system}, receiver $k$ observes
memoryless side information $Z_{k}^{n}$, and it produces a source
reconstruction $\hat{S}_{k}^{n}$ from the received signal $Y_{k}^{n}$
and side information $Z_{k}^{n}$.

\begin{defn}
An $n$-length Wyner-Ziv source-channel code is defined by the encoding
function $x^{n}:{\mathcal{S}}^{n}\mapsto{\mathcal{X}}^{n}$ and $K$
decoding functions $\hat{s}_{k}^{n}:\mathcal{Y}_{k}^{n}\times\mathcal{Z}_{k}^{n}\mapsto\hat{\mathcal{S}}_{k}^{n},1\leq k\leq K$.
\end{defn}
\begin{defn}
If there exists a sequence of Wyner-Ziv source-channel codes satisfying
\begin{equation}
\mathop{\limsup}\limits _{n\to\infty}\mathbb{E}d_{k}\left(S^{n},\hat{S}_{k}^{n}\right)\le D_{k},
\end{equation}
then we say that the distortion tuple $D_{[1:K]}$ is achievable.
\end{defn}
\begin{defn}
The admissible distortion region for the Wyner-Ziv broadcast problem
is defined as
\begin{align}
\mathcal{D}_{\mathsf{SI}}\triangleq & \left\{ D_{[1:K]}:D_{[1:K]}\textrm{ is achievable}\right\} .
\end{align}
\end{defn}
Shamai \emph{et al.} \cite[Thm. 2.1]{Shamai98} showed that for transmitting
a source over a point-to-point channel $p_{Y_{k}|X}$ with side information
$Z_{k}$ available at the decoder, the minimum achievable distortion
for the receiver $k$ satisfies $R_{S|Z_{k}}\left(D_{k}\right)=C_{k}$,
where $R_{S|Z_{k}}\left(\cdotp\right)$ is the Wyner-Ziv rate-distortion
function of the source $S$ given that the decoder observes $Z_{k}$
\cite{El Gamal}. Therefore, the optimal achievable distortion is
$D_{\mathsf{SI},k}^{*}=R_{S|Z_{k}}^{-1}\left(C_{k}\right).$ Obviously,{\footnotesize{}{}{}
}
\begin{equation}
\mathcal{D}_{\mathsf{SI}}\subseteq\mathcal{D}_{\mathsf{SI}}^{*}\triangleq\left\{ D_{[1:K]}:D_{k}\geq D_{\mathsf{SI},k}^{*},1\le k\le K\right\} ,\label{eq:wzouterbound}
\end{equation}
where $\mathcal{D}_{\mathsf{SI}}^{*}$ is named\emph{ single-user
Wyner-Ziv outer bound}.

Besides, we also consider the bandwidth mismatch case, whereby $m$
samples of a DMS are transmitted through $n$ uses of a DM-BC with
$l$ samples of side information available at each decoder. For simplicity,
we assume $m=l$. For this case, the bandwidth mismatch factor is
defined as $b=\frac{n}{m}$.

\subsection{Discrete Memoryless Wyner-Ziv Broadcast}

If consider $S$ and $Z_{[1:K]}$ as the input and outputs of a virtual
broadcast channel $p_{Z_{[1:K]}|S}$, then the Wyner-Ziv source broadcast
problem with the encoding function $x^{n}(s^{n})$ is equivalent to
the source broadcast problem of sending $S$ over $p_{Z_{[1:K]}|S}p_{Y_{[1:K]}|X}$
with the encoding function $(s^{n},x^{n}(s^{n}))$. Correspondingly,
the Wyner-Ziv source broadcast system can be considered as a (uncoded)
systematic source-channel coding system. Hence by setting the symbol-by-symbol
function as $\left(s,x\left(v_{[1:N]},s\right)\right)$, from Theorem
\ref{thm:AdmissibleRegion-GBC}, we obtain the following inner bound
for the Wyner-Ziv source broadcast problem.
\begin{align}
\mathcal{D}_{\mathsf{SI}}^{(i)}= & \Bigl\{ D_{[1:K]}:\exists p_{V_{[1:N]}|S},r_{[1:N]},x\left(v_{[1:N]},s\right),\nonumber \\
 & \hat{s}_{k}\left(v_{\mathcal{B}_{k}},y_{k},z_{k}\right),k\in[1:K]\text{ s.t.}\nonumber \\
 & \mathbb{E}d_{k}\left(S,\hat{S}_{k}\right)\le D_{k},k\in[1:K],\nonumber \\
 & \sum_{j\in\mathcal{J}}r_{j}>\sum_{j\in\mathcal{J}}H\left(V_{j}|V_{\mathcal{A}_{j}}\right)-H\left(V_{\mathcal{J}}|S\right)\nonumber \\
 & \textrm{for all }\mathcal{J}\subseteq[1:N]\textrm{ s.t. }\mathcal{J}\neq\emptyset\textrm{ and }\mathcal{A}_{j}\subseteq\mathcal{J},\forall j\in\mathcal{J},\nonumber \\
 & \sum_{j\in\mathcal{J}^{c}}r_{j}<\sum_{j\in\mathcal{J}^{c}}H\left(V_{j}|V_{\mathcal{A}_{j}}\right)-H\left(V_{\mathcal{J}^{c}}|Y_{k}Z_{k}V_{\mathcal{J}}\right),\nonumber \\
 & k\in[1:K]\textrm{ for all }\mathcal{J}\subseteq\mathcal{B}_{k}\textrm{ s.t. }\nonumber \\
 & \mathcal{J}^{c}\triangleq\mathcal{B}_{k}\backslash\mathcal{J}\neq\emptyset\textrm{ and }\mathcal{A}_{j}\subseteq\mathcal{J},\forall j\in\mathcal{J}\Bigr\}.\label{eq:-28-2-1}
\end{align}

In addition, regard $\left(U_{[1:L]},Z_{[1:K]}\right)$ as auxiliary
random variables following $p_{U_{[1:L]}|S}p_{Z_{[1:K]}|S}$ given
$S$, then following steps similar to the proof of the outer bound
$\mathcal{D}_{1}^{(o)}$ of Theorem \ref{thm:AdmissibleRegion-GBC},
we can obtain the following outer bound on $\mathcal{D}_{\mathsf{SI}}$.
\begin{align}
\mathcal{D}_{\mathsf{SI},1}^{(o)}= & \Bigl\{ D_{[1:K]}:\exists p_{\hat{S}_{[1:K]}|S,Z_{[1:K]}}\text{ s.t.}\nonumber \\
 & \mathbb{E}d_{k}\left(S,\hat{S}_{k}\right)\le D_{k},k\in[1:K],\nonumber \\
 & \text{and for any }p_{U_{[1:L]}|S,Z_{[1:K]}},\nonumber \\
 & \text{one can find }p_{\tilde{U}_{[1:L]},\tilde{Z}_{[1:K]},X}\textrm{ s.t. }\nonumber \\
 & I\left(\hat{S}_{\mathcal{A}};U_{\mathcal{B}}|U_{\mathcal{C}}Z_{\mathcal{A}}\right)\leq I\left(Y_{\mathcal{A}};\tilde{U}_{\mathcal{B}}|\tilde{U}_{\mathcal{C}}\tilde{Z}_{\mathcal{A}}\right)\nonumber \\
 & \textrm{for any }\mathcal{A}\subseteq\left[1:K\right],\mathcal{B},\mathcal{C}\subseteq\left[1:L\right]\Bigr\},
\end{align}

Similarly, following steps similar to the proof of the outer bound
$\mathcal{D}_{2}^{(o)}$ of Theorem \ref{thm:AdmissibleRegion-GBC},
we can prove another outer bound on $\mathcal{D}_{\mathsf{SI}}$.
\begin{align}
\mathcal{D}_{\mathsf{SI},2}^{(o)}= & \Bigl\{ D_{[1:K]}:\exists p_{X},\hat{s}_{k}^{n}\left(\tilde{y}_{k},z_{k}^{n}\right),k\in[1:K]\text{ s.t.}\nonumber \\
 & \mathbb{E}d_{k}\left(S,\hat{S}_{k}\right)\le D_{k},k\in[1:K],\nonumber \\
 & \text{and for any pmf }p_{U_{[1:L]}|Y_{[1:K]}},\nonumber \\
 & \text{one can find }p_{\tilde{Y}_{[1:K]}|S}p_{\tilde{U}_{[1:L]}|\tilde{Y}_{[1:K]}}\textrm{ s.t. }\nonumber \\
 & I\left(S^{n};\tilde{Y}_{\mathcal{B}}\tilde{U}_{\mathcal{B}'}|\tilde{Y}_{\mathcal{C}}\tilde{U}_{\mathcal{C}'}\right)\leq I\left(X;Y_{\mathcal{B}}U_{\mathcal{B}'}|Y_{\mathcal{C}}U_{\mathcal{C}'}\right)\nonumber \\
 & \textrm{for any }\mathcal{B},\mathcal{C}\subseteq\left[1:K\right],\mathcal{B}',\mathcal{C}'\subseteq\left[1:L\right]\Bigr\}.
\end{align}

Therefore, the following theorem holds. The proof is omitted.
\begin{thm}
\label{thm:AdmissibleRegionSI-GBC} For transmitting a DMS $S$ over
a DM-BC $p_{Y_{[1:K]}|X}$ with side information $Z_{k}$ at the decoder
$k$ ($k\in[1:K]$),
\begin{equation}
\mathcal{D}_{\mathsf{SI}}^{(i)}\subseteq\mathcal{D}_{\mathsf{SI}}\subseteq\mathcal{D}_{\mathsf{SI},1}^{(o)}\cap\mathcal{D}_{\mathsf{SI},2}^{(o)}.
\end{equation}
\end{thm}
\begin{rem}
\label{rem:Similar-to-Theorem}Similar to Theorem \ref{thm:AdmissibleRegion-GBC},
Theorem \ref{thm:AdmissibleRegionSI-GBC} could be extended to a Gaussian
or any other well-behaved continuous-alphabet source-channel pair.
It also can be extended to the problems of broadcasting Wyner-Ziv
correlated sources, and Wyner-Ziv source broadcast with channel input
cost.
\end{rem}

\subsection{Discrete Memoryless Wyner-Ziv Broadcast over Degraded Channel with
Degraded Side Information}

Theorem \ref{thm:AdmissibleRegionSI-GBC} can be used to derive an
inner bound and an outer bound for the degraded channel and degraded
side information case. Define
\begin{align}
\mathcal{D}_{\mathsf{SI-D}}^{(i)}= & \Bigl\{ D_{[1:K]}:\exists p_{V_{K}|S}p_{V_{K-1}|V_{K}}\cdots p_{V_{1}|V_{2}},x\left(v_{K},s\right),\nonumber \\
 & \hat{s}_{k}\left(v_{k},y_{k},z_{k}\right),k\in[1:K]\text{ s.t.}\nonumber \\
 & \mathbb{E}d_{k}\left(S,\hat{S}_{k}\right)\le D_{k},\nonumber \\
 & I\left(S;V_{k}\right)\leq\sum_{j=1}^{k}I\left(Y_{j}Z_{j};V_{j}|V_{j-1}\right),k\in[1:K],\nonumber \\
 & \text{where }V_{0}\triangleq\emptyset\Bigr\},
\end{align}
and
\begin{align}
 & \mathcal{D}_{\mathsf{SI-D}}^{(o)}=\Bigl\{ D_{[1:K]}:\exists p_{V_{K}|S}p_{V_{K-1}|V_{K}}\cdots p_{V_{1}|V_{2}},p_{X},\nonumber \\
 & \hat{s}_{k}\left(v_{k},z_{k}\right),k\in[1:K]\text{ s.t.}\nonumber \\
 & \mathbb{E}d_{k}\left(S,\hat{S}_{k}\right)\le D_{k},\nonumber \\
 & \left(I\left(V_{k};U_{k}|U_{k-1}Z_{k}\right):k\in[1:K]\right)\in\mathcal{R}_{\mathsf{DBC}}\left(p_{X}p_{Y_{[1:K]}|X}\right)\nonumber \\
 & \text{for any pmf }p_{U_{K-1}|S}p_{U_{K-2}|U_{K-1}}\cdots p_{U_{1}|U_{2}},\nonumber \\
 & U_{0}\triangleq\emptyset,U_{K}\triangleq S,\nonumber \\
 & \textrm{and }\left(I\left(V_{k};S|Z_{k}\right):k\in[1:K]\right)\in\mathcal{R}_{\mathsf{SRC}}\left(p_{X}p_{Y_{[1:K]}|X}\right)\Bigr\},
\end{align}
where $\mathcal{R}_{\mathsf{DBC}}\left(p_{X}p_{Y_{[1:K]}|X}\right)$
and $\mathcal{R}_{\mathsf{SRC}}\left(p_{X}p_{Y_{[1:K]}|X}\right)$
are given in \eqref{eq:DBCcapacity} and \eqref{eq:DBCcapacity2},
respectively. Then we have the following theorem. The proof is analogous
to that of Theorem \ref{thm:AdmissibleRegion-DBC}, and therefore
omitted.
\begin{thm}
\label{thm:AdmissibleRegionSI-DBC} For transmitting a DMS $S$ over
a degraded DM-BC $p_{Y_{[1:K]}|X}$ ($X\rightarrow Y_{K}\rightarrow Y_{K-1}\rightarrow\cdots\rightarrow Y_{1}$)
with degraded side information $Z_{k}$ ($S\rightarrow Z_{K}\rightarrow Z_{K-1}\rightarrow\cdots\rightarrow Z_{1}$)
at the decoder $k$ ($k\in[1:K]$),
\begin{equation}
\mathcal{D}_{\mathsf{SI-D}}^{(i)}\subseteq\mathcal{D}_{\mathsf{SI}}\subseteq\mathcal{D}_{\mathsf{SI-D}}^{(o)}.
\end{equation}
\end{thm}
\begin{rem}
$\mathcal{D}_{\mathsf{SI-D}}^{(o)}$ can be also expressed as
\begin{align}
\mathcal{D}_{\mathsf{SI-D}}^{(o)}= & \Bigl\{ D_{[1:K]}:\exists p_{V_{K}|S}p_{V_{K-1}|V_{K}}\cdots p_{V_{1}|V_{2}},\hat{s}_{k}\left(v_{k},z_{k}\right),\nonumber \\
 & k\in[1:K]\text{ s.t. }\mathbb{E}d_{k}\left(S,\hat{S}_{k}\right)\le D_{k},\nonumber \\
 & \mathcal{R}_{\mathsf{DBC-SI}}\left(p_{S}p_{Z_{[1:K]}|S}p_{V_{[1:K]}|S,Z_{[1:K]}}\right)\nonumber \\
 & \qquad\subseteq\mathcal{R}_{\mathsf{DBC}}\left(p_{X}p_{Y_{[1:K]}|X}\right),\nonumber \\
 & \mathcal{R}_{\mathsf{SRC-SI}}\left(p_{S}p_{Z_{[1:K]}|S}p_{V_{[1:K]}|S,Z_{[1:K]}}\right)\nonumber \\
 & \qquad\supseteq\mathcal{R}_{\mathsf{SRC}}\left(p_{X}p_{Y_{[1:K]}|X}\right)\Bigr\},
\end{align}
where
\begin{align}
 & \mathcal{R}_{\mathsf{DBC-SI}}\left(p_{X}p_{Z_{[1:K]}|X}p_{Y_{[1:K]}|X,Z_{[1:K]}}\right)\triangleq\nonumber \\
 & \Bigl\{ R_{[1:K]}:R_{k}\geq0,\exists p_{V_{K-1}|X}p_{V_{K-2}|V_{K-1}}\cdots p_{V_{1}|V_{2}}\text{ s.t.}\nonumber \\
 & \sum_{j=1}^{k}R_{j}\leq\sum_{j=1}^{k}I\left(Y_{j};V_{j}|V_{j-1}Z_{j}\right),\nonumber \\
 & k\in[1:K],\text{ where }V_{0}\triangleq\emptyset,V_{K}\triangleq X\Bigr\},\label{eq:DBCcapacity-3-2}
\end{align}
and
\begin{align}
 & \mathcal{R}_{\mathsf{SRC-SI}}\left(p_{X}p_{Z_{[1:K]}|X}p_{Y_{[1:K]}|X,Z_{[1:K]}}\right)\triangleq\nonumber \\
 & \Bigl\{ R_{[1:K]}:R_{k}\geq0,\sum_{j=1}^{k}R_{j}\geq I\left(X;Y_{k}|Z_{k}\right),k\in[1:K]\Bigr\}.
\end{align}
\end{rem}

\subsection{\label{sub:Wyner-Ziv-Binary-Source}Wyner-Ziv Binary Broadcast}

Consider sending a binary source $S\sim\textrm{Bern}\left(\frac{1}{2}\right)$
with the Hamming distortion measure $d_{k}(s,\hat{s})=d(s,\hat{s})\triangleq0,\textrm{ if }s=\hat{s};1,\textrm{ otherwise},$
over a binary broadcast channel $Y_{k}=X\oplus W_{k},1\le k\le K$
with $W_{k}\sim\textrm{Bern}\left(p_{k}\right),\frac{1}{2}\geq p_{1}\geq p_{2}\geq\cdots\geq p_{K}\geq0$.
Assume the side information $Z_{k}$ observed by the receiver $k$
satisfies $S=Z_{k}\oplus B_{k}$ with independent variables $Z_{k}\sim\textrm{Bern}\left(\frac{1}{2}\right)$
and $B_{k}\sim\textrm{Bern}\left(\beta_{k}\right)$. Assume the bandwidth
mismatch factor is $b$.

Let $V_{1}=\left(U_{1},X_{1}^{b}\right),V_{2}=\left(U_{2},X^{b}\right),V_{3}=\emptyset$
with $U_{1}$ and $U_{2}$ are independent of $X_{1}^{b}$ and $X^{b}$.
$S$, $U_{2}$ and $U_{1}$ satisfy the distribution $p_{S}p_{U_{2}|S}p_{U_{1}|U_{2}}$,
where
\begin{align}
 & \begin{array}{c}
\qquad\qquad\qquad0\qquad1\qquad2\\
p_{U_{2}|S}=\begin{array}{c}
0\\
1
\end{array}\left(\begin{array}{ccc}
q_{2}\bar{\alpha}_{2} & q_{2}\alpha_{2} & \bar{q}_{2}\\
q_{2}\alpha_{2} & q_{2}\bar{\alpha}_{2} & \bar{q}_{2}
\end{array}\right)
\end{array},\\
 & \begin{array}{c}
\qquad\qquad\qquad0\qquad1\qquad2\\
p_{U_{1}|U_{2}}=\begin{array}{c}
0\\
1\\
2
\end{array}\left(\begin{array}{ccc}
q'_{1}\bar{\alpha}'_{1} & q'_{1}\alpha'_{1} & \bar{q}'_{1}\\
q'_{1}\alpha'_{1} & q'_{1}\bar{\alpha}'_{1} & \bar{q}'_{1}\\
0 & 0 & 1
\end{array}\right)
\end{array},
\end{align}
with $0\leq q_{2},q'_{1}\leq1,0\leq\alpha_{2},\alpha'_{1}\leq\frac{1}{2}.$
$X^{b}$ and $X_{1}^{b}$ satisfy $X_{1}^{b}=X^{b}\oplus B^{b},$
$X^{b}\sim b\textrm{-dimensional Bern}(\frac{1}{2}),$ and $B^{b}\sim b\textrm{-dimensional Bern}(\theta)$
with $0\leq\theta\leq\frac{1}{2}.$ Denote $\alpha_{1}=\alpha_{2}\star\alpha'_{1},q{}_{1}=q{}_{2}q'_{1}$,
and set $x^{b}(v_{2},s)=x^{b}$ and for $k=1,2$,
\begin{equation}
\hat{s}_{k}\left(v_{k},y_{k}^{b},z_{k}\right)=\begin{cases}
z_{k}, & \textrm{if }\alpha_{k}\geq\beta_{k}\textrm{ or }\alpha_{k}<\beta_{k},u_{k}=2;\\
u_{k}, & \textrm{if }\alpha_{k}<\beta_{k},u_{k}=0,1.
\end{cases}
\end{equation}
Substituting these random variables and functions into $\mathcal{D}_{\mathsf{SI}}^{(i)}$
in Theorem \ref{thm:AdmissibleRegionSI-GBC} (for this case, the hybrid
coding reduces to a layered digital coding), we get the following
performance, which is tighter than that of the Layered Description
Scheme (LDS) \cite[Lem. 4]{Nayak}.
\begin{cor}[Layered Digital Coding]
\label{thm:bandwidthmismatch} For transmitting a binary source $S$
with the Hamming distortion measure over a $2$-user binary broadcast
channel with side information $Z_{k}$ at the decoder $k$ ($k\in[1:K]$),
\begin{align}
 & \mathcal{D}_{\mathsf{SI}}\supseteq\mathcal{D}_{\mathsf{LDC}}^{(i)}\triangleq\nonumber \\
 & \Bigl\{\left(D_{1},D_{2}\right):0\leq q_{1}\leq q_{2}\leq1,\nonumber \\
 & 0\leq\alpha_{2}\leq\alpha_{1}\leq\frac{1}{2},0\leq\theta\leq\frac{1}{2},\nonumber \\
 & q_{1}r(\alpha_{1},\beta_{1})\leq b\left(1-H_{2}(\theta\star p_{1})\right),\nonumber \\
 & q_{1}r(\alpha_{1},\beta_{2})\leq b\left(1-H_{2}(\theta\star p_{2})\right),\nonumber \\
 & q_{2}r(\alpha_{2},\beta_{2})\leq b\left(1-H_{2}(p_{2})\right),\nonumber \\
 & q_{1}r(\alpha_{1},\beta_{1})+\left(q_{2}r(\alpha_{2},\beta_{2})-q_{1}r(\alpha_{1},\beta_{2})\right)\nonumber \\
 & \qquad\leq b\left(1-H_{2}(\theta\star p_{1})\right)+b\left(H_{2}(\theta\star p_{2})-H_{2}(p_{2})\right),\nonumber \\
 & D_{i}\leq q_{i}\min\left\{ \alpha_{i},\beta_{i}\right\} +(1-q_{i})\beta_{i},i=1,2\Bigr\},
\end{align}
where
\begin{equation}
r(\alpha,\beta)=H_{2}(\alpha\star\beta)-H_{2}(\alpha),
\end{equation}
$\star$ denotes the binary operation given in \eqref{eq:star}, and
$H_{2}$ denotes the binary entropy function given in \eqref{eq:binaryentropy}.
\end{cor}
In addition, the outer bound of Theorem \ref{thm:AdmissibleRegionSI-DBC}
reduces to the following one for the Wyner-Ziv binary case. The proof
is given in Appendix \ref{sec:broadcast-BernoulliSI}.
\begin{thm}
\label{thm:AdmissibleRegionBBSI} For transmitting a binary source
$S$ with the Hamming distortion measure over a $K$-user binary broadcast
channel with degraded side information $Z_{k}$ ($\frac{1}{2}\geq\beta_{1}\geq\beta_{2}\geq\cdots\geq\beta_{K}\geq0$)
at the decoder $k$ ($k\in[1:K]$),
\begin{align}
 & \mathcal{D}_{\mathsf{SI}}\subseteq\mathcal{D}_{\mathsf{SI-D}}^{(o)}\triangleq\nonumber \\
 & \Bigl\{ D_{[1:K]}:\exists\:0\leq\alpha_{1},\alpha_{2},\cdots,\alpha_{K}\leq\frac{1}{2}\text{ s.t. }\nonumber \\
 & \alpha_{k}\leq D'_{k}\triangleq\min\left\{ D_{k},\beta_{k}\right\} ,k\in[1:K],\nonumber \\
 & \textrm{and for any values }\frac{1}{2}=\tau_{0}\geq\tau_{1}\geq\cdots\geq\tau_{K}=0,\nonumber \\
 & \frac{1}{b}\Bigl(\eta_{k}\bigl(H_{2}\left(\beta_{k}\star\tau_{k}\right)-H_{2}\left(\beta_{k}\star\tau_{k-1}\right)\nonumber \\
 & \qquad-\left(H_{4}\left(\alpha_{k},\beta_{k},\tau_{k}\right)-H_{4}\left(\alpha_{k},\beta_{k},\tau_{k-1}\right)\right)\bigr):\nonumber \\
 & \qquad k\in[1:K]\Bigr)\in\mathcal{R}_{\mathsf{BBC}}\Bigr\},
\end{align}
where $\mathcal{R}_{\mathsf{BBC}}$ denotes the capacity region of
the binary broadcast channel given in \eqref{eq:binarycapacity},
\textup{
\begin{equation}
\eta_{k}\triangleq\begin{cases}
\frac{\beta_{k}-D'_{k}}{\beta_{k}-\alpha_{k}}, & \textrm{if }\alpha_{k}<\beta_{k},\\
0, & \textrm{otherwise,}
\end{cases}
\end{equation}
\begin{align}
H_{4}\left(x,y,z\right)\triangleq & -\left(xyz+\overline{x}\overline{y}\overline{z}\right)\log\left(xyz+\overline{x}\overline{y}\overline{z}\right)\nonumber \\
 & -\left(x\overline{y}z+\overline{x}y\overline{z}\right)\log\left(x\overline{y}z+\overline{x}y\overline{z}\right)\nonumber \\
 & -\left(xy\overline{z}+\overline{x}\overline{y}z\right)\log\left(xy\overline{z}+\overline{x}\overline{y}z\right)\nonumber \\
 & -\left(x\overline{y}\overline{z}+\overline{x}yz\right)\log\left(x\overline{y}\overline{z}+\overline{x}yz\right),\label{eq:H4}
\end{align}
}and $\overline{x}\triangleq1-x$.
\end{thm}
The bounds in Corollary \ref{thm:bandwidthmismatch} and Theorem \ref{thm:AdmissibleRegionBBSI}
are shown in Fig. \ref{fig:HDAcoding-1}.

\subsection{Wyner-Ziv Gaussian Broadcast}

Consider sending a Gaussian source $S\sim\mathcal{N}\left(0,N_{S}\right)$
with the quadratic distortion measure $d_{k}(s,\hat{s})=d(s,\hat{s})\triangleq(s-\hat{s})^{2}$
over a power-constrained Gaussian broadcast channel $Y_{k}=X+W_{k},1\le k\le K$
with $\mathbb{E}\left(X^{2}\right)\leq P$ and $W_{k}\sim\mathcal{N}\left(0,N_{k}\right),N_{1}\geq N_{2}\geq\cdots\geq N_{K}$.
Assume the side information $Z_{k}$ observed by the receiver $k$
satisfies $S=Z_{k}+B_{k}$ with independent Gaussian variables $Z_{k}\sim\mathcal{N}\left(0,N_{S}-\beta_{k}\right)$
and $B_{k}\sim\mathcal{N}\left(0,\beta_{k}\right)$. Assume the bandwidth
mismatch factor is $b$. As stated in Remark \ref{rem:Similar-to-Theorem},
Theorem \ref{thm:AdmissibleRegionSI-GBC} (or \ref{thm:AdmissibleRegionSI-DBC})
holds for the Gaussian case as well. The inner bound of Theorem \ref{thm:AdmissibleRegionSI-DBC}
recovers the existing results in \cite{Nayak,Gao}, and the outer
bound of Theorem \ref{thm:AdmissibleRegionSI-DBC} generates the following
outer bound for the Wyner-Ziv Gaussian source broadcast problem. The
proof is given in Appendix \ref{sec:broadcast-GaussianSI}.
\begin{thm}
\label{thm:AdmissibleRegionGGSI}For transmitting a Gaussian source
$S$ over a Gaussian broadcast channel with degraded side information
$Z_{k}$ ($\beta_{1}\geq\beta_{2}\geq\cdots\geq\beta_{K}$) at the
decoder $k${\footnotesize{}{} }($k\in[1:K]$),
\begin{align}
 & \mathcal{D}_{\mathsf{SI}}\subseteq\mathcal{D}_{\mathsf{SI-D}}^{(o)}\triangleq\nonumber \\
 & \Bigl\{ D_{[1:K]}:\textrm{For any values }+\infty=\tau_{0}\geq\tau_{1}\geq\cdots\geq\tau_{K}=0,\nonumber \\
 & \frac{1}{b}\left(\frac{1}{2}\log\frac{\left(D_{k}+\tau_{k-1}\right)\left(\beta_{k}+\tau_{k}\right)}{\left(D_{k}+\tau_{k}\right)\left(\beta_{k}+\tau_{k-1}\right)}:k\in[1:K]\right)\in\mathcal{R}_{\mathsf{GBC}}\Bigr\},
\end{align}
where $\mathcal{R}_{\mathsf{GBC}}$ denotes the capacity of Gaussian
broadcast channel given by
\begin{align}
\mathcal{R}_{\mathsf{GBC}}= & \Bigl\{ R_{[1:K]}:R_{k}\geq0,k\in[1:K],N_{K+1}=0,\nonumber \\
 & \sum_{k=1}^{K}(N_{k}-N_{k+1})\exp(2\sum_{j=1}^{k}R_{j})\leq P+N_{1}\Bigr\}.\label{eq:Gaussiancapacity}
\end{align}
\end{thm}
The bound of Theorem \ref{thm:AdmissibleRegionGGSI} is shown in Fig.
\ref{fig:HDAcoding-GG}.

\section{Concluding Remarks}

In this paper, we focused on the joint source-channel coding problem
of sending a memoryless source over a memoryless broadcast channel,
and developed an inner bound and two outer bounds for this problem.
The inner bound is achieved by using a unified hybrid coding scheme,
and it can recover the best known performance of hybrid coding. Similarly,
our outer bounds can also recover the best known outer bound in the
literature. Besides, we extend the results to the Wyner-Ziv source
broadcast problem.

The inner bounds achieved by the proposed hybrid coding is established
by using generalized multivariate covering and packing lemmas, and
the outer bounds are derived by introducing auxiliary random variables
(at the sender side or receiver sides). These lemmas and tools are
expected to be exploited to derive more and stronger achievability
and converse results for network information theory.

\appendices{}

\section{Proof of Lemma \ref{lem:Covering}}

We follow similar steps to the proof of mutual covering lemma \cite{Minero}.
Let
\begin{equation}
\mathcal{B}=\bigl\{ m_{[1:k]}\in\prod_{i=1}^{k}[1:2^{nr_{i}}]:(U^{n},V_{0}^{n},V_{[1:k]}^{n}(m_{[1:k]}))\in\mathcal{T}_{\epsilon}^{\left(n\right)}\bigr\}.
\end{equation}
Then we only need to show $\lim_{n\rightarrow\infty}\mathbb{P}\left(|\mathcal{B}|=0\right)=0$.
On the other hand,
\begin{align}
 & \lim_{n\rightarrow\infty}\mathbb{P}\left(|\mathcal{B}|=0\right)\nonumber \\
 & =\lim_{n\rightarrow\infty}\sum_{u^{n},v_{0}^{n}}p_{U^{n},V_{0}^{n}}\left(u^{n},v_{0}^{n}\right)\mathbb{P}\left(|\mathcal{B}|=0|u^{n},v_{0}^{n}\right)\\
 & \leq\lim_{n\rightarrow\infty}\sum_{(u^{n},v_{0}^{n})\in\mathcal{T}_{\epsilon'}^{\left(n\right)}}p_{U^{n},V_{0}^{n}}\left(u^{n},v_{0}^{n}\right)\mathbb{P}\left(|\mathcal{B}|=0|u^{n},v_{0}^{n}\right)\nonumber \\
 & \qquad+\lim_{n\rightarrow\infty}\mathbb{P}\left((u^{n},v_{0}^{n})\notin\mathcal{T}_{\epsilon'}^{\left(n\right)}\right)\\
 & =\lim_{n\rightarrow\infty}\sum_{(u^{n},v_{0}^{n})\in\mathcal{T}_{\epsilon'}^{\left(n\right)}}p_{U^{n},V_{0}^{n}}\left(u^{n},v_{0}^{n}\right)\mathbb{P}\left(|\mathcal{B}|=0|u^{n},v_{0}^{n}\right)\label{eq:}
\end{align}
To prove $\lim_{n\rightarrow\infty}\mathbb{P}\left(|\mathcal{B}|=0\right)=0$,
it is sufficient to show $\lim_{n\rightarrow\infty}\mathbb{P}\left(|\mathcal{B}|=0|u^{n},v_{0}^{n}\right)=0$
for any $(u^{n},v_{0}^{n})\in\mathcal{T}_{\epsilon'}^{\left(n\right)}$.
Utilizing the Chebyshev lemma \cite[App. B]{El Gamal}, we can bound
the probability as

\begin{align}
 & \mathbb{P}\left(|\mathcal{B}|=0|u^{n},v_{0}^{n}\right)\nonumber \\
 & \leq\mathbb{P}\left((|\mathcal{B}|-\mathbb{E}|\mathcal{B}|)^{2}\geq(E|\mathcal{B}|)^{2}|u^{n},v_{0}^{n}\right)\\
 & \leq\frac{\textrm{Var}(|\mathcal{B}||u^{n},v_{0}^{n})}{(\mathbb{E}(|\mathcal{B}||u^{n},v_{0}^{n}))^{2}}.\label{eq:-8}
\end{align}

Next we prove the upper bound $\frac{\textrm{Var}(|\mathcal{B}||u^{n},v_{0}^{n})}{(\mathbb{E}(|\mathcal{B}||u^{n},v_{0}^{n}))^{2}}$
tends to zero as $n\rightarrow\infty$. Define
\begin{equation}
E\left(m_{[1:k]}\right)\triangleq\begin{cases}
1, & \textrm{if }(u^{n},v_{0}^{n},V_{[1:k]}^{n}(m_{[1:k]}))\in\mathcal{T}_{\epsilon}^{\left(n\right)};\\
0, & \textrm{otherwise},
\end{cases}
\end{equation}
for each $m_{[1:k]}\in\prod_{i=1}^{k}[1:2^{nr_{i}}]$, then $|\mathcal{B}|$
can be expressed as
\begin{equation}
|\mathcal{B}|=\sum_{m_{[1:k]}\in\prod_{i=1}^{k}[1:2^{nr_{i}}]}E\left(m_{[1:k]}\right).
\end{equation}
Denote
\begin{align}
p_{0} & =\mathbb{P}\left((u^{n},v_{0}^{n},V_{[1:k]}^{n}(m_{[1:k]}))\in\mathcal{T}_{\epsilon}^{\left(n\right)}|u^{n},v_{0}^{n}\right),\\
p_{\mathcal{I}} & =\mathbb{P}\Bigl((u^{n},v_{0}^{n},V_{[1:k]}^{n}(m_{[1:k]}))\in\mathcal{T}_{\epsilon}^{\left(n\right)},\nonumber \\
 & \qquad(u^{n},v_{0}^{n},V_{[1:k]}^{n}(m{}_{\mathcal{I}},m'_{\mathcal{I}^{c}}))\in\mathcal{T}_{\epsilon}^{\left(n\right)}|u^{n},v_{0}^{n}\Bigr),
\end{align}
for $m_{[1:k]}=\boldsymbol{1}$, and $m_{[1:k]}^{\prime}=\boldsymbol{2}$.
Obviously, $p_{[1:k]}=p_{0}$. Then
\begin{align}
 & \mathbb{E}(|\mathcal{B}||u^{n},v_{0}^{n})\nonumber \\
 & =\sum_{m_{[1:k]}}\mathbb{P}\left((u^{n},v_{0}^{n},V_{[1:k]}^{n}(m_{[1:k]}))\in\mathcal{T}_{\epsilon}^{\left(n\right)}|u^{n},v_{0}^{n}\right)\\
 & =2^{n\sum_{j=1}^{k}r_{j}}p_{0},
\end{align}
and
\begin{align}
 & \mathbb{E}(|\mathcal{B}|^{2}|u^{n},v_{0}^{n})\nonumber \\
 & =\sum_{\mathcal{I}\subseteq[1:k]}\sum_{m_{[1:k]}}\sum_{m'_{\mathcal{I}^{c}}:m'_{\mathcal{I}^{c}}\nLeftrightarrow m{}_{\mathcal{I}^{c}}}\nonumber \\
 & \qquad\mathbb{P}\Bigl((u^{n},v_{0}^{n},V_{[1:k]}^{n}(m_{[1:k]}))\in\mathcal{T}_{\epsilon}^{\left(n\right)},\nonumber \\
 & \qquad(u^{n},v_{0}^{n},V_{[1:k]}^{n}(m{}_{\mathcal{I}},m'_{\mathcal{I}^{c}}))\in\mathcal{T}_{\epsilon}^{\left(n\right)}|u^{n},v_{0}^{n}\Bigr)\\
 & =2^{n\sum_{j=1}^{k}r_{j}}p_{0}+\sum_{\mathcal{I}\subsetneqq[1:k]}\sum_{m_{[1:k]}}\sum_{m'_{\mathcal{I}^{c}}:m'_{\mathcal{I}^{c}}\nLeftrightarrow m{}_{\mathcal{I}^{c}}}\nonumber \\
 & \qquad\mathbb{P}\Bigl((u^{n},v_{0}^{n},V_{[1:k]}^{n}(m_{[1:k]}))\in\mathcal{T}_{\epsilon}^{\left(n\right)},\nonumber \\
 & \qquad(u^{n},v_{0}^{n},V_{[1:k]}^{n}(m{}_{\mathcal{I}},m'_{\mathcal{I}^{c}}))\in\mathcal{T}_{\epsilon}^{\left(n\right)}|u^{n},v_{0}^{n}\Bigr).
\end{align}
Define
\begin{equation}
\mathbb{J}\triangleq\left\{ \mathcal{J}\subsetneqq[1:k]:\textrm{ if }j\in\mathcal{J},\textrm{ then }\mathcal{A}_{j}\subseteq\mathcal{J}\right\} .
\end{equation}
Then any set $\mathcal{I}\subsetneqq[1:k]$ can transform into a $\mathcal{J}\left(\mathcal{I}\right)\in\mathbb{J}$
by removing all the elements $j$'s such that $\mathcal{A}_{j}\nsubseteq\mathcal{I}$.
According to generation of random codebook, we can observe that $p_{\mathcal{I}}=p_{\mathcal{J}\left(\mathcal{I}\right)}$.
Therefore,
\begin{align}
 & \sum_{\mathcal{I}\subsetneqq[1:k]}\sum_{m_{[1:k]}}\sum_{m'_{\mathcal{I}^{c}}:m'_{\mathcal{I}^{c}}\nLeftrightarrow m{}_{\mathcal{I}^{c}}}\mathbb{P}\Bigl((u^{n},v_{0}^{n},V_{[1:k]}^{n}(m_{[1:k]}))\in\mathcal{T}_{\epsilon}^{\left(n\right)},\nonumber \\
 & \qquad(u^{n},v_{0}^{n},V_{[1:k]}^{n}(m{}_{\mathcal{I}},m'_{\mathcal{I}^{c}}))\in\mathcal{T}_{\epsilon}^{\left(n\right)}|u^{n},v_{0}^{n}\Bigr)\nonumber \\
 & =\sum_{\mathcal{I}\subsetneqq[1:k]}\sum_{m_{[1:k]}}\sum_{m'_{\mathcal{I}^{c}}:m'_{\mathcal{I}^{c}}\nLeftrightarrow m{}_{\mathcal{I}^{c}}}p_{\mathcal{J}\left(\mathcal{I}\right)}\\
 & \leq\sum_{\mathcal{I}\subsetneqq[1:k]}2^{n\left(\sum_{j=1}^{k}r_{j}+\sum_{j\in\mathcal{I}^{c}}r_{j}\right)}p_{\mathcal{J}\left(\mathcal{I}\right)}\\
 & \leq\sum_{\mathcal{I}\subsetneqq[1:k]}2^{n\left(\sum_{j=1}^{k}r_{j}+\sum_{j\in\left(\mathcal{J}\left(\mathcal{I}\right)\right)^{c}}r_{j}\right)}p_{\mathcal{J}\left(\mathcal{I}\right)}\label{eq:-56}\\
 & \leq\sum_{\mathcal{J}\in\mathbb{J}}2^{k-|\mathcal{J}|}2^{n\left(\sum_{j=1}^{k}r_{j}+\sum_{j\in\mathcal{J}^{c}}r_{j}\right)}p_{\mathcal{J}}\label{eq:-58}\\
 & \leq\sum_{\mathcal{J}\in\mathbb{J}}2^{n\left(\sum_{j=1}^{k}r_{j}+\sum_{j\in\mathcal{J}^{c}}r_{j}+o(1)\right)}p_{\mathcal{J}},
\end{align}
where \eqref{eq:-56} follows from $\mathcal{J}\left(\mathcal{I}\right)\subseteq\mathcal{I}$,
\eqref{eq:-58} follows from that for each $\mathcal{J}\subseteq\mathbb{J}$,
there are at most $2^{k-|\mathcal{J}|}$ of $\mathcal{I}$'s that
could transform into $\mathcal{J}$, and $o(1)$ denotes a term that
vanishes as $n\rightarrow\infty$. Hence
\begin{align}
 & \textrm{Var}(|\mathcal{B}||u^{n},v_{0}^{n})\nonumber \\
 & \leq\mathbb{E}(|\mathcal{B}|^{2}|u^{n},v_{0}^{n})\\
 & \leq2^{n\sum_{j=1}^{k}r_{j}}p_{0}+\sum_{\mathcal{J}\in\mathbb{J}}2^{n\left(\sum_{j=1}^{k}r_{j}+\sum_{j\in\mathcal{J}^{c}}r_{j}+o(1)\right)}p_{\mathcal{J}}.
\end{align}
Furthermore we have
\begin{align}
 & \frac{\textrm{Var}(|\mathcal{B}||u^{n},v_{0}^{n})}{(\mathbb{E}(|\mathcal{B}||u^{n},v_{0}^{n}))^{2}}\nonumber \\
 & \leq\frac{2^{n\sum_{j=1}^{k}r_{j}}p_{0}+\sum_{\mathcal{J}\in\mathbb{J}}2^{n\left(\sum_{j=1}^{k}r_{j}+\sum_{j\in\mathcal{J}^{c}}r_{j}+o(1)\right)}p_{\mathcal{J}}}{\left(2^{n\sum_{j=1}^{k}r_{j}}p_{0}\right)^{2}}\\
 & =2^{-n\sum_{j=1}^{k}r_{j}}\frac{1}{p_{0}}+\sum_{\mathcal{J}\in\mathbb{J}}2^{n\left(-\sum_{j\in\mathcal{J}}r_{j}+o(1)\right)}\frac{p_{\mathcal{J}}}{p_{0}^{2}}.\label{eq:-9}
\end{align}
According to the generation process of random codebook, we can observe
that
\begin{align}
p_{0} & =\sum_{v_{[1:k]}^{n}:(u^{n},v_{0}^{n},v_{[1:k]}^{n})\in\mathcal{T}_{\epsilon}^{\left(n\right)}}\mathbb{P}\left(V_{[1:k]}^{n}(m_{[1:k]})=v_{[1:k]}^{n}|u^{n},v_{0}^{n}\right)\\
 & \geq2^{-n\left(\sum_{j=1}^{k}H\left(V_{j}|V_{\mathcal{A}_{j}}V_{0}\right)-H\left(V_{[1:k]}|UV_{0}\right)+2\delta\left(\epsilon\right)\right)},\label{eq:-60}
\end{align}
where \eqref{eq:-60} follows from that for any $(u^{n},v_{0}^{n},v_{[1:k]}^{n})\in\mathcal{T}_{\epsilon}^{\left(n\right)}$,
\begin{align}
 & \mathbb{P}\left(V_{[1:k]}^{n}(m_{[1:k]})=v_{[1:k]}^{n}|u^{n},v_{0}^{n}\right)\nonumber \\
 & \geq2^{-n\left(\sum_{j=1}^{k}H\left(V_{j}|V_{\mathcal{A}_{j}}V_{0}\right)+\delta\left(\epsilon\right)\right)},
\end{align}
and for any $\left(u^{n},v_{0}^{n}\right)\in\mathcal{T}_{\epsilon'}^{\left(n\right)}$,
\begin{align}
 & \left|\left\{ v_{[1:k]}^{n}:(u^{n},v_{0}^{n},v_{[1:k]}^{n})\in\mathcal{T}_{\epsilon}^{\left(n\right)}\right\} \right|\nonumber \\
 & \geq2^{n\left(H\left(V_{[1:k]}|UV_{0}\right)+\delta\left(\epsilon\right)\right)}.
\end{align}
Similarly, we also can get
\begin{align}
p_{\mathcal{J}} & \leq\exp\Bigl\{-n\Bigl(\sum_{j=1}^{k}H\left(V_{j}|V_{\mathcal{A}_{j}}V_{0}\right)+\sum_{j\in\mathcal{J}^{c}}H\left(V_{j}|V_{\mathcal{A}_{j}}V_{0}\right)\nonumber \\
 & \qquad-H\left(V_{[1:k]}|UV_{0}\right)-H\left(V_{\mathcal{J}^{c}}|UV_{0}V_{\mathcal{J}}\right)-4\delta\left(\epsilon\right)\Bigr)\Bigr\}.\label{eq:-61}
\end{align}
Substituting \eqref{eq:-60} and \eqref{eq:-61} into \eqref{eq:-9},
we have
\begin{align}
 & \frac{\textrm{Var}(|\mathcal{B}||u^{n},v_{0}^{n})}{(\mathbb{E}(|\mathcal{B}||u^{n},v_{0}^{n}))^{2}}\nonumber \\
 & \leq\exp\Bigl\{-n\Bigl(\sum_{j=1}^{k}r_{j}-\Bigl(\sum_{j=1}^{k}H\left(V_{j}|V_{\mathcal{A}_{j}}V_{0}\right)\nonumber \\
 & \qquad-H\left(V_{[1:k]}|UV_{0}\right)+2\delta\left(\epsilon\right)\Bigr)\Bigr)\Bigr\}\nonumber \\
 & \qquad+\sum_{\mathcal{J}\in\mathbb{J}}\exp\Bigl\{-n\Bigl(\sum_{j=1}^{k}r_{j}-\Bigl(\sum_{j\in\mathcal{J}}H\left(V_{j}|V_{\mathcal{A}_{j}}V_{0}\right)\nonumber \\
 & \qquad-H\left(V_{\mathcal{J}}|UV_{0}\right)+6\delta\left(\epsilon\right)+o(1)\Bigr)\Bigr)\Bigr\}.\label{eq:-10}
\end{align}
\eqref{eq:-10} tends to zero if
\begin{align}
 & \sum_{j=1}^{k}r_{j}>\sum_{j=1}^{k}H\left(V_{j}|V_{\mathcal{A}_{j}}V_{0}\right)-H\left(V_{[1:k]}|UV_{0}\right)+2\delta\left(\epsilon\right)\\
 & \sum_{j\in\mathcal{J}}r_{j}>\sum_{j\in\mathcal{J}}H\left(V_{j}|V_{\mathcal{A}_{j}}V_{0}\right)-H\left(V_{\mathcal{J}}|UV_{0}\right)+6\delta\left(\epsilon\right)+o(1),
\end{align}
i.e., $\sum_{j\in\mathcal{J}}r_{j}>\sum_{j\in\mathcal{J}}H\left(V_{j}|V_{\mathcal{A}_{j}}V_{0}\right)-H\left(V_{\mathcal{J}}|UV_{0}\right)+\delta^{\prime}\left(\epsilon\right)$
for some $\delta^{\prime}\left(\epsilon\right)$ that tends to zero
as $\epsilon\rightarrow0$. This completes the proof.

\section{Proof of Lemma \ref{lem:Packing}}

For any $\mathcal{J}$ such that $\mathcal{J}\neq\emptyset$ and if
$j\in\mathcal{J}$ then $\mathcal{A}_{j}\subseteq\mathcal{J}$,
\begin{align}
 & \mathbb{P}\left((U^{n},V_{0}^{n},V_{[1:k]}^{n}(m_{[1:k]}))\in\mathcal{T}_{\epsilon}^{\left(n\right)}\textrm{ for some }m_{[1:k]}\right)\nonumber \\
 & \leq\mathbb{P}\left((U^{n},V_{0}^{n},V_{\mathcal{J}}^{n}(m_{\mathcal{J}}))\in\mathcal{T}_{\epsilon}^{\left(n\right)}\textrm{ for some }m_{\mathcal{J}}\right)\\
 & =\sum_{u^{n},v_{0}^{n}}p_{U^{n},V_{0}^{n}}\left(u^{n},v_{0}^{n}\right)\nonumber \\
 & \qquad\times\mathbb{P}\left((u^{n},v_{0}^{n},V_{\mathcal{J}}^{n}(m_{\mathcal{J}}))\in\mathcal{T}_{\epsilon}^{\left(n\right)}\textrm{ for some }m_{\mathcal{J}}|u^{n},v_{0}^{n}\right)\\
 & \leq\sum_{u^{n},v_{0}^{n}}p_{U^{n},V_{0}^{n}}\left(u^{n},v_{0}^{n}\right)\nonumber \\
 & \qquad\times\sum_{m_{\mathcal{J}}}\mathbb{P}\left((u^{n},v_{0}^{n},V_{\mathcal{J}}^{n}(m_{\mathcal{J}}))\in\mathcal{T}_{\epsilon}^{\left(n\right)}|u^{n},v_{0}^{n}\right).\label{eq:-59}
\end{align}
Similar to \eqref{eq:-60}, we can obtain that
\begin{align}
 & \mathbb{P}\left((u^{n},v_{0}^{n},V_{\mathcal{J}}^{n}(m_{\mathcal{J}}))\in\mathcal{T}_{\epsilon}^{\left(n\right)}|u^{n},v_{0}^{n}\right)\nonumber \\
 & \leq2^{-n\left(\sum_{j\in\mathcal{J}}H\left(V_{j}|V_{\mathcal{A}_{j}}V_{0}\right)-H\left(V_{\mathcal{J}}|UV_{0}\right)-2\delta\left(\epsilon\right)\right)}.
\end{align}
Substituting it into \eqref{eq:-59}, we have
\begin{align}
 & \mathbb{P}\left((U^{n},V_{0}^{n},V_{[1:k]}^{n}(m_{[1:k]}))\in\mathcal{T}_{\epsilon}^{\left(n\right)}\textrm{ for some }m_{[1:k]}\right)\nonumber \\
 & \leq2^{n\left(\sum_{j\in\mathcal{J}}r_{j}-\left(\sum_{j\in\mathcal{J}}H\left(V_{j}|V_{\mathcal{A}_{j}}V_{0}\right)-H\left(V_{\mathcal{J}}|UV_{0}\right)-2\delta\left(\epsilon\right)\right)\right)}.\label{eq:-11}
\end{align}
\eqref{eq:-11} tends to zero if
\begin{align}
 & \sum_{j\in\mathcal{J}}r_{j}<\sum_{j\in\mathcal{J}}H\left(V_{j}|V_{\mathcal{A}_{j}}V_{0}\right)-H\left(V_{\mathcal{J}}|UV_{0}\right)-2\delta\left(\epsilon\right).
\end{align}
This completes the proof.

\section{Proof of Theorem \ref{thm:AdmissibleRegion-GBC}\label{sec:broadcast}}

\subsection{\label{sub:Inner-Bound}Inner Bound}

Actually the inner bound can be seen as a corollary to \cite[Thm. 1]{Lee}
by choosing a proper set of network topology, transit probability
and symbol-by-symbol functions. For completeness and clarity, next
we provide a direct description of the proposed hybrid coding scheme
and a direct proof for it.

\emph{Codebook Generation}: Fix the conditional pmf $p_{V_{[1:N]}|S}$,
vector $r_{[1:N]}$, encoding function $x\left(v_{[1:N]},s\right)$
and decoding functions $\hat{s}_{k}\left(v_{\mathcal{B}_{k}},y_{k}\right)$
that satisfy
\begin{align}
\mathbb{E}d_{k}\left(S,\hat{S}_{k}\right) & \le D_{k},1\le k\le K,\\
\sum_{j\in\mathcal{J}}r_{j} & >\sum_{j\in\mathcal{J}}H\left(V_{j}|V_{\mathcal{A}_{j}}\right)-H\left(V_{\mathcal{J}}|S\right)\nonumber \\
 & \qquad\textrm{for all }\mathcal{J}\subseteq[1:N]\textrm{ s.t. }\mathcal{J}\neq\emptyset\nonumber \\
 & \qquad\textrm{and }\mathcal{A}_{j}\subseteq\mathcal{J},\forall j\in\mathcal{J},\label{eq:-14}\\
\sum_{j\in\mathcal{J}^{c}}r_{j} & <\sum_{j\in\mathcal{J}^{c}}H\left(V_{j}|V_{\mathcal{A}_{j}}\right)-H\left(V_{\mathcal{J}^{c}}|Y_{k}V_{\mathcal{J}}\right),\nonumber \\
 & \qquad k\in[1:K]\textrm{ for all }\mathcal{J}\subseteq\mathcal{B}_{k}\textrm{ s.t. }\nonumber \\
 & \qquad\mathcal{J}^{c}\triangleq\mathcal{B}_{k}\backslash\mathcal{J}\neq\emptyset\textrm{ and }\mathcal{A}_{j}\subseteq\mathcal{J},\forall j\in\mathcal{J}.\label{eq:-16}
\end{align}
For each $j\in[1:N]$ and each $m_{\mathcal{A}_{j}}\in\prod_{i\in\mathcal{A}_{j}}[1:2^{nr_{i}}]$,
randomly and independently generate a set of sequences $v_{j}^{n}(m_{\mathcal{A}_{j}},m_{j}),m_{j}\in[1:2^{nr_{j}}],$
with each distributed according to $\prod_{i=1}^{n}p_{V_{j}|V_{\mathcal{A}_{j}}}(v_{j,i}|v_{\mathcal{A}_{j},i}(m_{\mathcal{A}_{j}}))$.
The codebook
\begin{equation}
\mathcal{C}=\left\{ \begin{array}{c}
v_{[1:N]}^{n}\left(m_{[1:N]}\right):m_{[1:N]}\in\prod_{i=1}^{N}[1:2^{nr_{i}}]\end{array}\right\} .
\end{equation}
is revealed to the encoder and all the decoders.

\emph{Encoding}: We use joint typicality encoding. Given $s^{n}$,
encoder finds the smallest index vector $m_{[1:N]}$ such that $\left(s^{n},v_{[1:N]}^{n}\left(m_{[1:N]}\right)\right)\in\mathcal{T}_{\epsilon}^{\left(n\right)}$.
If there is no such index vector, let $m_{[1:N]}=\mathbf{1}$. Then
the encoder transmits the signal
\begin{equation}
x_{i}=x\left(v_{[1:N],i}\left(m_{[1:N]}\right),s_{i}\right),1\leq i\leq n.
\end{equation}

\emph{Decoding}: We use joint typicality decoding. Let $\epsilon'>\epsilon$.
Upon receiving the signal $y_{k}^{n}$, the decoder of the receiver
$k$ finds the smallest index vector $\hat{m}_{\mathcal{B}_{k}}^{(k)}$
such that
\begin{equation}
(v_{\mathcal{B}_{k}}^{n}(\hat{m}_{\mathcal{B}_{k}}^{(k)}),y_{k}^{n})\in\mathcal{T}_{\epsilon'}^{\left(n\right)}.
\end{equation}
If there is no such index vector, let $\hat{m}_{\mathcal{B}_{k}}^{(k)}=\mathbf{1}$.
The decoder reconstructs the source as
\begin{equation}
\hat{s}_{k,i}=\hat{s}_{k}(v_{\mathcal{B}_{k},i}(\hat{m}_{\mathcal{B}_{k}}^{(k)}),y_{k,i}),1\leq i\leq n.
\end{equation}

\emph{Analysis of Expected Distortion}: We bound the distortions averaged
over $S^{n}$ and the random codebook $\mathcal{C}$. Define the ``error\textquotedblright{}
event
\begin{align}
\mathcal{E} & =\mathcal{E}_{1}\cup\left(\bigcup_{k}\mathcal{E}_{2,k}\right)\cup\left(\bigcup_{k}\mathcal{E}{}_{3,k}\right),
\end{align}
where
\begin{align}
\mathcal{E}_{1} & =\left\{ \left(S^{n},V_{[1:N]}^{n}\left(m_{[1:N]}\right)\right)\notin\mathcal{T}_{\epsilon}^{\left(n\right)}\textrm{ for all }m_{[1:N]}\right\} ,\\
\mathcal{E}_{2,k} & =\left\{ \left(S^{n},V_{[1:N]}^{n}\left(M_{[1:N]}\right),Y_{k}^{n}\right)\notin\mathcal{T}_{\epsilon'}^{\left(n\right)}\right\} ,\\
\mathcal{E}{}_{3,k} & =\Bigl\{\left(V_{\mathcal{B}_{k}}^{n}(m'_{\mathcal{B}_{k}}),Y_{k}^{n}\right)\in\mathcal{T}_{\epsilon'}^{\left(n\right)}\nonumber \\
 & \qquad\textrm{ for some }m'_{\mathcal{B}_{k}},m'_{\mathcal{B}_{k}}\neq M{}_{\mathcal{B}_{k}}\Bigr\},
\end{align}
for $1\le k\le K.$ Using the union bound, we have
\begin{align}
\mathbb{P}\left(\mathcal{E}\right) & \leq\mathbb{P}\left(\mathcal{E}_{1}\right)+\sum_{k=1}^{K}\mathbb{P}\left(\mathcal{E}_{1}^{c}\bigcap\mathcal{E}_{2,k}\right)+\sum_{k=1}^{K}\mathbb{P}\left(\mathcal{E}{}_{3,k}\right).\label{eq:proberror}
\end{align}

Now we claim that if \eqref{eq:-14} and \eqref{eq:-16} hold, then
$\mathbb{P}\left(\mathcal{E}\right)$ tends to zero as $n\to\infty$.
Before proving it, we show that this claim implies the inner bound
of Theorem \ref{thm:AdmissibleRegion-GBC}.

Define
\begin{equation}
\mathcal{E}_{4,k}=\left\{ \left(S^{n},V_{\mathcal{B}_{k}}^{n}(\hat{M}_{\mathcal{B}_{k}}^{(k)}),Y_{k}^{n}\right)\notin\mathcal{T}_{\epsilon'}^{\left(n\right)}\right\} .
\end{equation}
then we have $\mathcal{E}^{c}\subseteq\mathcal{E}_{4,k}^{c}$, i.e.,
$\mathcal{E}_{4,k}\subseteq\mathcal{E}$. This implies that $\mathbb{P}\left(\mathcal{E}_{4,k}\right)\leq\mathbb{P}\left(\mathcal{E}\right)\rightarrow0$
as $n\to\infty$. Then utilizing the typical average lemma \cite{El Gamal},
we have
\begin{align}
 & \limsup_{n\to\infty}\mathbb{E}d_{k}\left(S^{n},\hat{S}_{k}^{n}\right)\nonumber \\
 & =\limsup_{n\to\infty}\Bigl\{\mathbb{P}\left(\mathcal{E}_{4,k}\right)\mathbb{E}\left[d_{k}\left(S^{n},\hat{S}_{k}^{n}\right)|\mathcal{E}_{4,k}\right]\nonumber \\
 & \qquad+\mathbb{P}\left(\mathcal{E}_{4,k}^{c}\right)\mathbb{E}\left[d_{k}\left(S^{n},\hat{S}_{k}^{n}\right)|\mathcal{E}_{4,k}^{c}\right]\Bigr\}\\
 & =\limsup_{n\to\infty}\mathbb{E}\left[d_{k}\left(S^{n},\hat{S}_{k}^{n}\right)|\mathcal{E}_{4,k}^{c}\right]\\
 & \le\left(1+\epsilon'\right)\mathbb{E}d_{k}\left(S,\hat{S}_{k}\right)\\
 & \le\left(1+\epsilon'\right)D_{k}.
\end{align}
Therefore, the desired distortions are achieved for sufficiently small
$\epsilon'$.

Next we turn back to prove the claim above. Following from the multivariate
covering lemma (Lemma \ref{lem:Covering}), the first term of \eqref{eq:proberror},
$\mathbb{P}\left(\mathcal{E}_{1}\right)$, vanishes as $n\to\infty$,
and according to conditional typicality lemma \cite[Sec. 3.7]{El Gamal},
the second item tends to zero as $n\to\infty$.

Now we focus on the third term of \eqref{eq:proberror}. $\mathcal{E}{}_{3,k}$
can be writen as
\begin{align}
\mathcal{E}{}_{3,k} & =\bigcup_{\mathcal{I}\subseteq\mathcal{B}_{k}}\mathcal{E}{}_{3,k}^{\mathcal{I}},
\end{align}
where
\begin{align}
\mathcal{E}{}_{3,k}^{\mathcal{I}} & =\Bigl\{\left(V_{\mathcal{B}_{k}}^{n}\left(M_{\mathcal{I}},m'_{\mathcal{I}^{c}}\right),Y_{k}^{n}\right)\in\mathcal{T}_{\epsilon'}^{\left(n\right)}\nonumber \\
 & \qquad\textrm{ for some }m'_{\mathcal{I}^{c}},m'_{\mathcal{I}^{c}}\nLeftrightarrow M_{\mathcal{I}^{c}}\Bigr\},
\end{align}
with $\mathcal{I}^{c}\triangleq\mathcal{B}_{k}\backslash\mathcal{I}.$
Using the union bound we have
\begin{align}
\mathbb{P}\left(\mathcal{E}{}_{3,k}\right) & \leq\sum_{\mathcal{I}\subseteq\mathcal{B}_{k}}\mathbb{P}\left(\mathcal{E}{}_{3,k}^{\mathcal{I}}\right).\label{eq:proberror-1}
\end{align}
Each $\mathcal{B}_{k}$ has a finite number of subsets, hence we only
need to show for each $\mathcal{I}\subseteq\mathcal{B}_{k}$, $\mathbb{P}\left(\mathcal{E}{}_{3,k}^{\mathcal{I}}\right)$
vanishes as $n\to\infty$. To show this, it is needed to analyze the
correlation between coding index $M_{[1:N]}$ and nonchosen codewords.
Specifically, $M_{[1:N]}$ depends on the source sequence and the
entire codebook, and hence the standard packing lemma cannot be applied
directly. This problem has been resolved by the technique developed
in \cite{Minero,Lee}.
\begin{align}
 & \mathbb{P}\Bigl(\left(V_{\mathcal{B}_{k}}^{n}\left(M_{\mathcal{I}},m'_{\mathcal{I}^{c}}\right),Y_{k}^{n}\right)\in\mathcal{T}_{\epsilon'}^{\left(n\right)}\nonumber \\
 & \qquad\textrm{ for some }m'_{\mathcal{I}^{c}},m'_{\mathcal{I}^{c}}\nLeftrightarrow M_{\mathcal{I}^{c}}\Bigr)\nonumber \\
= & \sum_{m_{[1:N]}}\sum_{y_{k}^{n}}\mathbb{P}\Bigl(M_{[1:N]}=m_{[1:N]},Y_{k}^{n}=y_{k}^{n},\nonumber \\
 & \qquad\left(V_{\mathcal{B}_{k}}^{n}\left(m_{\mathcal{I}},m'_{\mathcal{I}^{c}}\right),y_{k}^{n}\right)\in\mathcal{T}_{\epsilon'}^{\left(n\right)}\nonumber \\
 & \textrm{\qquad\ for some }m'_{\mathcal{I}^{c}},m'_{\mathcal{I}^{c}}\nLeftrightarrow m_{\mathcal{I}^{c}}\Bigr)\\
\leq & \sum_{m_{[1:N]}}\sum_{y_{k}^{n}}\sum_{m'_{\mathcal{I}^{c}}:m'_{\mathcal{I}^{c}}\nLeftrightarrow m_{\mathcal{I}^{c}}}\mathbb{P}\Bigl(M_{[1:N]}=m_{[1:N]},\nonumber \\
 & \qquad Y_{k}^{n}=y_{k}^{n},\left(V_{\mathcal{B}_{k}}^{n}\left(m_{\mathcal{I}},m'_{\mathcal{I}^{c}}\right),y_{k}^{n}\right)\in\mathcal{T}_{\epsilon'}^{\left(n\right)}\Bigr)\label{eq:-5}
\end{align}
where \eqref{eq:-5} follows from the union bound.

Define a sub-codebook as
\begin{align}
\mathcal{C}_{(m_{\mathcal{I}},m_{\mathcal{I}^{c}}^{\prime})} & =\Bigl\{ V_{[1:N]}^{n}\left(m_{\mathcal{I}},m_{\mathcal{I}^{c}}^{\prime\prime},m_{\mathcal{B}_{k}^{c}}^{\prime\prime}\right):\nonumber \\
 & \qquad\forall\left(m_{\mathcal{I}^{c}}^{\prime\prime},m_{\mathcal{B}_{k}^{c}}^{\prime\prime}\right),m_{\mathcal{I}^{c}}^{\prime\prime}\nLeftrightarrow m_{\mathcal{I}^{c}}^{\prime}\Bigr\}.
\end{align}
Define another coding index as $\tilde{M}_{[1:N]}$ which is generated
by performing the same coding process as $M_{[1:N]}$ but on the codebook
$\mathcal{C}_{(m_{\mathcal{I}},m_{\mathcal{I}^{c}}^{\prime})}$, i.e.,
given the source sequence $s^{n}$, the encoder finds the smallest
index vector $\tilde{m}_{[1:N]}$ such that $\left(s^{n},v_{[1:N]}^{n}\left(\tilde{m}_{[1:N]}\right)\right)\in\mathcal{T}_{\epsilon}^{\left(n\right)}$;
if there is no such index vector, let $\tilde{m}_{[1:N]}=\mathbf{1}$.
Then according to the generation process of $M_{[1:N]}$ and $\tilde{M}_{[1:N]}$,
we have if $M_{[1:N]}=m_{[1:N]}$, then $\tilde{M}_{[1:N]}=m_{[1:N]}$.
Now continuing with \eqref{eq:-5}, we have
\begin{align}
 & \mathbb{P}\left(M_{[1:N]}=m_{[1:N]},Y_{k}^{n}=y_{k}^{n},\left(V_{\mathcal{B}_{k}}^{n}\left(m_{\mathcal{I}},m'_{\mathcal{I}^{c}}\right),y_{k}^{n}\right)\in\mathcal{T}_{\epsilon'}^{\left(n\right)}\right)\nonumber \\
 & =\sum_{v_{\mathcal{B}_{k}}^{\prime n},c,s^{n}}\mathbb{P}\Bigl(M_{[1:N]}=m_{[1:N]},\mathcal{C}_{(m_{\mathcal{I}},m_{\mathcal{I}^{c}}^{\prime})}=c,\nonumber \\
 & \qquad S^{n}=s^{n},V_{\mathcal{B}_{k}}^{n}\left(m_{\mathcal{I}},m'_{\mathcal{I}^{c}}\right)=v_{\mathcal{D}_{k}}^{\prime n}\Bigr)\nonumber \\
 & \qquad\times\prod_{i=1}^{n}p_{Y_{k}|X}\left(y_{k,i}|x\left(v_{[1:N],i}(m_{[1:N]}),s_{i}\right)\right)\nonumber \\
 & \qquad\times1\left\{ \left(v_{\mathcal{B}_{k}}^{\prime n},y_{k}^{n}\right)\in\mathcal{T}_{\epsilon'}^{\left(n\right)}\right\} \\
 & =\sum_{v_{\mathcal{B}_{k}}^{\prime n},c,s^{n}}\mathbb{P}\Bigl(M_{[1:N]}=m_{[1:N]},\tilde{M}_{[1:N]}=m_{[1:N]},\nonumber \\
 & \qquad\mathcal{C}_{(m_{\mathcal{I}},m_{\mathcal{I}^{c}}^{\prime})}=c,S^{n}=s^{n},V_{\mathcal{B}_{k}}^{n}\left(m_{\mathcal{I}},m'_{\mathcal{I}^{c}}\right)=v_{\mathcal{D}_{k}}^{\prime n}\Bigr)\nonumber \\
 & \qquad\times\prod_{i=1}^{n}p_{Y_{k}|X}\left(y_{k,i}|x\left(v_{[1:N],i}(m_{[1:N]}),s_{i}\right)\right)\nonumber \\
 & \qquad\times1\left\{ \left(v_{\mathcal{B}_{k}}^{\prime n},y_{k}^{n}\right)\in\mathcal{T}_{\epsilon'}^{\left(n\right)}\right\} \\
 & \leq\sum_{v_{\mathcal{B}_{k}}^{\prime n},c,s^{n}}\mathbb{P}\Bigl(\tilde{M}_{[1:N]}=m_{[1:N]},\mathcal{C}_{(m_{\mathcal{I}},m_{\mathcal{I}^{c}}^{\prime})}=c,\nonumber \\
 & \qquad S^{n}=s^{n},V_{\mathcal{B}_{k}}^{n}\left(m_{\mathcal{I}},m'_{\mathcal{I}^{c}}\right)=v_{\mathcal{D}_{k}}^{\prime n}\Bigr)\nonumber \\
 & \qquad\times\prod_{i=1}^{n}p_{Y_{k}|X}\left(y_{k,i}|x\left(v_{[1:N],i}(m_{[1:N]}),s_{i}\right)\right)\nonumber \\
 & \qquad\times1\left\{ \left(v_{\mathcal{B}_{k}}^{\prime n},y_{k}^{n}\right)\in\mathcal{T}_{\epsilon'}^{\left(n\right)}\right\} \\
 & =\sum_{v_{\mathcal{B}_{k}}^{\prime n},c,s^{n}}\mathbb{P}\left(\tilde{M}_{[1:N]}=m_{[1:N]},\mathcal{C}_{(m_{\mathcal{I}},m_{\mathcal{I}^{c}}^{\prime})}=c,S^{n}=s^{n}\right)\nonumber \\
 & \qquad\times\prod_{i=1}^{n}p_{Y_{k}|X}\left(y_{k,i}|x\left(v_{[1:N],i}(m_{[1:N]}),s_{i}\right)\right)\nonumber \\
 & \qquad\times\mathbb{P}\left(V_{\mathcal{B}_{k}}^{n}\left(m_{\mathcal{I}},m'_{\mathcal{I}^{c}}\right)=v_{\mathcal{B}_{k}}^{\prime n}|\mathcal{C}_{(m_{\mathcal{I}},m_{\mathcal{I}^{c}}^{\prime})}=c\right)\nonumber \\
 & \qquad\times1\left\{ \left(v_{\mathcal{B}_{k}}^{\prime n},y_{k}^{n}\right)\in\mathcal{T}_{\epsilon'}^{\left(n\right)}\right\} \label{eq:-7}
\end{align}
where $c=\Bigl\{ v_{[1:N]}^{n}\left(m_{\mathcal{I}},m_{\mathcal{I}^{c}}^{\prime\prime},m_{\mathcal{B}_{k}^{c}}^{\prime\prime}\right):\forall\left(m_{\mathcal{I}^{c}}^{\prime\prime},m_{\mathcal{B}_{k}^{c}}^{\prime\prime}\right),m_{\mathcal{I}^{c}}^{\prime\prime}\nLeftrightarrow m_{\mathcal{I}^{c}}^{\prime}\Bigr\}$,
and \eqref{eq:-7} follows from the fact that $V_{\mathcal{B}_{k}}^{n}\left(m_{\mathcal{I}},m'_{\mathcal{I}^{c}}\right)\rightarrow\mathcal{C}_{(m_{\mathcal{I}},m_{\mathcal{I}^{c}}^{\prime})}\rightarrow\left(S^{n},\tilde{M}_{[1:N]}\right)$
forms a Markov chain.

Define
\begin{equation}
\mathbb{J}\triangleq\left\{ \mathcal{J}\subseteq\mathcal{B}_{k}:\mathcal{A}_{j}\subseteq\mathcal{J},\forall j\in\mathcal{J}\right\} .
\end{equation}
Then any set $\mathcal{I}\subseteq\mathcal{B}_{k}$ can be transformed
into a $\mathcal{J}\left(\mathcal{I}\right)\in\mathbb{J}$ by removing
all the elements $j$'s such that $\mathcal{A}_{j}\nsubseteq\mathcal{I}$.
Denote $\mathcal{J}^{c}\triangleq\mathcal{B}_{k}\backslash\mathcal{J}.$
Then according to the generation process of the codebook, continuing
with \eqref{eq:-7}, we have
\begin{align}
 & \sum_{v_{\mathcal{B}_{k}}^{\prime n}}\mathbb{P}\left(V_{\mathcal{B}_{k}}^{n}\left(m_{\mathcal{I}},m'_{\mathcal{I}^{c}}\right)=v_{\mathcal{B}_{k}}^{\prime n}|\mathcal{C}_{(m_{\mathcal{I}},m_{\mathcal{I}^{c}}^{\prime})}=c\right)\nonumber \\
 & \qquad\times1\left\{ \left(v_{\mathcal{B}_{k}}^{\prime n},y_{k}^{n}\right)\in\mathcal{T}_{\epsilon'}^{\left(n\right)}\right\} \nonumber \\
 & =\sum_{v_{\mathcal{J}^{c}}^{\prime n}}\mathbb{P}\left(V_{\mathcal{J}^{c}}^{n}\left(m_{\mathcal{I}},m'_{\mathcal{I}^{c}}\right)=v_{\mathcal{J}^{c}}^{\prime n}|\mathcal{C}_{(m_{\mathcal{I}},m_{\mathcal{I}^{c}}^{\prime})}=c\right)\nonumber \\
 & \qquad\times1\left\{ \left(v_{\mathcal{J}}^{n}\left(m_{\mathcal{J}}\right),v_{\mathcal{J}^{c}}^{\prime n},y_{k}^{n}\right)\in\mathcal{T}_{\epsilon'}^{\left(n\right)}\right\} \\
 & =\sum_{v_{\mathcal{J}^{c}}^{\prime n}}\prod_{j\in\mathcal{J}^{c}}\prod_{i=1}^{n}p_{V_{j}|V_{\mathcal{A}_{j}}}\left(v_{j,i}^{\prime}|v_{\mathcal{A}_{j}\cap\mathcal{J},i}\left(m_{\mathcal{J}}\right),v_{\mathcal{A}_{j}\cap\mathcal{J}^{c},i}^{\prime}\right)\nonumber \\
 & \qquad\times1\left\{ \left(v_{\mathcal{J}}^{n}\left(m_{\mathcal{J}}\right),v_{\mathcal{J}^{c}}^{\prime n},y_{k}^{n}\right)\in\mathcal{T}_{\epsilon'}^{\left(n\right)}\right\} \label{eq:-17}\\
 & \leq2^{n\left(H\left(V_{\mathcal{J}^{c}}|Y_{k}V_{\mathcal{J}}\right)-\sum_{j\in\mathcal{J}^{c}}H\left(V_{j}|V_{\mathcal{A}_{j}}\right)+\left(|\mathcal{J}^{c}|+1\right)\delta\left(\epsilon'\right)\right)},\label{eq:-22}
\end{align}
where $\delta\left(\epsilon'\right)$ is a term that tends to zero
as $\epsilon'\rightarrow0$, and \eqref{eq:-22} follows from the
fact that $\prod_{i=1}^{n}p_{V_{j}|V_{\mathcal{A}_{j}}}\left(v_{j,i}|v_{\mathcal{A}_{j},i}\right)\leq2^{-n\left(H\left(V_{j}|V_{\mathcal{A}_{j}}\right)-\delta\left(\epsilon'\right)\right)}$
for any $\left(v_{j}^{n},v_{\mathcal{A}_{j}}^{n}\right)\in\mathcal{T}_{\epsilon'}^{\left(n\right)}$
and $\left|\left\{ v_{\mathcal{J}^{c}}^{\prime n}:\left(v_{\mathcal{J}}^{n},v_{\mathcal{J}^{c}}^{\prime n},y_{k}^{n}\right)\in\mathcal{T}_{\epsilon'}^{\left(n\right)}\right\} \right|\leq2^{n\left(H\left(V_{\mathcal{J}^{c}}|Y_{k}V_{\mathcal{J}}\right)+\delta\left(\epsilon'\right)\right)}$
for any $\left(y_{k}^{n},v_{\mathcal{J}}^{n}\right)$.

Combining \eqref{eq:-5}, \eqref{eq:-7} and \eqref{eq:-22} gives
\begin{align}
 & \mathbb{P}\bigl((V_{\mathcal{B}_{k}}^{n}(M_{\mathcal{I}},m'_{\mathcal{I}^{c}}),Y_{k}^{n})\in\mathcal{T}_{\epsilon'}^{\left(n\right)}\nonumber \\
 & \qquad\textrm{ for some }m'_{\mathcal{I}^{c}},m'_{\mathcal{I}^{c}}\nLeftrightarrow M_{\mathcal{I}^{c}}\bigr)\nonumber \\
 & \leq\exp\bigl\{ n\bigl(\sum_{j\in\mathcal{J}^{c}}r_{j}-\bigl(\sum_{j\in\mathcal{J}^{c}}H\left(V_{j}|V_{\mathcal{A}_{j}}\right)\nonumber \\
 & \qquad-H\left(V_{\mathcal{J}^{c}}|Y_{k}V_{\mathcal{J}}\right)-\left(|\mathcal{J}^{c}|+1\right)\delta\left(\epsilon'\right)\bigr)\bigr)\bigr\}.
\end{align}
Hence if $\sum_{j\in\mathcal{J}^{c}}r_{j}<\sum_{j\in\mathcal{J}^{c}}H\left(V_{j}|V_{\mathcal{A}_{j}}\right)-H\left(V_{\mathcal{J}^{c}}|Y_{k}V_{\mathcal{J}}\right)-\left(|\mathcal{J}^{c}|+1\right)\delta\left(\epsilon'\right)$
for all $\mathcal{J}\in\mathbb{J}$, then the third term of \eqref{eq:proberror}
tends to zero as $n\to\infty$. Letting $\epsilon'$ small enough,
this completes the proof of the inner bound.

It is worth noting that although the multivariate packing lemma (Lemma
\ref{lem:Packing}) has not been employed directly in the proof, the
derivation after \eqref{eq:-17} is essentially the same as that of
the multivariate packing lemma.

\subsection{Outer Bound $\mathcal{D}_{1}^{(o)}$}

For fixed $p_{U_{[1:L]}|S}$, we first introduce a set of auxiliary
random variables $U_{[1:L]}^{n}$ that follow the distribution $\prod_{i=1}^{n}p_{U_{[1:L]}|S}\left(u_{[1:L],i}|s_{i}\right)$.
Hence the Markov chains $U_{[1:L]}^{n}\rightarrow S^{n}\rightarrow X^{n}\rightarrow Y_{k}^{n}\rightarrow\hat{S}_{k}^{n},1\leq k\leq K$
hold. Consider that
\begin{align}
 & I\left(Y_{\mathcal{A}}^{n};U_{\mathcal{B}}^{n}|U_{\mathcal{C}}^{n}\right)\nonumber \\
 & =\sum_{t=1}^{n}I\left(Y_{\mathcal{A}}^{n};U_{\mathcal{B},t}|U_{\mathcal{C}}^{n}U_{\mathcal{B}}^{t-1}\right)\label{eq:-13}\\
 & =\sum_{t=1}^{n}H\left(U_{\mathcal{B},t}|U_{\mathcal{C}}^{n}U_{\mathcal{B}}^{t-1}\right)-H\left(U_{\mathcal{B},t}|U_{\mathcal{C}}^{n}U_{\mathcal{B}}^{t-1}Y_{\mathcal{A}}^{n}\right)\label{eq:-14-3-1-1}\\
 & =\sum_{t=1}^{n}H\left(U_{\mathcal{B},t}|U_{\mathcal{C},t}\right)-H\left(U_{\mathcal{B},t}|U_{\mathcal{C}}^{n}U_{\mathcal{B}}^{t-1}Y_{\mathcal{A}}^{n}\right)\label{eq:-3-1-1}\\
 & =\sum_{t=1}^{n}I\left(U_{\mathcal{B},t};U_{\mathcal{C}}^{n}U_{\mathcal{B}}^{t-1}Y_{\mathcal{A}}^{n}|U_{\mathcal{C},t}\right)\\
 & \geq\sum_{t=1}^{n}I\left(U_{\mathcal{B},t};\hat{S}_{\mathcal{A},t}|U_{\mathcal{C},t}\right)\\
 & =nI\left(U_{\mathcal{B},Q};\hat{S}_{\mathcal{A},Q}|U_{\mathcal{C},Q}Q\right)\label{eq:-2-1-1}\\
 & =nI\left(U_{\mathcal{B},Q};\hat{S}_{\mathcal{A},Q}Q|U_{\mathcal{C},Q}\right)\\
 & \geq nI\left(U_{\mathcal{B},Q};\hat{S}_{\mathcal{A},Q}|U_{\mathcal{C},Q}\right)\\
 & =nI\left(U_{\mathcal{B}};\hat{S}_{\mathcal{A}}|U_{\mathcal{C}}\right),\label{eq:lb}
\end{align}
where the time-sharing random variable $Q$ is defined to be uniformly
distributed $\left[1:n\right]$ and independent of all other random
variables, and in \eqref{eq:lb}, $U_{l}\triangleq U_{l,Q},\hat{S}_{k}\triangleq\hat{S}_{k,Q},1\leq l\leq L,1\leq k\leq K$.

On the other hand,
\begin{align}
 & I\left(Y_{\mathcal{A}}^{n};U_{\mathcal{B}}^{n}|U_{\mathcal{C}}^{n}\right)\nonumber \\
 & =\sum_{t=1}^{n}I\left(Y_{\mathcal{A},t};U_{\mathcal{B}}^{n}|U_{\mathcal{C}}^{n}Y_{\mathcal{A}}^{t-1}\right)\\
 & =nI\left(Y_{\mathcal{A},Q};U_{\mathcal{B}}^{n}|U_{\mathcal{C}}^{n}Y_{\mathcal{A}}^{Q-1}Q\right)\label{eq:-2-1-1-1-1-1}\\
 & =nI\left(Y_{\mathcal{A}};\tilde{U}_{\mathcal{B}}|\tilde{U}_{\mathcal{C}}\tilde{Y}_{\mathcal{A}}\right),\label{eq:ub}
\end{align}
Set $Y_{k}\triangleq Y_{k,Q},\tilde{U}_{l}\triangleq U_{l}^{n},\tilde{Y}_{k}\triangleq Y_{k}^{Q-1}Q,1\leq l\leq L,1\leq k\leq K$,
then combining \eqref{eq:lb} and \eqref{eq:ub} gives us the outer
bound $\mathcal{D}_{1}^{(o)}$.

\subsection{Outer Bound $\mathcal{D}_{2}^{(o)}$}

For fixed $p_{U_{[1:L]}|Y_{[1:K]}}$, we first introduce a set of
auxiliary random variables $U_{[1:L]}^{n}$ that follow distribution
$\prod_{i=1}^{n}p_{U_{[1:L]}|Y_{[1:K]}}\left(u_{[1:L],i}|y_{[1:K],i}\right)$.
Hence the Markov chains $S^{n}\rightarrow X^{n}\rightarrow Y_{[1:K]}^{n}\rightarrow U_{[1:L]}^{n}$
hold. Note that different from the proof of $\mathcal{R}_{1}^{(o)}$,
the auxiliary random variables $U_{[1:L]}^{n}$ here is introduced
at receiver sides, and $p_{Y_{[1:K]}U_{[1:L]}|X}=p_{U_{[1:L]}|Y_{[1:K]}}p_{Y_{[1:K]}|X}$
forms a new memoryless broadcast channel. Consider that
\begin{align}
 & I\left(S^{n};Y_{\mathcal{B}}^{n}U_{\mathcal{B}'}^{n}|Y_{\mathcal{C}}^{n}U_{\mathcal{C}'}^{n}\right)\nonumber \\
 & \leq I\left(X^{n};Y_{\mathcal{B}}^{n}U_{\mathcal{B}'}^{n}|Y_{\mathcal{C}}^{n}U_{\mathcal{C}'}^{n}\right)\\
 & =\sum_{t=1}^{n}I\left(X^{n};Y_{\mathcal{B},t}U_{\mathcal{B}',t}|Y_{\mathcal{C}}^{n}U_{\mathcal{C}'}^{n}Y_{\mathcal{B}}^{t-1}U_{\mathcal{B}'}^{t-1}\right)\\
 & =\sum_{t=1}^{n}H\left(Y_{\mathcal{B},t}U_{\mathcal{B}',t}|Y_{\mathcal{C}}^{n}U_{\mathcal{C}'}^{n}Y_{\mathcal{B}}^{t-1}U_{\mathcal{B}'}^{t-1}\right)\nonumber \\
 & \qquad-H\left(Y_{\mathcal{B},t}U_{\mathcal{B}',t}|Y_{\mathcal{C}}^{n}U_{\mathcal{C}'}^{n}Y_{\mathcal{B}}^{t-1}U_{\mathcal{B}'}^{t-1}X^{n}\right)\label{eq:-14-3-1-1-1}\\
 & \leq\sum_{t=1}^{n}H\left(Y_{\mathcal{B},t}U_{\mathcal{B}',t}|Y_{\mathcal{C},t}U_{\mathcal{C}',t}\right)\nonumber \\
 & \qquad-H\left(Y_{\mathcal{B},t}U_{\mathcal{B}',t}|Y_{\mathcal{C},t}U_{\mathcal{C}',t}X_{t}\right)\label{eq:-3-1-1-1}\\
 & =\sum_{t=1}^{n}I\left(Y_{\mathcal{B},t}U_{\mathcal{B}',t};X_{t}|Y_{\mathcal{C},t}U_{\mathcal{C}',t}\right)\\
 & =nI\left(Y_{\mathcal{B},Q}U_{\mathcal{B}',Q};X_{Q}|Y_{\mathcal{C},Q}U_{\mathcal{C}',Q}Q\right)\label{eq:-2-1-1-1}\\
 & =nH\left(Y_{\mathcal{B},Q}U_{\mathcal{B}',Q}|Y_{\mathcal{C},Q}U_{\mathcal{C}',Q}Q\right)\nonumber \\
 & \qquad-nH\left(Y_{\mathcal{B},Q}U_{\mathcal{B}',Q}|Y_{\mathcal{C},Q}U_{\mathcal{C}',Q}X_{Q}Q\right)\\
 & \leq nH\left(Y_{\mathcal{B},Q}U_{\mathcal{B}',Q}|Y_{\mathcal{C},Q}U_{\mathcal{C}',Q}\right)\nonumber \\
 & \qquad-nH\left(Y_{\mathcal{B},Q}U_{\mathcal{B}',Q}|Y_{\mathcal{C},Q}U_{\mathcal{C}',Q}X_{Q}\right)\label{eq:-12}\\
 & =nI\left(Y_{\mathcal{B},Q}U_{\mathcal{B}',Q};X_{Q}|Y_{\mathcal{C},Q}U_{\mathcal{C}',Q}\right)\\
 & =nI\left(X;Y_{\mathcal{B}}U_{\mathcal{B}'}|Y_{\mathcal{C}}U_{\mathcal{C}'}\right),\label{eq:-4-1-1-1}
\end{align}
where \eqref{eq:-3-1-1-1} follows from the Markov chain $U_{[1:L]}^{t-1}U_{[1:L],t+1}^{n}Y_{[1:K]}^{t-1}Y_{[1:K],t+1}^{n}X^{t-1}X_{t+1}^{n}\rightarrow X_{t}\rightarrow Y_{[1:K],t}U_{[1:L],t}$
and the fact conditioning reduces entropy, \eqref{eq:-12} follows
from the Markov chain $Q\rightarrow X_{Q}\rightarrow Y_{[1:K],Q}U_{[1:L],Q}$
and the fact conditioning reduces entropy, the time-sharing random
variable $Q$ is defined to be uniformly distributed $\left[1:n\right]$
and independent of all other random variables, and in \eqref{eq:-4-1-1-1},
$U_{l}\triangleq U_{l,Q},Y_{k}\triangleq Y_{k,Q},X\triangleq X_{Q},1\leq l\leq L,1\leq k\leq K$.

On the other hand,
\begin{align}
 & I\left(S^{n};Y_{\mathcal{B}}^{n}U_{\mathcal{B}'}^{n}|Y_{\mathcal{C}}^{n}U_{\mathcal{C}'}^{n}\right)\nonumber \\
 & =\sum_{t=1}^{n}I\left(S_{t};Y_{\mathcal{B}}^{n}U_{\mathcal{B}'}^{n}|Y_{\mathcal{C}}^{n}U_{\mathcal{C}'}^{n}S^{t-1}\right)\\
 & =nI\left(S_{Q};Y_{\mathcal{B}}^{n}U_{\mathcal{B}'}^{n}|Y_{\mathcal{C}}^{n}U_{\mathcal{C}'}^{n}S^{Q-1}Q\right)\label{eq:-2-1-1-1-1}\\
 & =nI\left(S;\tilde{Y}_{\mathcal{B}}\tilde{U}_{\mathcal{B}'}|\tilde{Y}_{\mathcal{C}}\tilde{U}_{\mathcal{C}'}\right),\label{eq:-4-1-1-1-1}
\end{align}
Set $S\triangleq S_{Q},\tilde{U}_{l}\triangleq U_{l}^{n}S^{Q-1}Q,\tilde{Y}_{k}\triangleq Y_{k}^{n}S^{Q-1}Q,1\leq l\leq L,1\leq k\leq K$,
then combining \eqref{eq:-4-1-1-1} and \eqref{eq:-4-1-1-1-1} gives
us the outer bound $\mathcal{D}_{2}^{(o)}$.

\section{Proof of Theorem \ref{thm:AdmissibleRegionBBSI}\label{sec:broadcast-BernoulliSI}}

Observe that if there is no information transmitted over the channel,
receiver $k$ could produce a reconstruction within the distortion
$\beta_{k}$. Hence we only need consider the case of $D_{[1:K]}$
with
\begin{equation}
D_{k}\leq\beta_{k},1\le k\le K.\label{eq:-55}
\end{equation}

For the Wyner-Ziv binary broadcast with bandwidth mismatch case (the
bandwidth mismatch factor $b$), Theorem \ref{thm:AdmissibleRegionSI-DBC}
states that if $D_{[1:K]}$ is achievable, then there exists some
pmf $p_{V_{K}|S}p_{V_{K-1}|V_{K}}\cdots p_{V_{1}|V_{2}}$ and functions
$\hat{s}_{k}\left(v_{k},z_{k}\right),1\le k\le K$ such that
\begin{equation}
\mathbb{E}d\left(S,\hat{S}_{k}\right)=\mathbb{P}\left(\hat{S}_{k}\oplus S=1\right)\le D_{k},\label{eq:-11-2}
\end{equation}
and for any pmf $p_{U_{K-1}|S}p_{U_{K-2}|U_{K-1}}\cdots p_{U_{1}|U_{2}}$,
\begin{equation}
\frac{1}{b}\left(I\left(V_{k};U_{k}|U_{k-1}Z_{k}\right):k\in[1:K]\right)\in\mathcal{R}_{\mathsf{BBC}}\label{eq:-12-2}
\end{equation}
holds, where the capacity of binary broadcast channel $\mathcal{R}_{\mathsf{BBC}}$
is given in \eqref{eq:binarycapacity} \cite{Bergmans}.

Define the sets
\begin{equation}
\mathcal{A}_{k}=\left\{ v_{k}:\hat{s}_{k}\left(v_{k},0\right)=\hat{s}_{k}\left(v_{k},1\right)\right\} ,1\le k\le K,
\end{equation}
so that their complements
\begin{equation}
\mathcal{A}_{k}^{c}=\left\{ v_{k}:\hat{s}_{k}\left(v_{k},0\right)\neq\hat{s}_{k}\left(v_{k},1\right)\right\} ,1\le k\le K.
\end{equation}
By hypothesis,
\begin{align}
 & \mathbb{E}d\left(S,\hat{S}_{k}\right)\nonumber \\
 & =\mathbb{P}(V_{k}\in\mathcal{A}_{k})\mathbb{E}\left[d\left(S,\hat{S}_{k}\right)|V_{k}\in\mathcal{A}_{k}\right]\nonumber \\
 & \qquad+\mathbb{P}(V_{k}\in\mathcal{A}_{k}^{c})\mathbb{E}\left[d\left(S,\hat{S}_{k}\right)|V_{k}\in\mathcal{A}_{k}^{c}\right]\\
 & \le D_{k}.\label{eq:-34}
\end{align}

We first show that
\begin{equation}
\mathbb{E}\left[d\left(S,\hat{S}_{k}\right)|V_{k}\in\mathcal{A}_{k}^{c}\right]\geq\beta_{k}.\label{eq:-26}
\end{equation}
To do this, we write
\begin{align}
 & \mathbb{E}\left[d\left(S,\hat{S}_{k}\right)|V_{k}\in\mathcal{A}_{k}^{c}\right]\nonumber \\
 & =\sum_{v_{k}\in\mathcal{A}_{k}^{c}}\frac{\mathbb{P}(V_{k}=v_{k})}{\mathbb{P}(V_{k}\in\mathcal{A}_{k}^{c})}\mathbb{E}\left[d\left(S,\hat{S}_{k}\right)|V_{k}=v_{k}\right].
\end{align}
If $v_{k}\in\mathcal{A}_{k}^{c}$ and $\hat{s}_{k}\left(v_{k},0\right)=0$
then $\hat{s}_{k}\left(v_{k},1\right)=1$. Therefore, for such $v_{k}$,
\begin{align}
 & \mathbb{E}\left[d\left(S,\hat{S}_{k}\right)|V_{k}=v_{k}\right]\nonumber \\
 & =\mathbb{P}\left(Z_{k}=0,S=1|V_{k}=v_{k}\right)\nonumber \\
 & \qquad+\mathbb{P}\left(Z_{k}=1,S=0|V_{k}=v_{k}\right)\\
 & =\mathbb{P}\left(S=1|V_{k}=v_{k}\right)\mathbb{P}\left(Z_{k}=0|S=1\right)\nonumber \\
 & \qquad+\mathbb{P}\left(S=0|V_{k}=v_{k}\right)\mathbb{P}\left(Z_{k}=1|S=0\right)\label{eq:-25}\\
 & =\beta_{k},\label{eq:-27}
\end{align}
where \eqref{eq:-25} follows that $Z_{k}\rightarrow S\rightarrow V_{k}$
forms a Markov chain. If $v_{k}\in\mathcal{A}_{k}^{c}$ but $\hat{s}_{k}\left(v_{k},0\right)=1$,
then for such $v_{k}$,
\begin{align}
\mathbb{E}\left[d\left(S,\hat{S}_{k}\right)|V_{k}=v_{k}\right] & =1-\beta_{k}\geq\beta_{k},\label{eq:-33}
\end{align}
since $\beta_{k}\leq\frac{1}{2}$. Therefore, \eqref{eq:-26} follows
from \eqref{eq:-27} and \eqref{eq:-33}.

Now we write
\begin{align}
 & \mathbb{E}\left[d\left(S,\hat{S}_{k}\right)|V_{k}\in\mathcal{A}_{k}\right]\nonumber \\
 & =\sum_{v_{k}\in\mathcal{A}_{k}}\frac{\mathbb{P}(V_{k}=v_{k})}{\mathbb{P}(V_{k}\in\mathcal{A}_{k})}\mathbb{E}\left[d\left(S,\hat{S}_{k}\right)|V_{k}=v_{k}\right],
\end{align}
and define $g_{k}\left(v_{k}\right)\triangleq\hat{s}_{k}\left(v_{k},0\right),\lambda_{v_{k}}\triangleq\frac{\mathbb{P}(V_{k}=v_{k})}{\mathbb{P}(V_{k}\in\mathcal{A}_{k})},\mu_{k}\triangleq\mathbb{P}(V_{k}\in\mathcal{A}_{k})$,
\begin{align}
d_{v_{k}} & \triangleq\mathbb{E}\left[d\left(S,\hat{S}_{k}\right)|V_{k}=v_{k}\right]\\
 & =\mathbb{P}\left(S\neq g_{k}\left(v_{k}\right)|V_{k}=v_{k}\right),
\end{align}
then utilizing \eqref{eq:-34} and \eqref{eq:-26}, we have
\begin{equation}
d'_{k}\triangleq\mu_{k}\sum_{v_{k}\in\mathcal{A}_{k}}\lambda_{v_{k}}d_{v_{k}}+\left(1-\mu_{k}\right)\beta_{k}\le D_{k}.\label{eq:-42}
\end{equation}

Next we will show
\begin{align}
 & I\left(V_{k};U_{k}|Z_{k}\right)\nonumber \\
 & \geq\frac{\beta_{k}-D_{k}}{\beta_{k}-\alpha_{k}}\bigl(H_{2}\left(\beta_{k}\star\tau_{k}\right)\nonumber \\
 & \qquad-\left(H_{4}\left(\alpha_{k},\beta_{k},\tau_{k}\right)-H_{2}\left(\alpha_{k}\star\beta_{k}\right)\right)\bigr).\label{eq:-6}
\end{align}

Choose $U_{K-1}=S\oplus E'_{K-1}$ and $U_{k}=U_{k+1}\oplus E'_{k},1\le k\le K-2$,
where $E'_{k}\sim\textrm{Bern}\left(\tau'_{k}\right)$ is independent
of all the other random variables. Define $E_{k}=E'_{K-1}\oplus E'_{K-2}\oplus\cdots\oplus E'_{k}\sim\textrm{Bern}\left(\tau_{k}\right)$
with $\tau_{k}=\tau'_{K-1}\star\tau'_{K-2}\star\cdots\star\tau'_{k}$.
Then
\begin{align}
 & I\left(V_{k};U_{k}|Z_{k}\right)\nonumber \\
 & =H\left(U_{k}|Z_{k}\right)-H\left(U_{k}|V_{k},Z_{k}\right)\\
 & =H_{2}\left(\beta_{k}\star\tau_{k}\right)-\mu_{k}\sum_{v_{k}\in\mathcal{A}_{k}}\lambda_{v_{k}}H\left(U_{k}|Z_{k},V_{k}=v_{k}\right)\nonumber \\
 & \qquad-\left(1-\mu_{k}\right)\sum_{v_{k}\in\mathcal{A}_{k}^{c}}\frac{\mathbb{P}(V_{k}=v_{k})}{\mathbb{P}(V_{k}\in\mathcal{A}_{k}^{c})}H\left(U_{k}|Z_{k},V_{k}=v_{k}\right).\label{eq:-36}
\end{align}
For fixed $v_{k}$, define a set of random variables $\left(V'_{k},S',U'_{k},Z'_{k}\right)\sim1\left\{ v'_{k}=v_{k}\right\} p_{SU_{k}Z_{k}|V_{k}}\left(s',u'_{k},z'_{k}|v'_{k}\right)$,
then $H\left(U'_{k}Z'_{k}|V'_{k}\right)=H\left(U_{k}Z_{k}|V_{k}=v_{k}\right)$
and $H\left(Z'_{k}|V'_{k}\right)=H\left(Z_{k}|V_{k}=v_{k}\right)$.
Since $p_{SU_{k}Z_{k}|V_{k}}$ satisfies
\begin{align}
 & p_{SU_{k}Z_{k}|V_{k}}\left(s',u'_{k},z'_{k}|v'_{k}\right)\nonumber \\
 & =p_{S|V_{k}}\left(s'|v'_{k}\right)p_{Z_{k}|S}\left(z'_{k}|s'\right)p_{U_{k}|S}\left(u'_{k}|s'\right),
\end{align}
it holds that $Z'_{k}=S'\oplus B{}_{k},U'_{k}=S'\oplus E{}_{k}$.
Hence $Z'_{k}\oplus U'_{k}=B{}_{k}\oplus E{}_{k}$.

For fixed $v_{k}$, consider
\begin{align}
 & H\left(U_{k}|Z_{k},V_{k}=v_{k}\right)\nonumber \\
 & =H\left(U_{k}Z_{k}|V_{k}=v_{k}\right)-H\left(Z_{k}|V_{k}=v_{k}\right)\\
 & =H\left(U'_{k}Z'_{k}|V'_{k}\right)-H\left(Z'_{k}|V'_{k}\right)\\
 & =H\left(U'_{k}|Z'_{k}V'_{k}\right)\\
 & =H\left(U'_{k}\oplus Z'_{k}|Z'_{k}V'_{k}\right)\\
 & =H\left(B_{k}\oplus E{}_{k}|Z'_{k}V'_{k}\right)\\
 & \leq H\left(B_{k}\oplus E{}_{k}\right)\\
 & =H_{2}\left(\beta_{k}\star\tau_{k}\right).\label{eq:-37}
\end{align}
Combine \eqref{eq:-36} and \eqref{eq:-37}, then we have
\begin{align}
 & I\left(V_{k};U_{k}|Z_{k}\right)\nonumber \\
 & \geq H_{2}\left(\beta_{k}\star\tau_{k}\right)-\mu_{k}\sum_{v_{k}\in\mathcal{A}_{k}}\lambda_{v_{k}}H\left(U_{k}|Z_{k},V_{k}=v_{k}\right)\nonumber \\
 & \qquad-\left(1-\mu_{k}\right)H_{2}\left(\beta_{k}\star\tau_{k}\right)\\
 & =\mu_{k}H_{2}\left(\beta_{k}\star\tau_{k}\right)-\mu_{k}\sum_{v_{k}\in\mathcal{A}_{k}}\lambda_{v_{k}}H\left(U_{k}|Z_{k},V_{k}=v_{k}\right).\label{eq:-38}
\end{align}

Now we consider the second term of \eqref{eq:-38}.
\begin{align}
 & \sum_{v_{k}\in\mathcal{A}_{k}}\lambda_{v_{k}}H\left(U_{k}|Z_{k},V_{k}=v_{k}\right)\nonumber \\
 & =\sum_{v_{k}\in\mathcal{A}_{k}}\lambda_{v_{k}}\left(H\left(U_{k}Z_{k}|V_{k}=v_{k}\right)-H\left(Z_{k}|V_{k}=v_{k}\right)\right)\\
 & =\sum_{v_{k}\in\mathcal{A}_{k}}\lambda_{v_{k}}\left(H_{4}\left(d_{v_{k}},\beta_{k},\tau_{k}\right)-H_{2}\left(d_{v_{k}}\star\beta_{k}\right)\right)\label{eq:-39}\\
 & =\sum_{v_{k}\in\mathcal{A}_{k}}\lambda_{v_{k}}G_{1}\left(d_{v_{k}},\beta_{k},\tau_{k}\right),\label{eq:-41}
\end{align}
where the function $H_{4}\left(x,y,z\right)$ is defined in \eqref{eq:H4}
and
\begin{equation}
G_{1}\left(x,y,z\right)\triangleq H_{4}\left(x,y,z\right)-H_{2}\left(x\star y\right).
\end{equation}
Equality \eqref{eq:-39} follows from calculating the entropies according
to the definition.

Now we show that $G_{1}\left(x,y,z\right)$ is concave in $x$. To
do this, we consider
\begin{align}
 & \frac{\partial^{2}}{\partial x^{2}}G_{1}\left(x,y,z\right)\nonumber \\
 & =-\frac{\left(yz-\overline{y}\overline{z}\right)^{2}}{xyz+\overline{x}\overline{y}\overline{z}}-\frac{\left(\overline{y}z-y\overline{z}\right)^{2}}{x\overline{y}z+\overline{x}y\overline{z}}-\frac{\left(y\overline{z}-\overline{y}z\right)^{2}}{xy\overline{z}+\overline{x}\overline{y}z}\nonumber \\
 & \qquad-\frac{\left(\overline{y}\overline{z}-yz\right)^{2}}{x\overline{y}\overline{z}+\overline{x}yz}+\frac{\left(y-\overline{y}\right)^{2}}{x\overline{y}+\overline{x}y}+\frac{\left(y-\overline{y}\right)^{2}}{xy+\overline{x}\overline{y}}\\
 & =-\left(\frac{\left(yz-\overline{y}\overline{z}\right)^{2}}{xyz+\overline{x}\overline{y}\overline{z}}+\frac{\left(y\overline{z}-\overline{y}z\right)^{2}}{xy\overline{z}+\overline{x}\overline{y}z}-\frac{\left(y-\overline{y}\right)^{2}}{xy+\overline{x}\overline{y}}\right)\nonumber \\
 & \qquad-\left(\frac{\left(\overline{y}z-y\overline{z}\right)^{2}}{x\overline{y}z+\overline{x}y\overline{z}}+\frac{\left(\overline{y}\overline{z}-yz\right)^{2}}{x\overline{y}\overline{z}+\overline{x}yz}-\frac{\left(y-\overline{y}\right)^{2}}{x\overline{y}+\overline{x}y}\right)\\
 & \leq0,\label{eq:-40}
\end{align}
where \eqref{eq:-40} follows from the following inequality
\begin{align}
\frac{a_{1}^{2}}{b_{1}}+\frac{a_{2}^{2}}{b_{2}} & =\frac{1}{b_{1}+b_{2}}\left(b_{1}+b_{2}\right)\left(\frac{a_{1}^{2}}{b_{1}}+\frac{a_{2}^{2}}{b_{2}}\right)\\
 & =\frac{1}{b_{1}+b_{2}}\left(a_{1}^{2}+a_{2}^{2}+\frac{b_{2}a_{1}^{2}}{b_{1}}+\frac{b_{1}a_{2}^{2}}{b_{2}}\right)\\
 & \geq\frac{1}{b_{1}+b_{2}}\left(a_{1}^{2}+a_{2}^{2}+2a_{1}a_{2}\right)\\
 & =\frac{\left(a_{1}+a_{2}\right)^{2}}{b_{1}+b_{2}},
\end{align}
for $b_{1},b_{2}>0$ and arbitrary real numbers $a_{1},a_{2}$. \eqref{eq:-40}
implies $G_{1}\left(x,y,z\right)$ is concave in $x$.

Then combining the concavity of $G_{1}\left(x,y,z\right)$ with \eqref{eq:-38}
and \eqref{eq:-41}, we have
\begin{align}
 & I\left(V_{k};U_{k}|Z_{k}\right)\nonumber \\
 & \geq\mu_{k}\left(H_{2}\left(\beta_{k}\star\tau_{k}\right)-G_{1}\left(\sum_{v_{k}\in\mathcal{A}_{k}}\lambda_{v_{k}}d_{v_{k}},\beta_{k},\tau_{k}\right)\right)\\
 & =\mu_{k}\left(H_{2}\left(\beta_{k}\star\tau_{k}\right)-G_{1}\left(\alpha_{k},\beta_{k},\tau_{k}\right)\right)
\end{align}
where
\begin{equation}
\alpha_{k}\triangleq\sum_{v_{k}\in\mathcal{A}_{k}}\lambda_{v_{k}}d_{v_{k}}.\label{eq:alphak}
\end{equation}
From \eqref{eq:-42}, $\alpha_{k}$ satisfies
\begin{equation}
\mu_{k}\alpha_{k}+\left(1-\mu_{k}\right)\beta_{k}\le D_{k}.\label{eq:-43}
\end{equation}
Combine \eqref{eq:-43} with $D_{k}\leq\beta_{k}$ (i.e., \eqref{eq:-55}),
then we have
\begin{equation}
0\leq\alpha_{k}\le D_{k}\leq\beta_{k}.\label{eq:-51}
\end{equation}
Therefore,
\begin{align}
 & I\left(V_{k};U_{k}|Z_{k}\right)\nonumber \\
 & \geq\frac{\beta_{k}-D_{k}}{\beta_{k}-\alpha_{k}}\left(H_{2}\left(\beta_{k}\star\tau_{k}\right)-G_{1}\left(\alpha_{k},\beta_{k},\tau_{k}\right)\right)\\
 & =\frac{\beta_{k}-D_{k}}{\beta_{k}-\alpha_{k}}\bigl(H_{2}\left(\beta_{k}\star\tau_{k}\right)\nonumber \\
 & \qquad-\left(H_{4}\left(\alpha_{k},\beta_{k},\tau_{k}\right)-H_{2}\left(\alpha_{k}\star\beta_{k}\right)\right)\bigr),\label{eq:-53-1}
\end{align}
i.e., \eqref{eq:-6} holds.

Next we will show
\begin{align}
 & I\left(V_{k};U_{k}|U_{k-1}Z_{k}\right)\nonumber \\
 & \geq\frac{\beta_{k}-D_{k}}{\beta_{k}-\alpha_{k}}\Bigl(H_{2}\left(\beta_{k}\star\tau_{k}\right)-H_{2}\left(\beta_{k}\star\tau_{k-1}\right)\nonumber \\
 & \qquad-\left(H_{4}\left(\alpha_{k},\beta_{k},\tau_{k}\right)-H_{4}\left(\alpha_{k},\beta_{k},\tau_{k-1}\right)\right)\Bigr).\label{eq:secinequity}
\end{align}

Consider
\begin{align}
 & I\left(V_{k};U_{k}|U_{k-1}Z_{k}\right)\nonumber \\
 & =H\left(U_{k}|U_{k-1}Z_{k}\right)-H\left(U_{k}|U_{k-1}Z_{k}V_{k}\right)\\
 & =H\left(U_{k-1}|U_{k}\right)+H\left(U_{k}|Z_{k}\right)\nonumber \\
 & \qquad-H\left(U_{k-1}|Z_{k}\right)-H\left(U_{k}|U_{k-1}Z_{k}V_{k}\right)\\
 & =H_{2}\left(\tau'_{k-1}\right)+H_{2}\left(\beta_{k}\star\tau_{k}\right)\nonumber \\
 & \qquad-H_{2}\left(\beta_{k}\star\tau_{k-1}\right)-H\left(U_{k}|U_{k-1}Z_{k}V_{k}\right).\label{eq:-46}
\end{align}
Write the last term as
\begin{align}
 & H\left(U_{k}|U_{k-1}Z_{k}V_{k}\right)\nonumber \\
 & =-\left(1-\mu_{k}\right)\sum_{v_{k}\in\mathcal{A}_{k}^{c}}\frac{\mathbb{P}(V_{k}=v_{k})}{\mathbb{P}(V_{k}\in\mathcal{A}_{k}^{c})}H\left(U_{k}|U_{k-1},Z_{k},V_{k}=v_{k}\right)\nonumber \\
 & \qquad-\mu_{k}\sum_{v_{k}\in\mathcal{A}_{k}}\lambda_{v_{k}}H\left(U_{k}|U_{k-1},Z_{k},V_{k}=v_{k}\right).\label{eq:-47}
\end{align}
For fixed $v_{k}$, define $\left(V'_{k},S',U'_{k},U'_{k-1},Z'_{k}\right)\sim1\left\{ v'_{k}=v_{k}\right\} p_{SU_{k}U_{k-1}Z_{k}|V_{k}}\left(s',u'_{k},u'_{k-1},z'_{k}|v'_{k}\right)$.
Since
\begin{align}
 & p_{SU_{k}U_{k-1}Z_{k}|V_{k}}\left(s',u'_{k},u'_{k-1},z'_{k}|v'_{k}\right)\nonumber \\
 & =p_{S|V_{k}}\left(s'|v'_{k}\right)p_{Z_{k}|S}\left(z'_{k}|s'\right)p_{U_{k}|S}\left(u'_{k}|s'\right)\nonumber \\
 & \qquad\times p_{U_{k-1}|U_{k}}\left(u'_{k-1}|u'_{k}\right),
\end{align}
we have $Z'_{k}=S'\oplus B{}_{k},U'_{k}=S'\oplus E{}_{k},U'_{k-1}=U'_{k}\oplus E'{}_{k-1}$.
Hence $Z'_{k}\oplus U'_{k}=B{}_{k}\oplus E{}_{k},Z'_{k}\oplus U'_{k-1}=B{}_{k}\oplus E{}_{k-1}$.
Similar to the derivation for $H\left(U_{k}|U_{k-1},V_{k}=v_{k}\right)$,
we can write
\begin{align}
 & H\left(U_{k}|U_{k-1},Z_{k},V_{k}=v_{k}\right)\nonumber \\
 & =H\left(U'_{k}|U'_{k-1}Z'_{k}V'_{k}\right)\\
 & =H\left(U'_{k}\oplus Z'_{k}|U'_{k-1}\oplus Z'_{k},Z'_{k},V'_{k}\right)\\
 & \leq H\left(U'_{k}\oplus Z'_{k}|U'_{k-1}\oplus Z'_{k}\right)\\
 & =H\left(B_{k}\oplus E{}_{k}|B{}_{k}\oplus E{}_{k-1}\right)\\
 & =H\left(B_{k}\oplus E{}_{k}\right)+H\left(B_{k}\oplus E{}_{k-1}|B{}_{k}\oplus E{}_{k}\right)\nonumber \\
 & \qquad-H\left(B_{k}\oplus E{}_{k-1}\right)\\
 & =H_{2}\left(\tau'_{k-1}\right)+H_{2}\left(\beta_{k}\star\tau_{k}\right)-H_{2}\left(\beta_{k}\star\tau_{k-1}\right).\label{eq:-45}
\end{align}
Combine \eqref{eq:-46}, \eqref{eq:-47} and \eqref{eq:-45}, then
we have
\begin{align}
 & I\left(V_{k};U_{k}|U_{k-1}Z_{k}\right)\nonumber \\
 & \geq\mu_{k}\left(H_{2}\left(\tau'_{k-1}\right)+H_{2}\left(\beta_{k}\star\tau_{k}\right)-H_{2}\left(\beta_{k}\star\tau_{k-1}\right)\right)\nonumber \\
 & \qquad-\mu_{k}\sum_{v_{k}\in\mathcal{A}_{k}}\lambda_{v_{k}}H\left(U_{k}|U_{k-1},Z_{k},V_{k}=v_{k}\right).\label{eq:-48}
\end{align}
Consider the last term of \eqref{eq:-48},
\begin{align}
 & \sum_{v_{k}\in\mathcal{A}_{k}}\lambda_{v_{k}}H\left(U_{k}|U_{k-1},Z_{k},V_{k}=v_{k}\right)\nonumber \\
 & =\sum_{v_{k}\in\mathcal{A}_{k}}\lambda_{v_{k}}\Bigl(H\left(U_{k}|Z_{k},V_{k}=v_{k}\right)\nonumber \\
 & \qquad+H\left(U_{k-1}|U_{k},Z_{k},V_{k}=v_{k}\right)\nonumber \\
 & \qquad-H\left(U_{k-1}|Z_{k},V_{k}=v_{k}\right)\Bigr)\\
 & =\sum_{v_{k}\in\mathcal{A}_{k}}\lambda_{v_{k}}\Bigl(H\left(U_{k},Z_{k}|V_{k}=v_{k}\right)+H_{2}\left(\tau'_{k-1}\right)\nonumber \\
 & \qquad-H\left(U_{k-1},Z_{k}|V_{k}=v_{k}\right)\Bigr)\\
 & =H_{2}\left(\tau'_{k-1}\right)\nonumber \\
 & \qquad+\sum_{v_{k}\in\mathcal{A}_{k}}\lambda_{v_{k}}\left(H_{4}\left(d_{v_{k}},\beta_{k},\tau_{k}\right)-H_{4}\left(d_{v_{k}},\beta_{k},\tau_{k-1}\right)\right)\label{eq:-49}\\
 & =H_{2}\left(\tau'_{k-1}\right)+\sum_{v_{k}\in\mathcal{A}_{k}}\lambda_{v_{k}}G_{2}\left(d_{v_{k}},\beta_{k},\tau_{k},\tau_{k-1}\right),\label{eq:-50}
\end{align}
where \eqref{eq:-49} is by directly calculating the entropies according
to the definition, and
\begin{equation}
G_{2}\left(x,y,z,t\right)\triangleq H_{4}\left(x,y,z\right)-H_{4}\left(x,y,t\right).
\end{equation}
Note that function $G_{1}\left(x,y,z\right)$ is a special case of
function $G_{2}\left(x,y,z,t\right)$ given $t=\frac{1}{2}$, i.e.,
\begin{equation}
G_{1}\left(x,y,z\right)=G_{2}\left(x,y,z,\frac{1}{2}\right).
\end{equation}

Now we show that $G_{2}\left(x,y,z,t\right)$ is concave in $x$ when
$0\leq z\leq t\leq\frac{1}{2}$, which generalizes the concavity of
$G_{1}\left(x,y,z\right)$. To do this, we consider
\begin{align}
 & \frac{\partial^{2}}{\partial x^{2}}G_{2}\left(x,y,z,t\right)\nonumber \\
 & =-\frac{\left(yz-\overline{y}\overline{z}\right)^{2}}{xyz+\overline{x}\overline{y}\overline{z}}-\frac{\left(\overline{y}z-y\overline{z}\right)^{2}}{x\overline{y}z+\overline{x}y\overline{z}}-\frac{\left(y\overline{z}-\overline{y}z\right)^{2}}{xy\overline{z}+\overline{x}\overline{y}z}-\frac{\left(\overline{y}\overline{z}-yz\right)^{2}}{x\overline{y}\overline{z}+\overline{x}yz}\nonumber \\
 & \quad+\frac{\left(yt-\overline{y}\overline{t}\right)^{2}}{xyt+\overline{x}\overline{y}\overline{t}}+\frac{\left(\overline{y}t-y\overline{t}\right)^{2}}{x\overline{y}t+\overline{x}y\overline{t}}+\frac{\left(y\overline{t}-\overline{y}t\right)^{2}}{xy\overline{t}+\overline{x}\overline{y}t}+\frac{\left(\overline{y}\overline{t}-yt\right)^{2}}{x\overline{y}\overline{t}+\overline{x}yt},
\end{align}
and
\begin{align}
 & \frac{\partial}{\partial t}\left(\frac{\partial^{2}}{\partial x^{2}}G_{2}\left(x,y,z,t\right)\right)\nonumber \\
 & =\frac{\partial}{\partial t}\left(\frac{\left(yt-\overline{y}\overline{t}\right)^{2}}{xyt+\overline{x}\overline{y}\overline{t}}+\frac{\left(y\overline{t}-\overline{y}t\right)^{2}}{xy\overline{t}+\overline{x}\overline{y}t}\right)\nonumber \\
 & \qquad+\frac{\partial}{\partial t}\left(\frac{\left(\overline{y}t-y\overline{t}\right)^{2}}{x\overline{y}t+\overline{x}y\overline{t}}+\frac{\left(\overline{y}\overline{t}-yt\right)^{2}}{x\overline{y}\overline{t}+\overline{x}yt}\right)\\
 & =\frac{-y^{2}\cdot\overline{y}^{2}\cdot\left(xy+\overline{x}\overline{y}\right)\cdot\left(1-2t\right)}{\left(xyt+\overline{x}\overline{y}\overline{t}\right)^{2}\left(xy\overline{t}+\overline{x}\overline{y}t\right)^{2}}\nonumber \\
 & \qquad+\frac{-y^{2}\cdot\overline{y}^{2}\cdot\left(x\overline{y}+\overline{x}y\right)\cdot\left(1-2t\right)}{\left(x\overline{y}t+\overline{x}y\overline{t}\right)^{2}\left(x\overline{y}\overline{t}+\overline{x}yt\right)^{2}}.
\end{align}
Hence for $0\leq t\leq\frac{1}{2}$,
\begin{align}
\frac{\partial}{\partial t}\left(\frac{\partial^{2}}{\partial x^{2}}G_{2}\left(x,y,z,t\right)\right) & \leq0,
\end{align}
i.e., $\frac{\partial^{2}}{\partial x^{2}}G_{2}\left(x,y,z,t\right)$
is decreasing in $t$. Then we have for $0\leq z\leq t\leq\frac{1}{2}$,
\begin{equation}
\frac{\partial^{2}}{\partial x^{2}}G_{2}\left(x,y,z,t\right)\leq\frac{\partial^{2}}{\partial x^{2}}G_{2}\left(x,y,z,z\right)=0.
\end{equation}
It implies $G_{2}\left(x,y,z,t\right)$ is concave in $x$ when $0\leq z\leq t\leq\frac{1}{2}$.

Combining \eqref{eq:-48} and \eqref{eq:-50}, and utilizing the concavity
of $G_{2}\left(x,y,z,t\right)$, we have
\begin{align}
 & I\left(V_{k};U_{k}|U_{k-1}Z_{k}\right)\nonumber \\
 & \geq\mu_{k}\Bigl(H_{2}\left(\beta_{k}\star\tau_{k}\right)-H_{2}\left(\beta_{k}\star\tau_{k-1}\right)\nonumber \\
 & \qquad-G_{2}\bigl(\sum_{v_{k}\in\mathcal{A}_{k}}\lambda_{v_{k}}d_{v_{k}},\beta_{k},\tau_{k},\tau_{k-1}\bigr)\Bigr)\\
 & =\mu_{k}\Bigl(H_{2}\left(\beta_{k}\star\tau_{k}\right)-H_{2}\left(\beta_{k}\star\tau_{k-1}\right)\nonumber \\
 & \qquad-G_{2}\left(\alpha_{k},\beta_{k},\tau_{k},\tau_{k-1}\right)\Bigr)
\end{align}
where $\alpha_{k}$ is given by \eqref{eq:alphak} and satisfies \eqref{eq:-43}
and \eqref{eq:-51}. Therefore,
\begin{align}
 & I\left(V_{k};U_{k}|U_{k-1}Z_{k}\right)\nonumber \\
 & \geq\frac{\beta_{k}-D_{k}}{\beta_{k}-\alpha_{k}}\Bigl(H_{2}\left(\beta_{k}\star\tau_{k}\right)-H_{2}\left(\beta_{k}\star\tau_{k-1}\right)\nonumber \\
 & \qquad-G_{2}\left(\alpha_{k},\beta_{k},\tau_{k},\tau_{k-1}\right)\Bigr)\\
 & =\frac{\beta_{k}-D_{k}}{\beta_{k}-\alpha_{k}}\Bigl(H_{2}\left(\beta_{k}\star\tau_{k}\right)-H_{2}\left(\beta_{k}\star\tau_{k-1}\right)\nonumber \\
 & \qquad-\left(H_{4}\left(\alpha_{k},\beta_{k},\tau_{k}\right)-H_{4}\left(\alpha_{k},\beta_{k},\tau_{k-1}\right)\right)\Bigr),\label{eq:-52}
\end{align}
i.e., \eqref{eq:secinequity} holds.

Combining \eqref{eq:-12-2}, \eqref{eq:-6} and \eqref{eq:secinequity}
gives Theorem \ref{thm:AdmissibleRegionBBSI}.

\section{Proof of Theorem \ref{thm:AdmissibleRegionGGSI}\label{sec:broadcast-GaussianSI}}

For the Wyner-Ziv Gaussian broadcast with bandwidth mismatch case
(the bandwidth mismatch factor $b$), Theorem \ref{thm:AdmissibleRegionSI-DBC}
states that if $D_{[1:K]}$ is achievable, then there exist some pmf
$p_{V_{K}|S}p_{V_{K-1}|V_{K}}\cdots p_{V_{1}|V_{2}}$ and functions
$\hat{s}_{k}\left(v_{k},z_{k}\right),1\le k\le K$ such that
\begin{equation}
\mathbb{E}d\left(S,\hat{S}_{k}\right)\le D_{k},\label{eq:-11-1}
\end{equation}
and for any pmf $p_{U_{K-1}|S}p_{U_{K-2}|U_{K-1}}\cdots p_{U_{1}|U_{2}}$,
\begin{equation}
\frac{1}{b}\left(I\left(V_{k};U_{k}|U_{k-1}Z_{k}\right):k\in[1:K]\right)\in\mathcal{R}_{\mathsf{GBC}}\label{eq:-12-1}
\end{equation}
holds, where the capacity of Gaussian broadcast channel $\mathcal{R}_{\mathsf{GBC}}$
is given in \eqref{eq:Gaussiancapacity}.

Choose $U_{K-1}=S+E'_{K-1}$ and $U_{k}=U_{k+1}+E'_{k},1\le k\le K-2$,
where $E'_{k}\sim\mathcal{N}\left(0,\tau'_{k}\right)$ is independent
of all the other random variables. Define $E_{k}=\sum_{j=k}^{K-1}E'_{j}\sim\mathcal{N}\left(0,\tau{}_{k}\right)$
with $\tau_{k}=\sum_{j=k}^{K-1}\tau'_{j}$. Then
\begin{align}
 & I\left(V_{1};U_{1}|Z_{1}\right)\nonumber \\
 & \geq I\left(\hat{S}_{1};U_{1}|Z_{1}\right)\\
 & =h\left(U_{1}|Z_{1}\right)-h\left(U_{1}|\hat{S}_{1}Z_{1}\right)\\
 & =h\left(U_{1}|Z_{1}\right)-h\left(U_{1}-\hat{S}_{1}|\hat{S}_{1}Z_{1}\right)\\
 & \geq h\left(U_{1}|Z_{1}\right)-h\left(U_{1}-\hat{S}_{1}\right)\\
 & \geq\frac{1}{2}\log\left(2\pi e\left(\beta_{1}+\tau_{1}\right)\right)-\frac{1}{2}\log\left(2\pi e\left(D_{1}+\tau_{1}\right)\right)\label{eq:-15}\\
 & =\frac{1}{2}\log\frac{\beta_{1}+\tau_{1}}{D_{1}+\tau_{1}},\label{eq:-24}
\end{align}
where \eqref{eq:-15} follows from the fact that a Gaussian distribution
maximizes the differential entropy for a given second moment.

On the other hand,
\begin{align}
 & I\left(V_{k};U_{k}|U_{k-1}Z_{k}\right)\nonumber \\
 & \geq I\left(\hat{S}_{k};U_{k}|U_{k-1}Z_{k}\right)\\
 & =I\left(\hat{S}_{k};U_{k}|Z_{k}\right)-I\left(\hat{S}_{k};U_{k-1}|Z_{k}\right)\\
 & =h\left(U_{k}|Z_{k}\right)-h\left(U_{k-1}|Z_{k}\right)\nonumber \\
 & \qquad+h\left(U_{k-1}|Z_{k}\hat{S}_{k}\right)-h\left(U_{k}|Z_{k}\hat{S}_{k}\right).\label{eq:-18}
\end{align}
The first two terms of \eqref{eq:-18}
\begin{equation}
h\left(U_{k}|Z_{k}\right)-h\left(U_{k-1}|Z_{k}\right)=\frac{1}{2}\log\frac{\beta_{k}+\tau_{k}}{\beta_{k}+\tau_{k-1}}.\label{eq:-21}
\end{equation}
The last two terms of \eqref{eq:-18}
\begin{align}
 & h\left(U_{k-1}|Z_{k}\hat{S}_{k}\right)-h\left(U_{k}|Z_{k}\hat{S}_{k}\right)\nonumber \\
 & =h\left(U_{k-1}|Z_{k}\hat{S}_{k}\right)-h\left(U_{k}|Z_{k}\hat{S}_{k}E'_{k-1}\right)\\
 & =h\left(U_{k-1}|Z_{k}\hat{S}_{k}\right)-h\left(U_{k-1}|Z_{k}\hat{S}_{k}E'_{k-1}\right)\\
 & =I\left(U_{k-1};E'_{k-1}|Z_{k}\hat{S}_{k}\right)\\
 & =h\left(E'_{k-1}\right)-h\left(E'_{k-1}|Z_{k}\hat{S}_{k}U_{k-1}\right)\\
 & =h\left(E'_{k-1}\right)-h\left(E'_{k-1}|Z_{k},\hat{S}_{k},U_{k-1}-\hat{S}_{k}\right)\\
 & \geq h\left(E'_{k-1}\right)-h\left(E'_{k-1}|U_{k-1}-\hat{S}_{k}\right)\\
 & =I\left(E'_{k-1};S-\hat{S}_{k}+E_{k}+E'_{k-1}\right)\\
 & \geq\frac{1}{2}\log\frac{D_{k}+\tau_{k-1}}{D_{k}+\tau_{k}},\label{eq:-20}
\end{align}
where \eqref{eq:-20} is by applying the mutual information game result
that Gaussian noise is the worst additive noise under a variance constraint
\cite[p. 298, Problem 9.21]{Cover91} and taking $E'_{k-1}$ as the
channel input.

Combining \eqref{eq:-18}, \eqref{eq:-21} and \eqref{eq:-20}, we
have
\begin{align}
I\left(V_{k};U_{k}|U_{k-1}Z_{k}\right) & \geq\frac{1}{2}\log\frac{\left(\beta_{k}+\tau_{k}\right)\left(D_{k}+\tau_{k-1}\right)}{\left(\beta_{k}+\tau_{k-1}\right)\left(D_{k}+\tau_{k}\right)}.\label{eq:-23}
\end{align}

\eqref{eq:-12-1}, \eqref{eq:-24} and \eqref{eq:-23} imply Theorem
\ref{thm:AdmissibleRegionGGSI}.

\section*{Acknowledgements}

The authors would like to thank Prof. Jun Chen for sharing submitted
versions of his papers \cite{Khezeli14,Khezeli} with us.

\begin{IEEEbiographynophoto}{Lei Yu}
received the B.E. and Ph.D. degrees, both in electronic engineering,
from University of Science and Technology of China (USTC) in 2010
and 2015, respectively. From 2015 to 2017, he was a postdoctoral researcher
at the Department of Electronic Engineering and Information Science
(EEIS), USTC. Currently, he is a research fellow at the Department
of Electrical and Computer Engineering, National University of Singapore.
His research interests include information theory, probability theory,
and security.
\end{IEEEbiographynophoto}

\begin{IEEEbiographynophoto}{Houqiang Li}
(SM'12) received the B.S., M.Eng., and Ph.D. degrees in electronic
engineering from the University of Science and Technology of China,
Hefei, China, in 1992, 1997, and 2000, respectively, where he is currently
a Professor with the Department of Electronic Engineering and Information
Science.

His research interests include video coding and communication, multimedia
search, image/video analysis. He has authored and co-authored over
100 papers in journals and conferences. He served as an Associate
Editor of IEEE Transactions
on Circuits and Systems for Video Technology from 2010 to 2013. He was the recipient of the Best Paper
Award for Visual Communications and Image Processing (VCIP) in 2012,
the recipient of the Best Paper Award for International Conference
on Internet Multimedia Computing and Service (ICIMCS) in 2012, the
recipient of the Best Paper Award for the International Conference
on Mobile and Ubiquitous Multimedia from ACM (ACM MUM) in 2011, and
a senior author of the Best Student Paper of the 5th International
Mobile Multimedia Communications Conference (MobiMedia) in 2009.
\end{IEEEbiographynophoto}

\begin{IEEEbiographynophoto}{Weiping Li}
(F'00) received his B.S. degree from University of Science and Technology
of China (USTC) in 1982, and his M.S. and Ph.D. degrees from Stanford
University in 1983 and 1988 respectively, all in electrical engineering.
He was an Assistant Professor, Associate Professor with Tenure, and
Professor of Lehigh University from 1987 to 2001. He worked in several
high-tech companies in the Silicon Valley with technical and management
responsibilities from 1998 to 2010. He has been a Professor in USTC
since 2010. He served as the Editor-in-Chief of IEEE Transactions
on Circuits and Systems for Video Technology, a founding member of
the Board of Directors of MPEG-4 Industry Forum, and several other
positions in IEEE and SPIE. He is an IEEE Fellow.
\end{IEEEbiographynophoto}

\end{document}